\documentclass[twocolumn]{aastex62}
\usepackage{amsmath}
\usepackage{amssymb}
\usepackage{mathbbol}
\usepackage{dsfont}
\usepackage{amsfonts} 
\usepackage{color}
\usepackage{amsopn}
\usepackage{graphicx,longtable,times,threeparttable,multirow}
\usepackage{longtable,times,threeparttable,multirow}
\usepackage{float}

\DeclareGraphicsExtensions{.eps,.eps.gz,.epsi, .jpg, .png}

\submitjournal{ApJ Supplement}
\shorttitle{LAMOST DR5 stellar labels with $data$-$driven$ $Payne$}
\shortauthors{Xiang et al.}

\begin{document}
\title
{Abundance Estimates for 16 Elements in 6 Million Stars from LAMOST DR5 Low-Resolution Spectra}

\author{Maosheng Xiang}
\affil{Max-Planck Institute for Astronomy, K\"onigstuhl 17, D-69117 Heidelberg, Germany}
\email{{\rm email: } {\em mxiang@mpia.de}}
\author{Yuan-Sen Ting}\thanks{Hubble fellow}
\affil{Institute for Advanced Study, Princeton, NJ 08540, USA}
\affil{Department of Astrophysical Sciences, Princeton University, Princeton, NJ 08544, USA}
\affil{Observatories of the Carnegie Institution of Washington, 813 Santa
Barbara Street, Pasadena, CA 91101, USA}
\author{Hans-Walter Rix}
\affil{Max-Planck Institute for Astronomy, K\"onigstuhl 17, D-69117 Heidelberg, Germany}
\author{Nathan Sandford}
\affil{Department of Astronomy and Theoretical Astrophysics, University of California Berkeley, Berkeley, CA 94720, USA}
\author{Sven Buder}
\affil{Max-Planck Institute for Astronomy, K\"onigstuhl 17, D-69117 Heidelberg, Germany}
\author{Karin Lind}
\affil{Max-Planck Institute for Astronomy, K\"onigstuhl 17, D-69117 Heidelberg, Germany}
\affil{Department of Physics and Astronomy, Uppsala University, Box 516, SE-751 20 Uppsala, Sweden} 
\author{Xiao-Wei Liu}
\affil{South-Western Institute for Astronomy Research, Yunnan University, Kunming 650500, P. R. China}
\author{Jian-Rong Shi}
\affil{Key Laboratory of Optical Astronomy, National Astronomical Observatories, Chinese Academy of Sciences, Beijing 100012, P. R. China}
\affil{University of Chinese Academy of Sciences, Beijing 100049, P. R. China}
\author{Hua-Wei Zhang}
\affil{Department of Astronomy, Peking University, Beijing 100871, P. R. China}

\begin{abstract}{We present the determination of stellar parameters and individual elemental abundances for 6 million stars from $\sim$8 million low-resolution ($R\sim1800$) spectra from LAMOST DR5. This is based on a modeling approach that we dub $The$ $Data$--$Driven$ $Payne$ ($DD$--$Payne$), which inherits essential ingredients from both {\it The Payne} \citep{Ting2019} and $The$ $Cannon$ \citep{Ness2015}. It is a data-driven model that incorporates constraints from theoretical spectral models to ensure the derived abundance estimates are physically sensible. Stars in LAMOST DR5 that are in common with either GALAH DR2 or APOGEE DR14 are used to train a model that delivers stellar parameters ($T_{\rm eff}$, $\log g$, $V_{\rm mic}$) and abundances for 16 elements (C, N, O, Na, Mg, Al, Si, Ca, Ti, Cr, Mn, Fe, Co, Ni, Cu, and Ba) when applied to LAMOST spectra. Cross-validation and repeat observations suggest that, for ${\rm S/N}_{\rm pix}\ge 50$, the typical internal abundance precision is 0.03--0.1\,dex for the majority of these elements, with 0.2--0.3\,dex for Cu and Ba, and the internal precision of $T_{\rm eff}$ and $\log g$ is better than 30\,K and 0.07\,dex, respectively. Abundance systematics at the $\sim$0.1\,dex level are present in these estimates, but are inherited from the high-resolution surveys' training labels.  For some elements, GALAH provides more robust training labels, for others, APOGEE. We provide flags to guide the quality of the label determination and to identify binary/multiple stars in LAMOST DR5. The abundance catalogs are publicly accessible via \href{url}{http://dr5.lamost.org/doc/vac}.}
\end{abstract}

\section{Introduction} \label{introduction}
Galactic archaeology is at present being propelled by a number of completed or ongoing large-scale spectroscopic surveys: e.g. RAVE \citep{Steinmetz2006}, SEGUE \citep{Yanny2009}, LAMOST \citep{Deng2012, Zhao2012}, Gaia-ESO \citep{Gilmore2012}, GALAH \citep{De_Silva2015}, APOGEE \citep{Majewski2017}, and the $Gaia$ Radial Velocity Spectrometer \citep{Cropper2018}; as well as upcoming surveys such as SDSS-V \citep{Kollmeier2017}, 4MOST \citep{Feltzing2018, deJong2019}, and WEAVE \citep{Dalton2014}. These spectroscopic surveys, combined with the astrometric and photometric information from the $Gaia$ mission \citep{Prusti2016, Brown2018}, make it possible for us to obtain precise and accurate information for millions of stars in phase space or orbit space, along with estimates of age, mass, metallicity, and abundances for many elements, providing unprecedented opportunities to unravel the assemblage and evolution history of our Galaxy \citep[see e.g.,][]{Rix2013, Ting2015, Xiang2017, Xiang2018, Frankel2018, BH19}.   

These very same spectroscopic survey data sets are posing great data-analysis challenges in  rigorously and efficiently deriving stellar labels that are both precise and accurate. Beyond traditional spectroscopic data analysis, which is based on comparison with theoretical model spectra \citep[e.g., see a review by][]{Jofre2018, Nissen2018}, various data-driven methods have been recently put forward and are suggested to be effective ways to derive precise stellar parameters \citep{Re_Fiorentin2007, Bu2015, Li2015, LiuC2015, Yang2015, Wang2019} and elemental abundances \citep{Ness2015, Casey2016, Rix2016, Ho2017, Ting2017b, Ting2017a, Xiang2017a, Leung2019a, Zhang2019}. However, a potential shortcoming of data-driven model is that the determination of stellar labels is foremost a mathematical ``prediction'' or ``inference'', rather than a physically motivated ``measurement''; this limits the scientific interpretability of the results. This limitation can be important for low-resolution spectra due to blended features which could lead to the possibility of ``inferring'' elemental abundances from their astrophysical correlations. To address this drawback, \citet{Ting2017a} proposed to regularize the training process using priors of theoretical gradient spectra from stellar atmospheric models, enforcing that the models determine elemental abundances from the expected spectral features of individual elements. This idea, combined with {\it The Payne}, which is a flexible and efficient tool for simultaneous determination of numerous stellar labels with full spectral fitting \citep{Ting2019}, has been applied to LAMOST DR3 as a proof of concept, yielding 14 elemental abundances from a subset of LAMOST low-resolution ($R\sim$1800) spectra \citep{Ting2017a}. 

In this work, we further develop the approach laid out in \citet{Ting2017a} and apply it to the fifth data release of LAMOST (LAMOST DR5) to derive robust and physically sensible stellar labels, including effective temperature ($T_{\rm eff}$), surface gravity ($\log g$), micro-turbulence velocity ($V_{\rm mic}$), and abundance ratio [X/Fe] for 16 individual elements: C, N, O, Na, Mg, Al, Si, Ca, Ti, Cr, Mn, Fe, Co, Ni, Cu, and Ba. We adopt LAMOST stars that are in common with either GALAH or APOGEE as training sets for building a spectral model.  This model utilizes the neural network spectral interpolator and the fitting technique from the {\it The Payne}, and at the same time, it adopts physical gradient spectra from the Kurucz spectral model \citep{Kurucz1970, Kurucz1993, Kurucz2005} to regularize the training process. On that basis, we derive stellar labels for the entire LAMOST DR5 data set. Considering the modelling approach is a combination of the data-driven approach and {\it The Payne}, we will refer to it as the $Data$--$Driven$ $Payne$ or as the $Cannon$--$Payne$ $hybrid$ to acknowledge the pioneering contribution of $The$ $Cannon$ \citep{Ness2015} to data-driven spectroscopic data analysis. 

The LAMOST Galactic survey \citep{Deng2012, Zhao2012, Liu2014, Liu2015} is the first dedicated spectroscopic survey to obtain spectra of $\mathcal{O}(10^7)$ stars.\,\footnote{http://dr6.lamost.org, http://dr7.lamost.org} In its latest and final data release of the LAMOST Phase I (2011--2017) surveys, the LAMOST DR5 has released 9,027,634 optical ($\lambda$3700--9000{\AA}) spectra with $R\sim1800$, of which more than 90 percent are stellar spectra. The LAMOST DR5 provides classifications and radial velocity $V_{\rm r}$ measurements for the spectra. For about 5 million of them, LAMOST DR5 also provides the basic stellar parameters $T_{\rm eff}$, $\log g$, and [Fe/H] derived with the LAMOST stellar parameter pipeline \citep[LASP;][]{Wu2011, Luo2015}. Meanwhile, there is also a planned value-added catalog providing stellar labels derived with the LAMOST stellar parameter pipeline at Peking University \citep[LSP3;][]{Xiang2015b, Li2016, Xiang2017a}, including $V_{\rm r}$, $T_{\rm eff}$, $\log g$, [Fe/H], [$\alpha$/Fe], [C/Fe], [N/Fe], E(B-V), and distance for the LAMOST DR5 stars (Huang et al. in prep.).

Despite LAMOST's low spectral resolution ($R\sim1800$) it has been argued, and in part verified, by \citet{Ting2017b, Ting2017a, Ting2018b} that physically sensible abundances for $\gtrsim$10 individual elements should be derivable from its spectra with astrophysically interesting precision. This is what this paper sets out to do for LAMOST DR5. The results will greatly broaden the science products from LAMOST data, especially for Galactic archaeology.

The paper is organized as follows. Section\,\ref{method} introduces the $DD$--$Payne$ method in detail. Section\,\ref{trainingset} presents the training sets, and Section\,\ref{verification} presents the verification of the LAMOST DR5 labels with cross-validation data sets and repeat observations. Section\,\ref{lamostdr5abundancecatalog} shows some key results of the LAMOST DR5 labels, and introduces the error estimation and flag assignment. 
A summary is presented in Section\,\ref{summary}.
     
\section{Method: The {\it data-driven Payne}} \label{method}
{\it The Payne} is an efficient tool for determining numerous stellar labels simultaneously from the full observed spectra by fitting {\it ab initio} spectral models \citep{Ting2019}. At the core of {\it The Payne} is a flexible, non-parametric interpolator built with a neural network (NN) that has a functional form of
\begin{equation}
  f_{\lambda} = w \cdot \sigma\left(\tilde{w}_{\lambda}^i\sigma\left(w_{{\lambda}i}^kl_k + b_{{\lambda}i}\right)+\tilde{b}\right)+\bar{f}_{\lambda},
\end{equation}
where $\sigma(x)= 1/(1+e^{-x})$ is the Sigmoid function and $l$ are the stellar labels, i.e., stellar atmospheric parameters and elemental abundances. $\mathbf{w}$ and $\mathbf{b}$ are coefficient arrays to be optimized, the index $i$ is the number of neurons, and $k$ is the number of labels. Following \citet{Ting2019}, here the index summation is written in the Einstein convention, and the network function has two hidden layers, but one can extend the function to more hidden layers. It has been shown \citep{Ting2019} that {\it The Payne} is able to generate accurate model spectra (with typical uncertainty of $10^{-3}$ -- $10^{-2}$) in a wide range of parameters over 20-dimensional label space utilizing only a few thousand $\mathcal{O}(1000)$ training spectra of either high resolution (e.g., $R\sim22,500$ for  APOGEE) or low resolution (e.g. $R\sim1800$ for LAMOST).  
 
As {\it The Payne} fits {\it ab initio} synthetic spectra to data, the accuracy of the resultant stellar label is largely dependent on the accuracy of the synthetic model spectra. Their accuracy depends on a number of factors, such as the choice of a 1D hydrostatic or 3D hydrodynamic atmosphere, LTE or non-LTE line formation, atomic and molecular line list, etc. To overcome the impact from imperfect line list, one practical way is to build a spectral mask to isolate wavelength regions synthesized with poor accuracy for well-known reference stars \citep[see e.g.,][]{Ting2019}. Such a solution has been suggested to be efficient for high-resolution spectra, whereas it is expected to be less effective for low-resolution spectra because of the serious blending of spectral lines that may lead to the exclusion of too many informative pixels if a strong masking strategy is imposed. Fitting the observed spectra to synthetic spectra may also be affected by other systematic errors, which may be especially significant for the LAMOST spectra because of the complex instrument and observational conditions (from lunar to dark background). 

For a large-scale, low-resolution spectroscopic survey like the LAMOST survey, an alternative way to further make use of {\it The Payne} for robust and accurate determinations of stellar labels is to combine it with a data-driven approach, similar to \citet{Ting2017a}. This is the technique we adopt in the current work. The method inherits the neural network spectral interpolating algorithm and the spectral fitting technique from {\it The Payne}. However, instead of training on synthetic spectra, the neural network model is trained on LAMOST spectra using training sets built from stars common between LAMOST and the high-resolution spectroscopic surveys (GALAH and APOGEE; Section\,\ref{trainingset}), assuming the latter provide accurate stellar labels. To ensure that these determinations of stellar labels are physically motivated rather than simply a reflection of astrophysical correlations among different stellar labels, we introduce priors on the gradient spectra from stellar models. Specifically, for the optimization of the neural network, we adopt a loss function 
\begin{equation}
\begin{aligned}
&  \mathcal{L}(\{f_{\rm obs}(\lambda)\} | \mathbf{w},\mathbf{b}) =  \frac{1}{N_S}\sum_{i=1}^{N_S} 
         \frac{(f(\lambda | \boldsymbol{\ell}_{{\rm obs},i}) - f_{{\rm obs},i}(\lambda))^2}{\sigma^2_{{\rm obs},i}(\lambda)}  \\
 &   +\sum_{j=1}^{N_r} \mathbf{D}_{\rm scale} \cdot \sum_{k=1}^{N_l} | f^{\prime}(\lambda | \boldsymbol{\ell}_{\rm ref}) - f^{\prime}_{\rm {ab\,initio}}(\lambda | \boldsymbol{\ell}_{\rm ref})|
\end{aligned} 
\end{equation}
where $f_{\rm obs}(\lambda)$ and $\boldsymbol{\ell}_{\rm obs}$ are respectively the normalized flux of LAMOST spectra and the stellar labels from high-resolution spectra for the training stars. Similar to \citet{Ho2017}, we normalize the LAMOST and Kurucz spectra with local ``continuum" derived by smoothening the spectra with Gaussian kernels of 50\,{\AA} in width. Note that we only require the spectra to be normalized on a consistent scale, not necessarily to be normalized to the real/accurate continuum. $N_S$, $N_r$ and $N_l$ are the number of training spectra, reference stars and stellar labels, respectively. 

The regularization term is the absolute difference of gradient spectra $\partial{f(\lambda)}/\partial{l}$ between the data-driven model $f^{\prime}$ and the {\it ab initio} Kurucz model $f^{\prime}_{\rm {ab\,initio}}$ for a number of reference stars with $\boldsymbol{\ell}_{\rm ref}$. $\mathbf{D}_{\rm scale}$ is a vector of parameters which defines the strength of the prior term for each label (last term in Eq.\,2). It is set to be a vector so that we can define different strengths for different labels. Note that this loss function is slightly different from that of \citet{Ting2017a}, in which they adopted the logarithmic value of the relative difference of gradient spectra as the prior. Here we adopt the absolute difference to avoid numerical fluctuations due to small value of $f^{\prime}_{\rm {ab\,initio}}$. 

With appropriate constraints from the priors of physical gradient spectra, our neural network model is able to generate ``physically plausible" gradient spectra, thus ensuring that the stellar label determination is physically motivated. We still caution that our method may be affected by the imperfection of the model spectra, as our method conditions on the model gradient spectra. However, since we use only the flux gradients rather than the absolute flux from the model spectra, the impact of model uncertainty may be less severe as we expect that the stellar-model-based gradient spectra are less uncertain than the absolute flux. This should be   true considering that, in the case of unsaturated lines, the transition probability $\log gf$, which is an important source of uncertainty for the line list, has a simple monotonic relation with the logarithmic line strength, meaning that imperfect values of $\log gf$ adopted by the Kurucz model should not cause dramatic problematic gradient spectra.  

\begin{table}
\caption{List of fiducial reference stars of which the gradient spectra are used to regularize the training process. The $T_{\rm eff}$ and $\log g$ values are selected from stellar isochrones and span the HR diagram uniformly. The numbers in brackets show the step size adopted to evaluate the gradient spectra numerically.}
\label{table:table1}
\begin{tabular}{cclrl}
\hline
 No. &  $T_{\rm eff}$ ($\Delta$) & $\log g$ ($\Delta$) & [Fe/H]  ($\Delta$) & [X/Fe] ($\Delta$)  \\
\hline        
 1 &  5778 (200) & 4.53 ($-$0.25)  &  0.0 (0.5)     & 0.0 (0.5) \\
 2 &  6380 (200) & 4.06 ($+$0.25)       & $-$0.5 (0.5) & 0.0 (0.5)\\
 3 &  4791 (200) & 2.52 ($+$0.50)       &  0.0 (0.5)     & 0.0 (0.5)\\
 4 &  5097 (200) & 2.16 ($+$0.50)       & $-$1.5 (0.5) & 0.0 (0.5)\\
 5 &  5012 (200) & 4.62 ($-$0.25)  &  0.0 (0.5)     & 0.0 (0.5)\\
 6 &  4991 (200) & 4.68 ($-$0.25)  & $-$0.5 (0.5) & 0.0 (0.5)\\
 7 &  6500 (200) & 4.02 ($+$0.50)       & $-$1.5 (0.5) & 0.0 (0.5)\\
 8 &  5854 (200) & 4.68 ($-$0.25)  & $-$2.5 (0.5) & 0.0 (0.5)\\
 9 &  5183 (200) & 2.04 ($+$0.50)       & $-$2.5 (0.5) & 0.0 (0.5)\\
 10 & 5671 (200) & 3.56 ($+$0.25)      & $-$1.0 (0.5) & 0.0 (0.5)\\
 11 & 5200 (200) & 3.55 ($+$0.25)      &  0.0 (0.5)     & 0.0 (0.5)\\
 12 & 6003 (200) & 3.57 ($+$0.25)      & $-$2.0 (0.5) & 0.0 (0.5)\\
 13 & 4784 (200) & 4.77 ($-$0.25) & $-$1.5 (0.5) & 0.0 (0.5)\\
 14 & 4369 (200) & 1.65 ($+$0.50)        & $-$0.5 (0.5) & 0.0 (0.5)\\
 15 & 7183 (200) & 4.18 ($-$0.25) & $-$0.5 (0.5) & 0.0 (0.5)\\
 16 & 7059 (200) & 4.19 ($-$0.25) & $-$2.5 (0.5) & 0.0 (0.5)\\
 \hline
\end{tabular}
\begin{tablenotes}
\item[1]{All fiducial stars are assumed to have a micro-turbulence velocity $V_{\rm mic}$ of 1.5\,km/s and a step size $\Delta V_{\rm mic}$ of 1.0\,km/s.}
\end{tablenotes}
\end{table}

Sixteen reference stars with physical gradient spectra from the Kurucz spectral model \citep{Kurucz1970, Kurucz1993, Kurucz2005} are adopted for the regularization term. These reference stars cover a wide range in the parameter space, from $-2.5$ to $0.5$ in [Fe/H] and 4000 to 7000\,K in $T_{\rm eff}$. Their parameters ($T_{\rm eff}$, $\log g$ [Fe/H], [X/Fe]), as well as the step ${\Delta}l$ used to derive the physical gradient spectra are listed in Table\,\ref{table:table1}. More precisely, to calculate the gradient spectra $f^{\prime}$, we adopt the $\Delta{f}$, the difference of spectra for two sets of parameters differed by $\Delta{l}$. We explored a variety of values for $\mathbf{D}_{\rm scale}$, and decided on a value of 5 for $T_{\rm eff}$, $\log g$, and $V_{\rm mic}$ and 50 for {\rm [Fe/H]} and [X/Fe]. The choice is largely empirical. We found that a strong enough constraint for [X/Fe] is necessary to guarantee that our neural network model accurately reproduces the gradient spectra. However, imposing strong constraints on the gradient spectra of $T_{\rm eff}$ and $\log g$ may cause noticeable systematic bias in $T_{\rm eff}$ and $\log g$ determinations for stars near the boundary of the parameter space, particularly for cool dwarfs and metal-poor giants. This is not unexpected --- the gradient spectra of $T_{\rm eff}$ and $\log g$ from the Kurucz model atmosphere is likely to suffer non-negligible uncertainties at these regimes.  

The ATLAS12 and SYNTHE codes \citep{Kurucz2005, Castelli2005} are used for synthesizing the Kurucz spectral models. We adopt the solar abundance scale from \citet{Asplund2009}. The synthesized spectra at $R=300,000$ are convolved to the LAMOST resolution assuming the mean line spread function (LSF) of LAMOST spectra, and are normalized in the same way as the observed spectra. We have derived the LSF for all individual LAMOST spectra using arc lines and sky emission lines assuming Gaussian line profiles. The LSF is found to vary with wavelength, and also vary from fiber to fiber and plate to plate. Typical fiber-to-fiber variation of the LSF is sub-angstrom \citep[see e.g., Fig.\,28 of][]{Xiang2015b}. In this work, we adopt the wavelength-dependent LSF averaged from all the individual LAMOST spectra, while leaving the more precise analysis considering the fiber-to-fiber and plate-to-plate variations of LSF to a future work.

\begin{figure*}[htp]
\centering
\includegraphics[width=180mm]{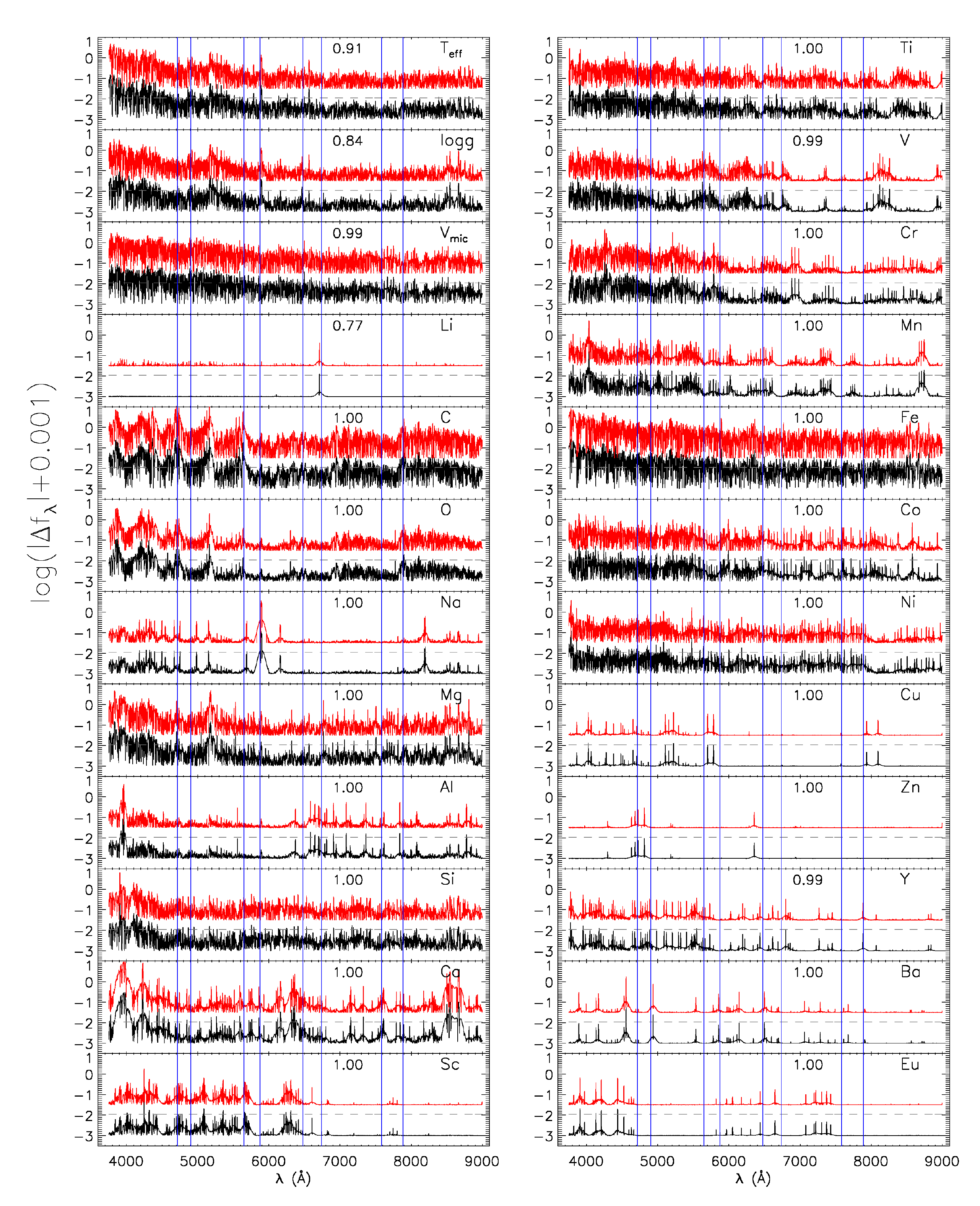}
\caption{Comparison of gradient spectra from the $DD$--$Payne$ using the LAMOST--GALAH training set (red) with those of the Kurucz spectral model (black) for a fiducial star with $T_{\rm eff}$ = 4821\,K, $\log g$ = 2.5, [Fe/H] = 0, and [X/Fe] = 0. The gradient spectra are generated based on normalized spectra, utilizing the same local-continuum normalization algorithm as that of \citet{Ho2017}. The very broad features presented in the gradient spectra, for instance, the one at $\sim$5900{\AA} in {\rm Na}, are effects due to the continuum-normalization. Correlation coefficients between the $DD$--$Payne$ and the Kurucz gradient spectra are marked at the top of each panel. The results show that, for this reference stellar label, the $DD$--$Payne$ reproduces the Kurucz model gradient spectra very well, demonstrating that the $DD$--$Payne$ measures stellar labels (in particular elemental abundances) from {\it ab initio} features, instead of drawing from astrophysical correlations among stellar labels. Blue vertical lines mark the wavelength windows of GALAH.}
\label{fig:Fig1}
\end{figure*}

To demonstrate how the gradient-spectrum prior works, Fig.\,\ref{fig:Fig1} shows a comparison of gradient spectra generated by the $DD$--$Payne$ using the LAMOST--GALAH training set (Section\,\ref{trainingset}) with those of the Kurucz spectral model for one of the reference stars in Table\,\ref{table:table1}. The excellent consistency of gradient spectra for all the labels indicates that the training process of the $DD$--$Payne$ works as expected. The figure also demonstrates that for most of the presented labels, there are plenty of spectral features in the LAMOST wavelength range (3700--9000{\AA}), even though they could be weak or blended. The numerous gradient features provide the possibility to derive robust abundance from the LAMOST spectra when fitting for the full spectra. Finally, it is well known that oxygen only has limited features in the optical. We note that most of the features shown in the oxygen gradient spectrum is not directly from {\rm O} lines but from {\rm C} and {\rm N} features via the {\rm CNO} atomic-molecular network. \citet{Ting2018b} has shown that the abundance of oxygen can be derived from these features, independent of the abundance of carbon and nitrogen.

We stress that, depending on the capacity of the neural network model, overfitting could still be a problem due to the non-convex nature of neural network optimization. In other words, while the gradient spectra of the reference stars are reproduced well, it is still possible that the $DD$--$Payne$ may fail to yield realistic gradient spectra for stars that are not located close to the reference stars in the $\sim$20-dimensional label space. Therefore, to evaluate the quality of the label determinations, we generate the $DD$--$Payne$ gradient spectra for each of the LAMOST DR5 stars based on the derived labels, and calculate the correlations with the Kurucz gradient spectra from the closest reference star. Here the closest reference star is defined with a distance metric $D:=\sqrt{(\Delta T_{\rm eff}/100{\rm K})^2+(\Delta\log g/0.2)^2+(\Delta{\rm [Fe/H]}/0.1)^2}$.

\begin{figure*}[htp]
\centering
\includegraphics[width=180mm]{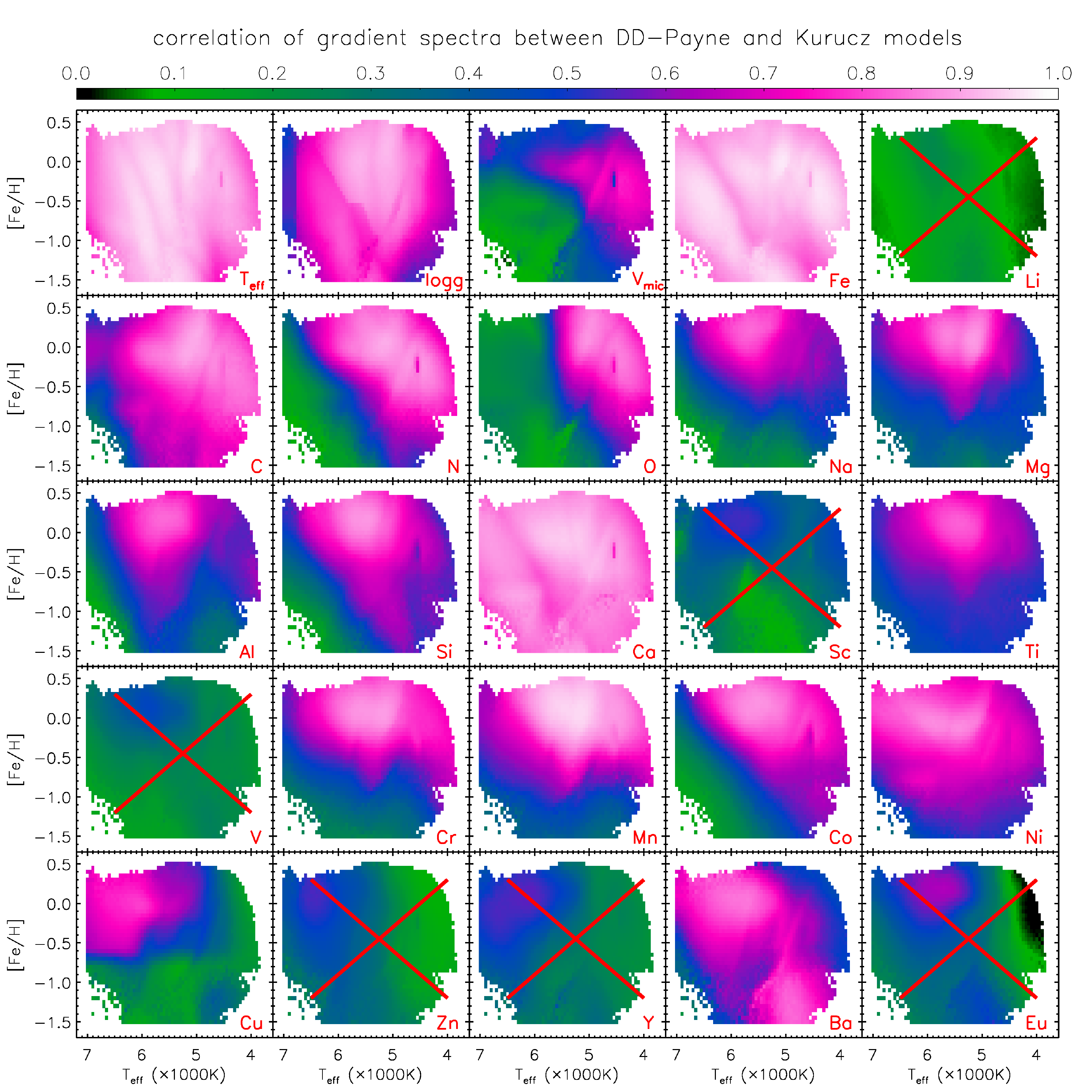}
\caption{Color-coded correlation coefficients between the gradient spectra of the best-fit $DD$--$Payne$ model and those of the Kurucz models for the closest reference star. The correlation is shown for LAMOST DR5 stars across the $T_{\rm eff}$--[Fe/H] plane, and the color scale indicates the median value of correlation coefficients for all stars in the bin. For Li, Na, Mg, Al, Si, Sc, Ti, V, Cr, Mn, Co, Zn, Y, Ba, and Eu, the $DD$-$Payne$ model was built from the LAMOST--GALAH training set, while for $T_{\rm eff}$, log\,$g$, $V_{\rm mic}$, C, N, O, Ca, Fe, Ni, and Cu, from the LAMOST--APOGEE training set (see Section\,\ref{trainingset}). Small values ($\lesssim 0.5$) indicate that much of the model's abundance ``predictions'' are determined with information from parts of the spectrum that are deemed uninformative by the theoretical models. For Li, Sc, V, Zn, Y, and Eu, when extrapolating from the reference stellar labels (see e.g., Fig.\,\ref{fig:Fig1}), the $DD$--$Payne$ models for most stars show little gradient correlation with the Kurucz models. The lack of gradient correlation implies that abundances for those elements are likely derived from indirect astrophysical correlations with other labels; they are therefore eliminated from the current work for further analysis.}
\label{fig:Fig2}
\end{figure*}

Fig.\,\ref{fig:Fig2} presents the median value of the correlation coefficients for stars across the $T_{\rm eff}$ and [Fe/H] plane. It shows that the values can change strongly among different elements and vary, to a different degree, with $T_{\rm eff}$ and [Fe/H] for individual elements. The variation is probably a consequence of the intrinsic change of the strength of the spectral features, and may also be partly due to imperfections of the current method. The method is limited by the imperfect abundance precision of the training stars in the metal-poor case and the simplistic neural network architecture which might not be fully adequate for such a high-dimensional space which covers a wide range of parameter values ($3800<T_{\rm eff}<7000$\,K, $0<\log g<5$, $-2.5<{\rm [Fe/H]}<0.5$). For elements Li, Sc, V, Y, and Eu, the figure demonstrates that only a few stars show good correlations (e.g. $>0.5$), indicating that the $DD$--$Payne$ abundance for these elements are most likely derived through astrophysical correlations with other labels. Therefore, we choose to exclude them for further analysis. For the remaining labels, we assign flags to the results based on their location in the $T_{\rm eff}$, $\log g$ and [Fe/H] space. We cut the results if the correlation coefficient is smaller than a predefined critical value (see Section\,\ref{consistenctofgradient}). Finally, to better visualize the $DD$--$Payne$ gradient spectra for a given value of correlation coefficient, we will show in the Appendix the gradient spectra of stars that have correlation coefficients of 0.6 and 0.9 as examples. 

\begin{figure*}[htp]
\centering
\includegraphics[width=180mm]{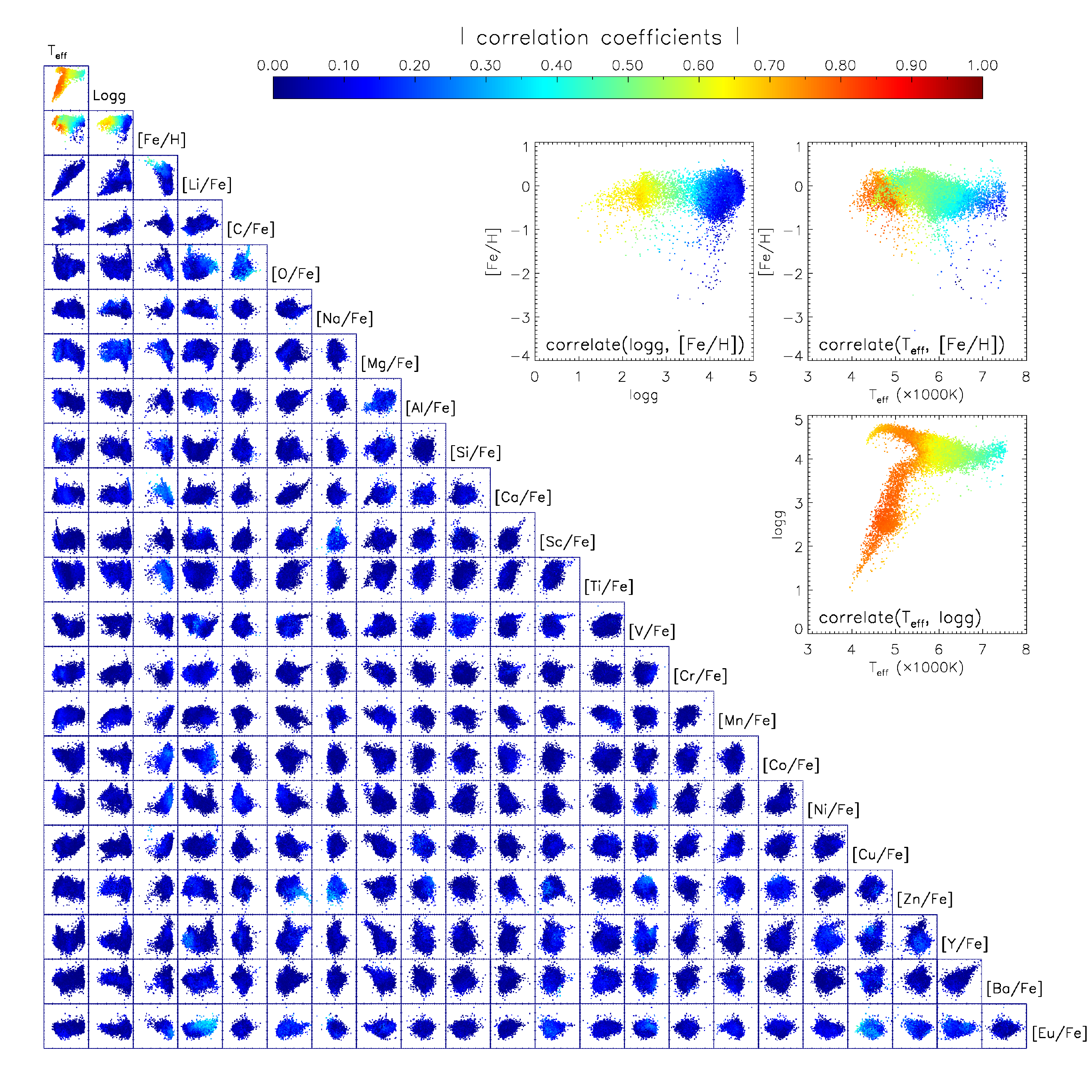}
\caption{Covariances among different labels of the gradient spectra resulting from the $DD$--$Payne$ models. The spectral models are trained using the LAMOST--GALAH/APOGEE overlapping sample (see Fig.\,2), and the empirical gradients are evaluated individually at the estimated stellar label value for each star. The plot shows a random subsample of 15,000 stars drawn from the LAMOST DR5 dataset. Each dot in the figure represents a star and their corresponding stellar label values (e.g., the $T_{\rm eff}-\log g$ subplot shows the Kiel diagram). The color scale indicates the value of correlation coefficients between two labels. The panels among $T_{\rm eff}$, $\log g$ and [Fe/H] are magnified at the top-right corner. The results show the well-known strong correlations between $T_{\rm eff}$ and $\log g$, and between $T_{\rm eff}$ and [Fe/H] for stars across the whole parameter space, as well as between $\log g$ and [Fe/H] for giants, which cause slight degeneracy for the parameter determination. However, the color scale of the figure illustrates that the correlations among [X/Fe] are small. Furthermore, the distributions of data for all subplots involving only [X/Fe] do not exhibit strong trends, indicating that we are inferring elemental abundances through {\it ab initio} features, instead of astrophysical correlations.}
\label{fig:Fig3}
\end{figure*}

With the neural network serving as spectral interpolator, the $DD$--$Payne$ adopts a $\chi^2$ minimization algorithm to determine all stellar labels simultaneously from any given target spectrum. Given the low resolution of the LAMOST spectra, the labels may be degenerate or very co-variant, if the spectral features of two labels are encoded in the same pixels. This can be quantified by the cross-label covariances of the model spectral gradients at the best fit labels of each star.  To explore the level of such covariances, we plot these label covariances in Fig.\,\ref{fig:Fig3}. The figure shows the result for a sample of randomly drawn LAMOST stars covering a wide range of stellar parameters. The figure shows that there are indeed strong correlations ($>0.5$) between $T_{\rm eff}$ and $\log g$, $T_{\rm eff}$ and [Fe/H] for stars in almost the full parameter space, and between $\log g$ and [Fe/H] for giants, as has been well-known \citep[see e.g.,][]{Ting2017b}. However, we do not observe strong correlations for [X/Fe]. These results are consistent with the conclusion of \citet{Ting2017b}, which is based on examinations of H-band synthetic spectra at $R=1000$. In our case, such correlations are even smaller ($\lesssim0.1$), likely a consequence of different wavelength windows (full optical wavelength range instead of the restricted H-band).  

\section{The training sets} \label{trainingset}
\begin{figure*}
\centering
\includegraphics[width=180mm]{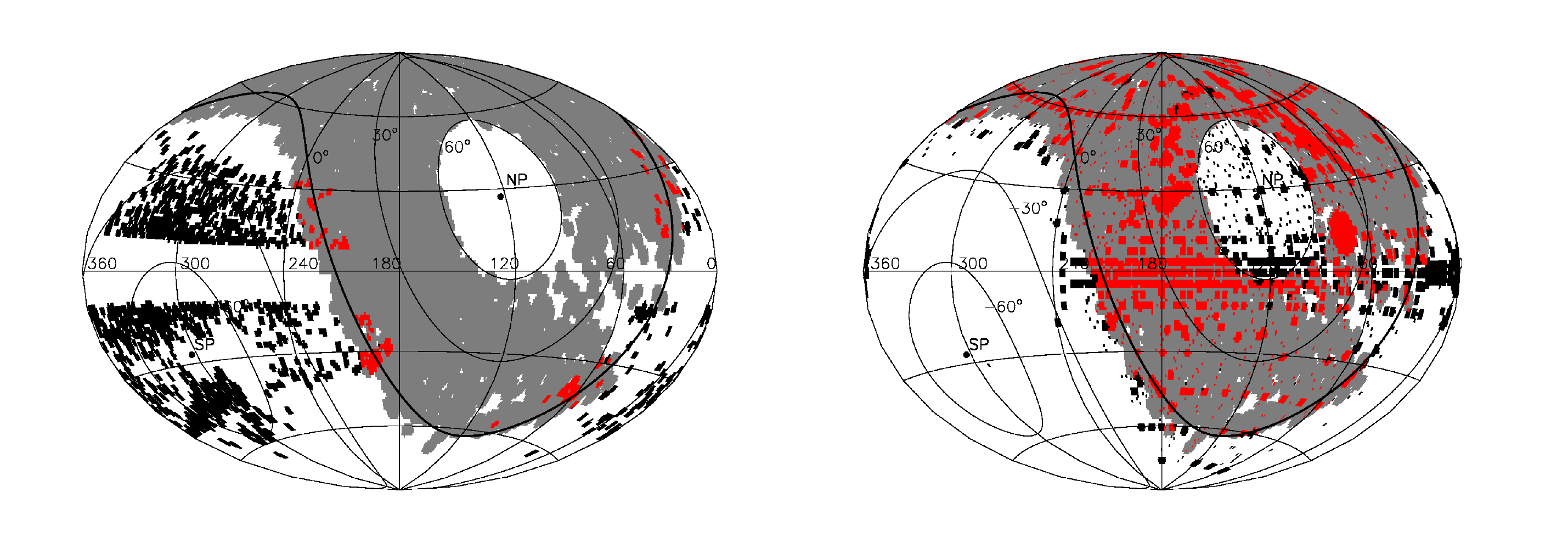}
\caption{{\em Left:} Footprints of LAMOST DR5 (grey) and GALAH DR2 (black) in Galactic coordinates ($l$, $b$) centered at the Galactic anti-center ($l=180^\circ$, $b=0^\circ$). 
Common areas between LAMOST and GALAH are shown in red. In total, there are 14,619 stars in common. {\em Right:} Footprints of LAMOST DR5 (grey) and APOGEE DR14 (black).  Common areas between LAMOST and APOGEE are shown in red. There are 77,249 stars in common.}
\label{fig:Fig4}
\end{figure*}

A good training set is crucial for obtaining accurate stellar labels with the current method, which is largely data-driven. The training set should have reliable determinations of stellar labels and have sufficient stars distributed widely throughout the label space. The GALAH \citep{De_Silva2015} and APOGEE surveys \citep{Majewski2017} fit these requirements as they are both high-resolution surveys and share a sufficient number of common stars with LAMOST, as shown in Fig.\,\ref{fig:Fig4}.
 
Since the GALAH and APOGEE surveys are implemented in different wavelength ranges and yield abundance for different elements, we make use of both surveys separately to define our training sets to obtain abundances of more elements from LAMOST spectra as well as to assess potential systematic uncertainties of the elemental abundance estimates. For each star, we derive two sets of stellar labels utilizing a LAMOST--GALAH training set and a LAMOST--APOGEE training set. We  compare the two sets of labels and provide a recommended set of results for individual labels.

\subsection{The LAMOST--GALAH training set} \label{lamostgalah-trainingset}
The Galactic Archaeology with HERMES (GALAH) survey is a high-resolution ($R\sim28,000$) spectroscopic survey using the High Efficiency and Resolution Multi-Element Spectrograph (HERMES) on the Anglo-Australian Telescope \citep{De_Silva2015}. In its second public data release, GALAH DR2 provides stellar parameters $T_{\rm eff}$, $\log g$, $V_{\rm mic}$, $V_{\rm macro}$, and abundances of 23 elements for 342,682 stars \citep{Buder2018}. The stellar parameters and abundances are derived with the data-driven method of $The$ $Cannon$ \citep{Ness2015} using a small sample of GALAH stars as the training set. The stellar labels of the training set are determined with the spectrum synthesis analysis tool Spectroscopy Made Easy \citep[SME;][]{Piskunov2017}. For the abundances determination, masks are adopted to pick clean lines with feasible strength. Non-LTE line formation has been incorporated for {\rm Li, O, Na, Mg, Al, Si, and Fe} for spectra synthesis. The accuracy of the derived stellar abundances depends on the location of stars in the parameter space because the line strength varies strongly \citep[see Fig.\,6 of][]{Buder2018}. For elements whose spectral line depth is too small (for instance, $<2\sigma$ respect to the continuum), the derived abundances may be problematic. GALAH DR2 provides flags to mark the quality of the derived stellar abundances. The flags are given in bitmask format, with 0 = no flag, i.e., recommended; 1 = `Line strength below $2\sigma$ = upper limit'; 2 = `The $Cannon$ starts to extrapolate'; 3 = 1 + 2, both 1 and 2 raised; 4 = The $\chi^2$ of the best fitting model spectrum is significantly higher; 5 = 1 + 4, both 1 and 4 raised, similarly for 6 and 7 \citep[see more details in][]{Buder2018}.  
  
A cross-identification of LAMOST DR5 with GALAH DR2 yields 19,989 common spectra of 14,619 unique stars. To select the training stars, we adopt the criteria 
\begin{equation}
\left\{
\begin{array}{lr}
   {\rm S/N_{LAMOST}} > 50, {\rm\, S/N_{GALAH}}>20;  {\rm\, if\,} {\rm [Fe/H]}>-0.6, \\
   {\rm S/N_{LAMOST}} > 30, {\rm\, S/N_{GALAH}}>10; {\rm\, if\,} {\rm [Fe/H]}<-0.6, \\
   {\rm [Mg/Fe]}>-0.3, \\
   {\rm flag}_{\rm [Mg/Fe]} \leq 3,  \\
   {\rm [X/Fe]} {\rm \,available\, for\, all\, X=Li, C, O, Na, Mg, Al, Si,} \\
                                 {\rm Ca, Sc, Ti, V, Cr, Mn, Co, Ni, Cu, Zn, Y, Ba, and\, Eu}. \\           
\end{array}
\right.
\end{equation}
Here the ${\rm S/N}_{\rm LAMOST}$ is the S/N per pixel in $g$-band of the LAMOST spectra. The $g$-band S/N from LAMOST is adopted throughout this paper. For ${\rm S/N_{GALAH}}$, we refer to the lowest S/N of the four wavelength windows in GALAH, meaning that we require the S/N for all the four wavelength windows to be larger than 20. To ensure a significant number of metal-poor stars in the training set, we adopt a less stringent S/N cut at the metal-poor region (${\rm [Fe/H]}<-0.6$\,dex). We have discarded stars with ${\rm [Mg/Fe]}<-0.3$\,dex as these [Mg/Fe] are likely unrealistic. We found most of them to be metal-rich (${\rm [Fe/H]}\sim0$) stars, which are expected not to have such low values of [Mg/Fe] in the Milky Way chemical evolution point of view. Rather than making a cut on quality flag for all the elemental abundances, we employ a cut only for the [Mg/Fe] flag. This choice is a balance between the quality of the stellar parameters and the number of training stars since the more stringent the cut on the quality flag, the fewer number of stars we will have as a training set. We further discard main-sequence binary or multiple stars from the training set. The binary/multiple stars are identified based on comparison of the $Gaia$ astrometric parallax with the spectroscopic parallax. The spectroscopic parallax is derived with the inferred distance modulus using absolute magnitudes from the LAMOST spectra (see Section\,\ref{binary} for more details). The training sample ultimately contains 4557 stars.

Table\,\ref{table:table2} lists the number of training stars that have different GALAH flags for abundances. For the majority of elements, GALAH detected the spectral lines (flag = 0 or flag = 2) for most ($\gtrsim80$\%) of the training stars. However, for {\rm Li, C, Co and Eu}, the majority of stars do not have strong enough features in the GALAH spectra to acquire accurate abundance determination. For these elements, the stars with significant (2$\sigma$) line detection by GALAH cover only a relatively small part of the parameter space. The majority of stars are without significant line detection and are only upper limit \citep{Buder2018}. 
\begin{table*}
\centering
\caption{GALAH DR2 abundance flags for the LAMOST--GALAH training stars.}
\label{table:table2}
\begin{tabular}{llllll}
\hline
 Label &  flag = 0 & flag = 1 & flag = 2 & flag = 3 & flag$>3$ \\
\hline        
 {\rm Li} & 46(1.0\%) & 58(1.3\%) & 998(21.9\%) & 1041(22.8\%) & 2414(53.0\%)  \\
 {\rm C} &  494(10.8\%) & 383(8.4\%) & 185(4.1\%) & 24(0.5\%) & 3471(76.2\%)   \\
 {\rm O} & 3837(84.2\%) & 283(6.2\%) & 61(1.3\%) & 136(3.0\%) & 240(5.3\%) \\
 {\rm Na} & 4073(89.4\%) & 175(3.8\%) & 119(2.6\%) & 140(3.1\%) & 50(1.1\%)\\
 {\rm Mg} & 4081(89.6\%) & 157(3.4\%) & 145(3.2\%) & 174(3.8\%) & 0(0.0\%) \\
 {\rm Al} & 2499(54.8\%) & 272(6.0\%) & 1177(25.8\%) & 155(3.4\%) & 454(10.0\%) \\
 {\rm Si} & 3652(80.1\%) & 347(7.6\%) & 241(5.3\%) & 302(6.6\%) & 15(0.3\%) \\
{\rm  Sc} & 3912(85.8\%) & 180(3.9\%) & 285(6.3\%) & 148(3.2\%) & 32(0.7\%) \\
{\rm  Ti} & 3651(80.1\%) & 329(7.2\%) & 383(8.4\%) & 169(3.7\%) & 25(0.5\%) \\
{\rm  V} & 2683(58.9\%)  & 565(12.4\%) & 987(21.7\%) & 207(4.5\%)  & 115(2.5\%) \\
{\rm  Cr} & 3493(76.7\%) & 464(10.2\%) & 413(9.1\%) & 159(3.5\%)   & 28(0.6\%) \\
{\rm  Mn} & 3700(81.2\%) & 238(5.2\%)  & 420(9.2\%) & 168(3.7\%)  & 31(0.7\%) \\
{\rm  Co} & 136(3.0\%)  & 375(8.2\%)  & 1368(30.0\%) & 218(4.8\%) & 2460(54.0\%) \\
{\rm  Ni} & 3461(75.9\%)  & 336(7.4\%) & 486(10.7\%) & 178(3.9\%) & 96(2.1\%) \\
{\rm  Cu} & 2024(44.4\%)  & 330(7.2\%) & 1635(35.9\%) & 180(3.9\%) & 388(8.5\%) \\
{\rm  Zn} & 4060(89.1\%)  & 284(6.2\%) & 24(0.5\%) & 138(3.0\%) & 51(1.1\%) \\
{\rm  Y} & 3802(83.4\%) & 461(10.1\%) & 115(2.5\%) & 167(3.7\%) & 12(0.3\%) \\
{\rm  Ba} & 2969(65.2\%) & 330(7.2\%) & 869(19.1\%) & 319(7.0\%) & 70(1.5\%) \\
{\rm  Eu} & 451(9.9\%) & 193(4.2\%) & 544(11.9\%) & 44(1.0\%) & 3325(73.0\%) \\
\hline
\end{tabular}
\begin{tablenotes}
\item[]{Summary of GALAH DR2 abundance flags: 0 = no flag, i.e., recommended; 1 = `Line strength below $2\sigma$ = upper limit'; 2 = `The $Cannon$ starts to extrapolate'; 3 = 1+2 = both 1 and 2 raised; 4 = `The $\chi^2$ of the best fitting model spectrum is significantly higher'; 5 = 1 + 4 = both 1 and 4 raised; 6 = 2 + 4 = both 1 and 4 raised; 7 = 3 + 4 = both 3 and 4 raised.} 
\end{tablenotes}
\end{table*}

Specifically, for Li, only $\sim$1000 stars, mostly dwarfs with $5500\lesssim{T}_{\rm eff}\lesssim6500$\,K or metal-rich giants, have tentative line detection (flag = 0 or flag = 2). Among them, only 46 stars are not flagged (flag = 0), while the others have Li abundance determined by extrapolation with the $Cannon$ (flag = 2). For C, only metal-rich dwarfs have robust carbon line detection. For Co, most of the metal-rich giant stars have reliable Co line detection. However, for dwarfs, only $\sim$1/5 of them, mostly metal-rich stars, have useful line detection. As for Eu, all stars with no flag (flag = 0) are giants. About two-thirds of the giant stars in the training set have robust detection of Eu lines. There are also $\sim$100 dwarf stars whose Eu lines are detected, but the abundances are determined by extrapolation with the $Cannon$. For other elements, there are also general trends that more metal-poor stars have a higher probability of failing to detect significant spectral lines by the GALAH pipeline. 

\begin{figure*}
\centering
\includegraphics[width=140mm]{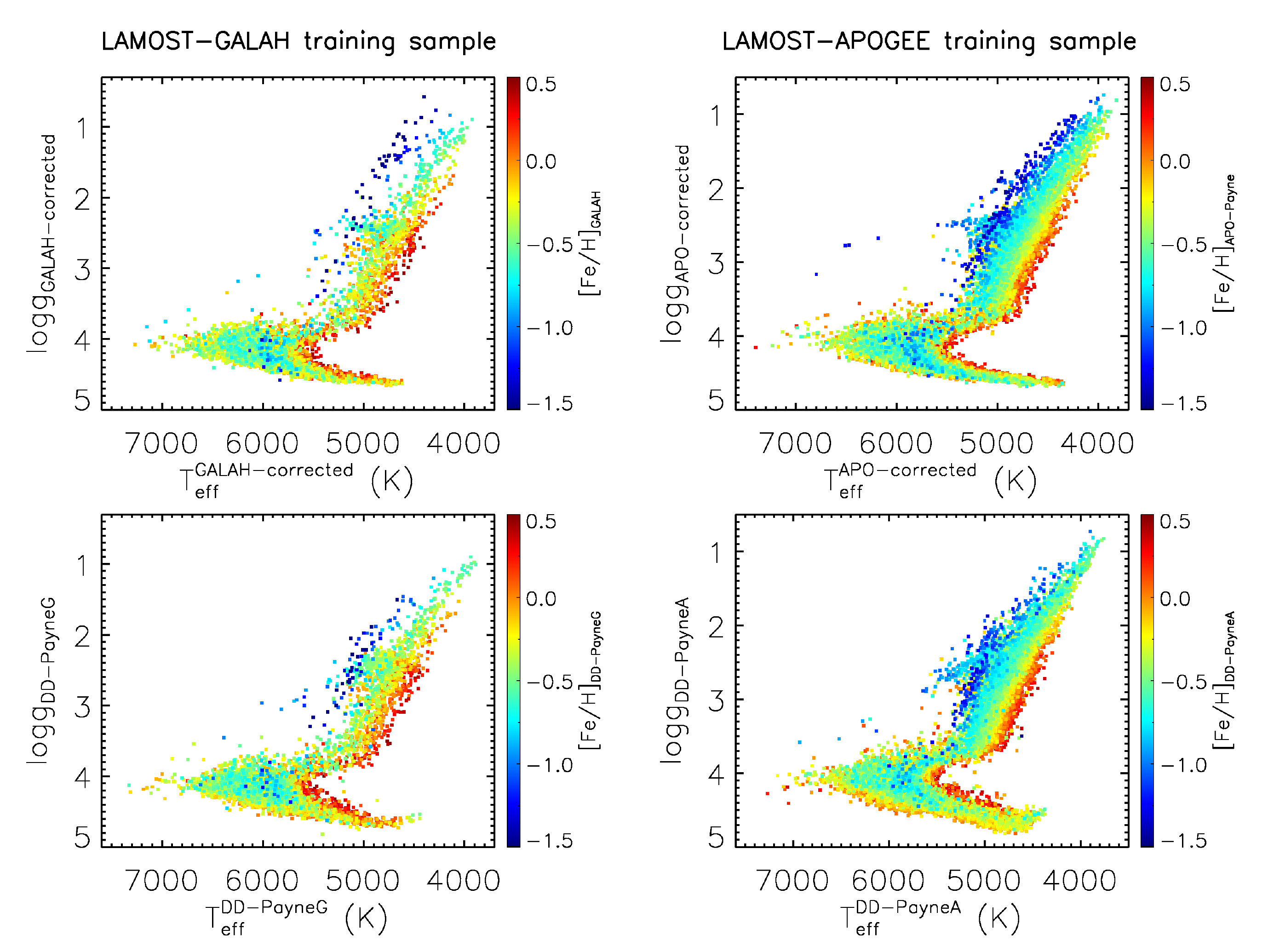}
\caption{Distributions of the $DD$--$Payne$ training samples in the $T_{\rm eff}$--$\log g$ plane. {\em Top left:} Distribution of the LAMOST--GALAH training stars in the Kiel diagram. The $T_{\rm eff}$ and $\log g$ values are from a corrected version of the GALAH DR2 values exploiting the Gaia parallax and multi-band photometry as extra constraints (see Section \ref{trainingset} and Appendix \ref{appendixA}). {\em Bottom left:} Same as the {\em top left} panel, but showing the $T_{\rm eff}$ and $\log g$ values from the $DD$--$Payne$ fits to the LAMOST spectra, i.e., a self-validation of the $DD$--$Payne$ method using the LAMOST--GALAH training set. For both panels, the stars are color-coded by the  [Fe/H] values of GALAH DR2.
{\em Right panels:} Similar to the {\em left} panels, but for the LAMOST--APOGEE training set. The training set is constructed by cross-matching LAMOST DR5 with the APOGEE--$Payne$ catalog \citep{Ting2019}.}
\label{fig:Fig5}
\end{figure*}

The stellar parameters of the GALAH DR2, estimated before the publication of Gaia DR2, could have considerable uncertainties (e.g., $>\,$0.1\,dex in $\log g$). We correct for the $T_{\rm eff}$ and $\log g$ of our training stars by using extra constraints from the multi-band photometry and Gaia parallax, which are now readily available. The correction is done with a Bayesian approach, where the Gaia parallax and multi-band photometry are adopted as the observed quantities to generate the likelihood through the comparison with stellar isochrones. The GALAH stellar parameters themselves are adopted as part of the priors. A detailed introduction of the method is presented in the Appendix\,\ref{appendixA}. The top-left panel of Fig.\,\ref{fig:Fig5} plots the distribution of the training stars in the $T_{\rm eff}$--$\log g$ diagram. The training set has a reasonable coverage of $4500 \lesssim T_{\rm eff} \lesssim 7000$\,K for dwarfs and $T_{\rm eff}>4000$\,K for giants. Most of the dwarf stars have ${\rm [Fe/H]}>-1.0$; there are more metal-poor giant stars than dwarfs. We emphasize that although we have corrected for the $T_{\rm eff}$ and $\log g$ of our training set, the elemental abundances of the training stars are not modified. Therefore, our results will inherit the systematics of GALAH DR2. 

We derived 23 labels with the LAMOST--GALAH training set, namely $T_{\rm eff}$, $\log g$, [Fe/H], and abundance ratios [X/Fe] for elements {\rm Li, C, O, Na, Mg, Al, Si, Ca, Sc, Ti, V, Cr, Mn, Co, Ni, Cu, Zn, Y, Ba, and Eu}. The choice of elements is based on two considerations: (a) The element should be informative enough. It should contain prominent enough features in its gradient spectra, which allow for a robust abundance determination. (b) We want to determine the abundances for as many elements as possible, especially elements from different yield channels (such as C, N, $\alpha$-elements, s- and r-neutron capture elements) to maximize the science potentials. However, as has been shown in Section\,\ref{method}, not all abundances are derived directly from the right spectral features. Based on examinations of the $DD$--$Payne$ gradient spectra as presented in Section\,2, we find that the abundance ratios of Li, Sc, V, Zn, Y, Cu, and Eu for most of the LAMOST stars are probably a consequence of astrophysical correlations rather than physical determinations. Thus, we decide to discard the results for these elements.

\subsection{The LAMOST--APOGEE training sets} \label{lamostapogee-trainingset}
The Apache Point Observatory Galactic Evolution Experiment (APOGEE) is a high-resolution ($R\sim22,500$) infrared (1.51--1.70\,$\mu$m) sky survey \citep{Majewski2017, Zasowski2017}. 
The catalog of the APOGEE data release 14 (DR14) provides stellar parameters and abundances of 20 elements for 277,371 stars \citep{Holtzman2018}. The APOGEE stellar labels for giant stars are relatively well-calibrated/validated and widely used for extensive science cases. But the stellar labels for dwarfs still need to be further verified considering that the derived $\log g$ for dwarf stars do not follow the isochrones \citep{Jonsson2018}.   

An independent set of stellar labels for APOGEE DR14 stars are those of \citet{Ting2019}, which provides stellar parameters and abundances for 15 elements ({\rm C, N, O, Mg, Al, Si, S, K, Ca, Ti, Cr, Mn, Fe, Ni, and Cu}) derived with {\it The Payne}. All stellar labels from {\it The Payne} are derived simultaneously by fitting the APOGEE spectra to self-consistently computed (solving for the atmospheric structures and radiative transfer for {\em all} elements) Kurucz (LTE) model spectra with recently improved line lists (Cargile et al. in prep.). \citet{Ting2019} showed that, even without calibration, both the dwarf and giant stars follow well the isochrones in the HR diagram, and an examination with cluster member stars suggests the abundances, especially that of C and N, may have better precision and accuracy \citep{Ting2019, Nataf2019}. The catalog of \citet[][hereafter the APOGEE--$Payne$]{Ting2019} contains 220,000 stars with ${\rm [Fe/H]}>-1.5$\,dex, and flags are provided to indicate the quality of the stellar label determinations.

In this work, we adopt the APOGEE labels from {\it The Payne} as our training set mainly because we want to make full use of the dwarf stars, which are currently not well calibrated in APOGEE DR14. A cross-identification of the LAMOST DR5 with the APOGEE--$Payne$ catalog yields 77,249 stars in common (130,590 spectra). To define the training set, we further adopt the criteria
\begin{equation}
\left\{
\begin{array}{lr}
   {\rm S/N_{LAMOST}} > 50, \\
   {\rm S/N_{APOGEE}}>80, \\
   {\rm [Mg/Fe]}>-0.3, \\
   {\rm qflag = ``good"},  \\
   {\rm [X/H]} {\rm \,available\, for\, all\, X=C, N, O, Mg, Al, Si,} \\
                                 {\rm Ca, Sc, Ti, Cr, Mn, Ni, and\, Cu}. \\          
\end{array}
\right.
\end{equation}
These criteria lead to a sample of $\sim$26,000 stars. We further exclude main-sequence binary/multiple stars as was done for the LAMOST--GALAH training set. We also discard stars that have more than ten bad pixels (see Section\,\ref{spectralnormalandmask} for the definition of a bad pixel) in their LAMOST spectra. From the remaining stars, we select 15,000 stars as our training stars. The others are included in a cross-validation data set to verify the method. To select the training stars, we keep all stars with ${\rm [Fe/H]}\leq-0.8$\,dex because they are rarer, and randomly sampling stars with ${\rm [Fe/H]}>-0.8$\,dex.

The top-right panel of Fig.\,\ref{fig:Fig5} shows the distribution of the LAMOST--APOGEE training stars in the $T_{\rm eff}$--$\log g$ diagram. The training set has a wide coverage: $4300\,$K $\,\lesssim T_{\rm eff} \lesssim 7000$\,K for dwarfs and $3800\,$K $\,\lesssim T_{\rm eff} \lesssim 7000$\,K for giants. Similar to the LAMOST--GALAH training set, we have opted to correct the APOGEE--$Payne$ $T_{\rm eff}$ and $\log g$ of our training stars by considering extra constraints from the Gaia parallax and multi-band photometry. We train 16 labels with this LAMOST--APOGEE training set, namely $T_{\rm eff}$, $\log g$, $V_{\rm mic}$, [Fe/H], and [X/Fe] for X = C, N, O, Mg, Al, Si, Ca, Ti, Cr, Mn, Ni, and Cu. As already shown in Section\,\ref{method}, for all 16 labels, the $DD$--$Payne$ neural network built with the LAMOST--APOGEE training set reproduces the theoretical gradient spectra for most of the LAMOST stars, indicating that all the labels are derived in a physical way rather than through astrophysical correlations.

As discussed in detail in Section\,\ref{verification}, we found that for both the LAMOST--GALAH training set and the LAMOST--APOGEE training set, the $DD$--$Payne$ tends to systematically overestimate the [Fe/H] in the metal-poor case (${\rm [Fe/H]}\lesssim-0.8$\,dex). This is likely owing to the non-uniform distribution of training stars in the parameter space --- there are much more metal-rich stars than metal-poor ones, so the latter has small contribution to the $\chi^2$ optimization in the training process, and the labels are thus biased due to imperfections of the training process, for instance, failing to find the global minimum. 

In order to derive accurate labels for metal-poor stars, we define a specific training set of metal-poor stars from the above LAMOST--APOGEE training set. We only select stars with ${\rm [Fe/H]}<-0.5$\,dex, which leads to a sample of 2097 stars for this metal-poor training set. Results from this metal-poor training set are then combined with those from the above overall LAMOST--APOGEE training set. In particular, for metal-rich stars (${\rm [Fe/H]}>-0.6$\,dex), the results from the overall training set are adopted, where [Fe/H] refers to those estimates from the overall LAMOST--APOGEE training set. In the transition regime of $-1.0<{\rm [Fe/H]}<-0.6$\,dex, we take the weighted mean value for each label for combining the results from the two training sets, and the weight is a linear function from 0 to 1 between the transitional metallicity boundaries. For stars with ${\rm [Fe/H]}<-1.0$\,dex, results from the metal-poor training set are adopted. However, the derived labels for stars whose [Fe/H] are far beyond the lower limit of the training set ($-1.5$\,dex) should be used with cautious. 

In the following, when we refer to the labels derived from the LAMOST--APOGEE training set we mean this combined set of results. We do not define a metal-poor training set for the LAMOST--GALAH training set because the number of metal-poor stars in the LAMOST--GALAH common sample is too small to construct a separate training sample.

\section{Verification on LAMOST DR5} \label{verification}
We apply the $DD$--$Payne$ method to the LAMOST DR5 data set, which contains 9,017,844 spectra. According to the classification of the LAMOST 1D pipeline, 8,171,443 spectra are from stars, 153,090 galaxies, 51,133 quasars, and 642,178 unknown objects. In the following, we exclude spectra of galaxies, quasars and unknown objects, and apply our method to stellar spectra only. 

\subsection{Spectra normalization and mask} \label{spectralnormalandmask}
As mentioned in Section\,\ref{method}, all the LAMOST spectra are normalized following the method in \citet{Ho2017}, i.e., dividing the spectrum by a continuum derived through smoothening the spectrum with a Gaussian kernel of 50\,{\AA} in width. We mask all bad pixels due to instrument problems or cosmic ray contaminations using the LAMOST ``PIXAMSK". We also mask out wavelength regions of the telluric bands: 6270--6330{\AA}, 6800--6990{\AA}, 7100--7320{\AA}, 7520--7740{\AA}, and 8050-8350{\AA}, the very blue ($<3900${\AA}), the very red ($>8880${\AA}) wavelength region, and the dichroic region 5720--6060{\AA}, because they have lower spectra quality. However, we do keep the 5880--5910{\AA} region because it contains crucial information of sodium abundance. We caution that for a good fraction of the DR5 sample, the flux in this wavelength range could be problematic due to poor flux calibration. In light of this limitation, we provide a $\chi^2$ value computed specifically from this wavelength window for each spectrum. One should be cautious about the derived ${\rm Na}$ abundance if the quoted $\chi^2$ is large. Note that because the flux of ${\rm Na}$ lines may be also affected by the interstellar absorption, especially for stars in high extinction region, the derived ${\rm Na}$ abundance could suffer extra uncertainty which is not considered in the current work.       

We further mask all pixels that may be inaccurately trained by comparing solar spectrum predicted by the $DD$--$Payne$ with the observed LAMOST solar spectrum. We derive the LAMOST solar spectrum from the twilight flat. We define that a pixel is poorly trained if the difference between the interpolated model and the observed solar spectrum is larger than 0.05. This masks 4 percent of total number of pixels. We choose this conservative value (0.05) because the LAMOST twilight flat is not accurate enough to represent the true solar spectrum. In fact, we have found that residuals between the normal solar-like spectra and the best-fitting $DD$--$Payne$ models are generally smaller than residuals between the twilight solar spectrum and the $DD$--$Payne$ model for the solar spectrum. We have also tested different values from 0.01 to infinity (no mask) for the criteria. We found that a stricter mask does not significantly improve the result. However, imposing a more stringent mask than what we adopted here will substantially increase the uncertainties of abundances for several elements due to the reduction of informative pixels. 

\subsection{Verification with training and cross-validation data sets} \label{cross-validation}
A comprehensive examination of the derived labels should be based on comparisons with independent data sets that have accurate label determinations across a broad swath of label space. 
However, currently, a perfect ``ground truth" data set unfortunately does not exist. Due to a number of reasons, such as imperfect stellar atmospheric models, imperfect atomic molecular database, and different pipeline algorithms, there are still considerable mismatch and systematic patterns among results of different high-resolution surveys \citep[see e.g.,][]{Jofre2018, Griffith2019}. As will be shown in Section\,\ref{systematics}, there are systematic mismatches between GALAH and APOGEE--$Payne$ abundances for a few elements. In this section, we will first examine our results with self-validations of the training sets and with cross-validations using validation data sets. 

Self-validations are shown in Fig.\,\ref{fig:Fig5} for both the LAMOST--GALAH and the LAMOST--APOGEE training set, distributions of the training stars in the $T_{\rm eff}$--$\log g$ plane are well reproduced by the $DD$--$Payne$, indicating that the training process does not introduce significant systematic errors for the majority of stars. However, for metal-poor giants, $\log g$ is overestimated for both training sets. On the other hand, for cool ($T_{\rm eff} <4800$\,K) dwarfs, $\log g$ is underestimated. These metal-poor giants and cool dwarfs are located near the boundary of the parameter space, where the training stars are few and have sparse distributions, which might explain these slight systematic errors.

\begin{figure*}[htp]
\centering
\includegraphics[width=180mm]{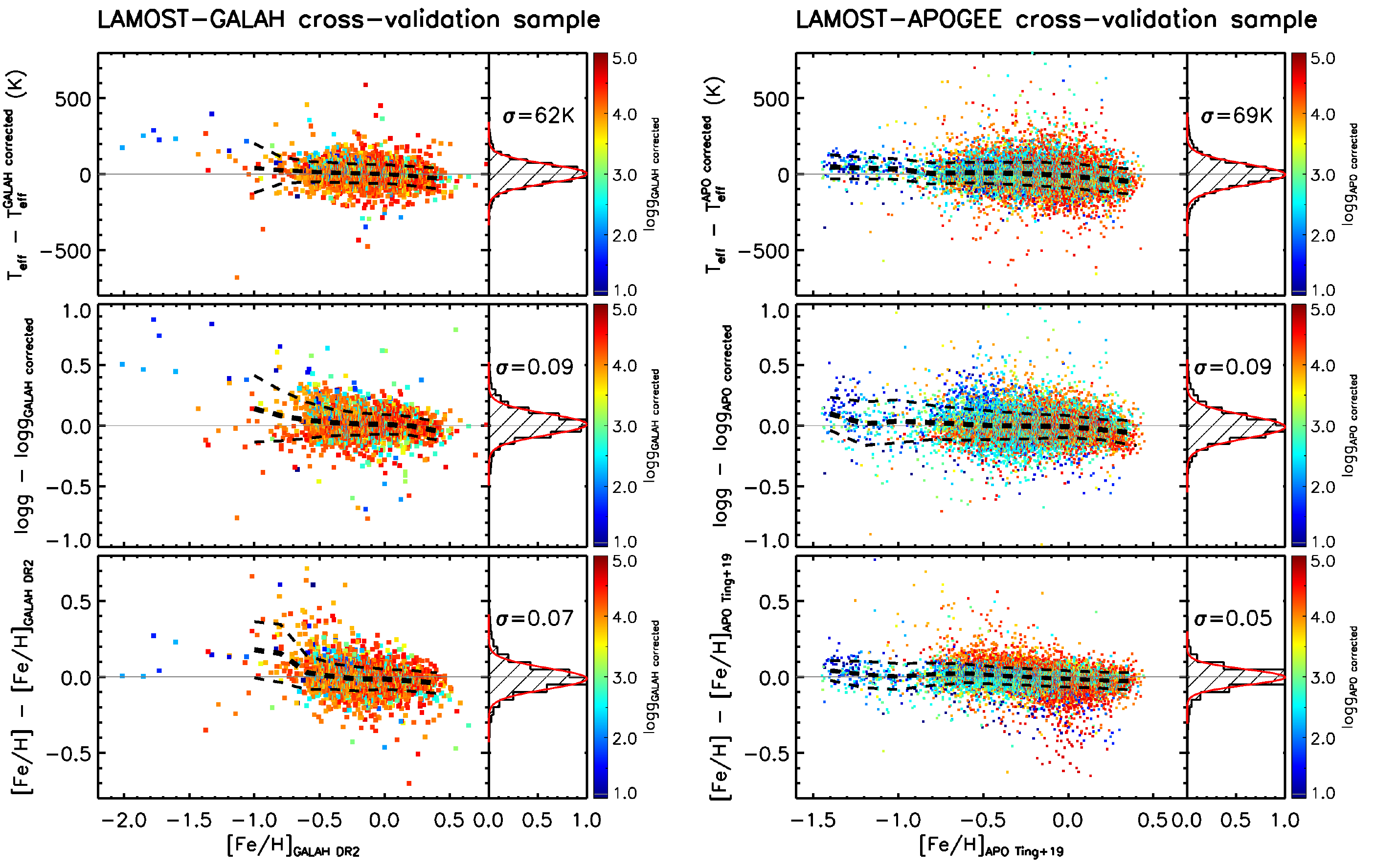}
\caption{Cross-validations of $T_{\rm eff}$, $\log g$, and [Fe/H] for $DD$--$Payne$ results derived using the LAMOST--GALAH training set ({\em left}) and results derived using the LAMOST--APOGEE training set ({\em right}). The cross-validation data sets are not used during the training of the $DD$--$Payne$. For $T_{\rm eff}$ and $\log g$, we adopt a corrected version of the GALAH and APOGEE values, using Gaia parallax and multi-band photometry as extra constraints. For [Fe/H], the GALAH DR2 and APOGEE--$Payne$ \citep[``Ting+19";][]{Ting2019} catalog values are adopted as the reference values. The individual stars are color-coded by the $\log g$ reference values. Dashed lines in each panel delineate the median and $\pm$1$\sigma$ values of the parameter differences of the recovery as a function of [Fe/H]. The number in the top-right corner of each panel marks the standard deviation. For the LAMOST--GALAH results, the figure shows good consistency for stars with ${\rm [Fe/H]}\gtrsim-0.7$\,dex. However, due to the lack of training set at the metal-poor end, systematic deviation occurs at the low metallicity end. For the LAMOST--APOGEE results, since the metal-rich and the metal-poor stars are trained separately (see Section\,\ref{lamostapogee-trainingset}), systematic deviation is small across the whole [Fe/H] range, from $-1.5$\,dex to 0.4\,dex.}
\label{fig:Fig6}
\end{figure*}

For the cross-validation data sets, we have 4621 LAMOST--GALAH common stars with ${\rm S/N}>40$ to validate results from the LAMOST--GALAH training set, and 13,667 LAMOST--APOGEE common stars with ${\rm S/N}>40$ to validate results from the LAMOST--APOGEE training set. Binary and multiple stars were discarded in the same way as for the training set. Fig.\,\ref{fig:Fig6} plots the difference of the $DD$--$Payne$ stellar parameters, for which the $T_{\rm eff}$ and $\log g$ refer to the corrected values through Gaia parallax and multi-band photometry. The figure demonstrates that the $DD$--$Payne$ results are in good agreement with the reference values for the bulk of the stars. The standard deviation is 61\,K, 0.09\,dex, and 0.07\,dex, respectively, for $T_{\rm eff}$, log\,$g$, and [Fe/H] derived using the LAMOST--GALAH training set, and 69\,K, 0.09\,dex, and 0.05\,dex, respectively, for $T_{\rm eff}$, $\log g$, and [Fe/H] derived using the LAMOST--APOGEE training set. However, for the metal-poor stars (${\rm [Fe/H]}\lesssim-0.7$\,dex), $T_{\rm eff}$ and $\log g$ from the LAMOST--GALAH training set are overestimated by up to 200\,K and 0.5\,dex, respectively. The [Fe/H] is also overestimated by up to 0.2\,dex, a consequence of the small number and sparse distribution of metal-poor training stars. On the other hand, for the LAMOST--APOGEE training set, the systematic error for metal-poor stars is not significant. As discussed, unlike for the LAMOST--GALAH training set, the results for metal-poor stars are more accurate when using the LAMOST--APOGEE training set because these values are derived with the $DD$--$Payne$ specifically trained using metal-poor training stars (Section\,\ref{lamostapogee-trainingset}).  

\begin{figure*}
\centering
\includegraphics[width=160mm]{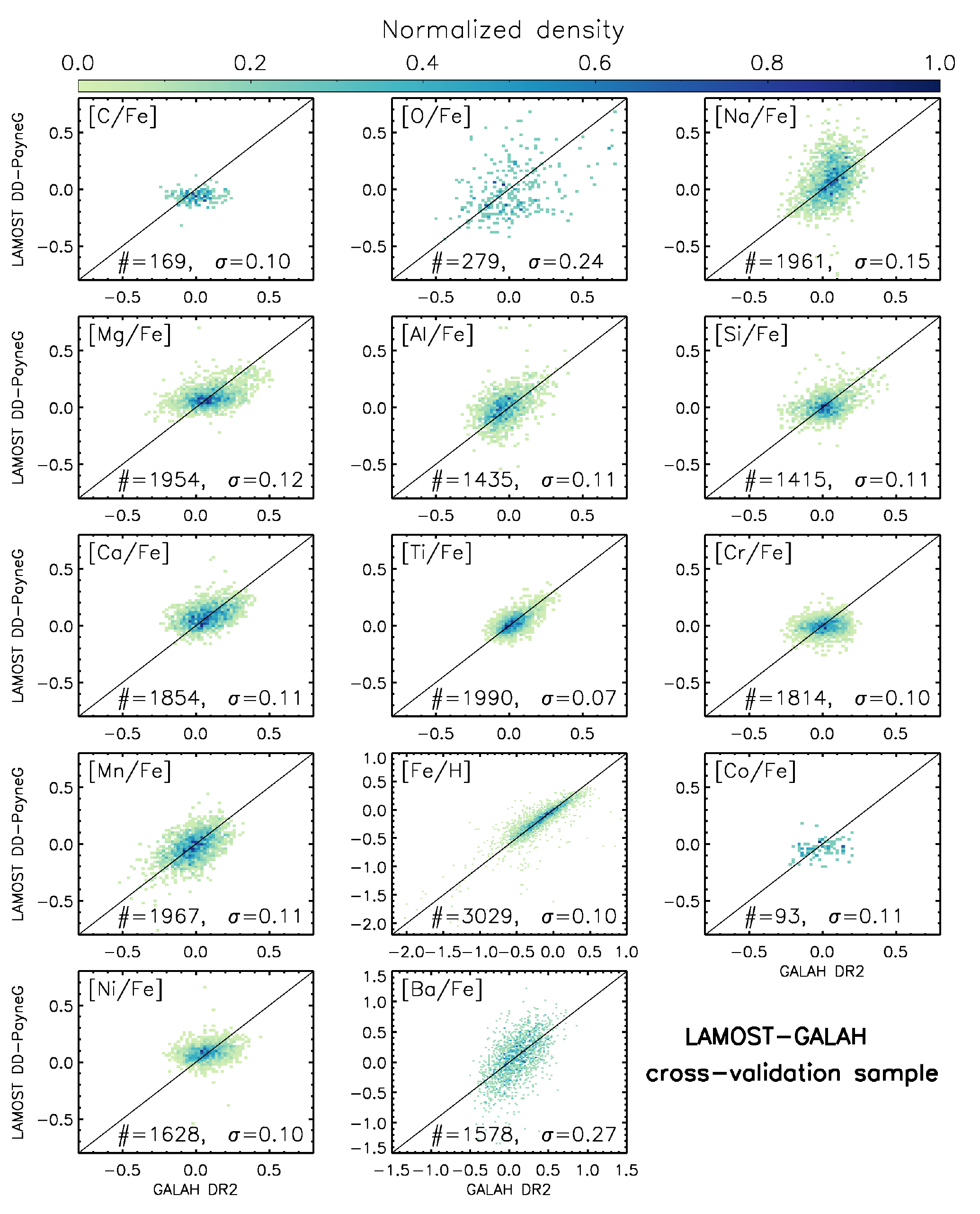}
\caption{Cross-validations of the $DD$--$Payne$ elemental abundances derived with the LAMOST--GALAH training set. The number of stars adopted for cross-validation, as well as the standard deviation between our estimates from the LAMOST spectra and the GALAH DR2 values, are marked in each panel. We discard stars whose $DD$--$Payne$ abundances are potentially contaminated by additional information from astrophysical correlations, instead of being determined from {\it ab initio} spectral features. We only adopt stars with their GALAH DR2 values deemed reliable by GALAH (flag = 0). As a result, the number of stars varies among different elements. For the majority of elements, the standard deviation, which indicates the $DD$--$Payne$ determination uncertainty, is about 0.1\,dex, with the exception of [O/Fe] and [Ba/Fe].}
\label{fig:Fig7}
\end{figure*}

Fig.\,\ref{fig:Fig7} shows that the $DD$--$Payne$ abundances derived using the LAMOST--GALAH training set are consistent with the GALAH DR2 values. To have a more robust estimate, we have discarded our LAMOST estimates for stars of which the abundances are potentially determined by astrophysical correlations (see Section\,\ref{consistenctofgradient}). Further, we only consider the GALAH DR2 abundances that are more reliable (flag = 0 in GALAH). For C, Na, Mg, Al, Si, Ca, Ti, Cr, Mn, Fe, Co, and Ni, the standard deviation in [X/Fe] between LAMOST and GALAH is on the order of $\sim$0.1\,dex, showing decent agreement. For O and Ba, the dispersion is larger ($>\,$ 0.2\,dex), likely due to the relatively large uncertainty of the training labels and/or the weaker and fewer spectral features of these elements in the LAMOST spectra. We note that, since uncertainties in both data sets could contribute to the standard deviation, the standard deviation should be regarded as an estimate of the upper limit of the uncertainty.

\begin{figure*}
\centering
\includegraphics[width=160mm]{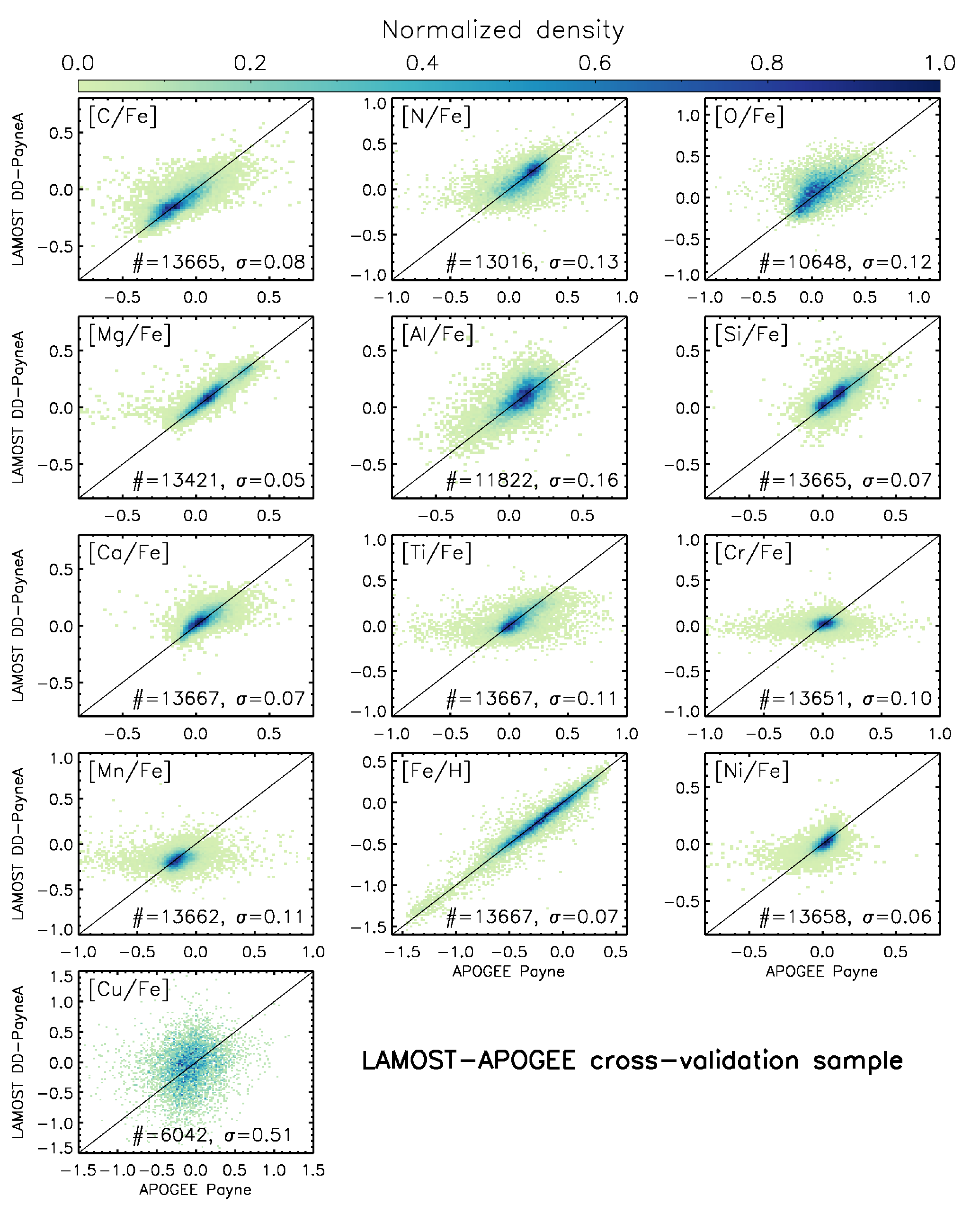}
\caption{Similar to Fig.\,7, but here we show the cross-validations of the $DD$--$Payne$ abundances derived using the LAMOST--APOGEE training set. For all elements except for Cu, the abundance difference has a standard deviation of $\sim$0.1\,dex or smaller. But we note that, as determining elemental abundances from dwarf stars in the APOGEE H-band is challenging, some outliers and scatters could be contributed by errors from the APOGEE--$Payne$ estimates \citep{Ting2019}, and do not necessarily come from $DD$--$Payne$ (e.g., ${\rm [Mn/Fe]}\sim-1$ and ${\rm [Ni/Fe]}\sim-1$\,dex from APOGEE--$Payne$).}
\label{fig:Fig8}
\end{figure*}

Similar to Fig.\,\ref{fig:Fig7}, Fig.\,\ref{fig:Fig8} shows the cross-validation results for the LAMOST--APOGEE training set. Fig.\,\ref{fig:Fig8} illustrates that the $DD$--$Payne$ abundances  are consistent with the APOGEE--$Payne$ values. The standard deviation is $\lesssim\,$0.1\,dex for almost all of the elements except for Cu. [Cu/Fe] has a dispersion of 0.51\,dex, mainly owing to the fact that Cu lines are extremely weak in the LAMOST spectra. For a few elements such as Mg, Ti, Cr, and Mn, there are a small fraction of stars of which the APOGEE--$Payne$ abundances deviate significantly from the LAMOST $DD$--$Payne$ values. Some of them have extremely low abundances, e.g. ${\rm [Mg/Fe]\sim-1.0}$, from the APOGEE--$Payne$. For such stars, it is hard to speculate, but it is possible that the APOGEE--$Payne$ values are in error. 

\subsection{Repeat observations} \label{repeatobservation}
Besides cross-validating with high-resolution results, another test would be to check the consistency of the stellar labels from repeat observations. Fortunately,  about a quarter of the LAMOST DR5 spectra are repeat observations of the same stars. Therefore, we can have a detailed examination on the robustness of the label determination using this extensive repeat observation data set. Furthermore, to make sure that our uncertainty estimates are robust, we consider only repeat observations that were carried out on different nights and discard those carried out on the same night. Observation within the same night could underestimate the uncertainties as they are observed under similar instrument conditions. For example, they cannot fully reflect the variation of the line spread function.

\begin{figure*}
\centering
\includegraphics[width=180mm]{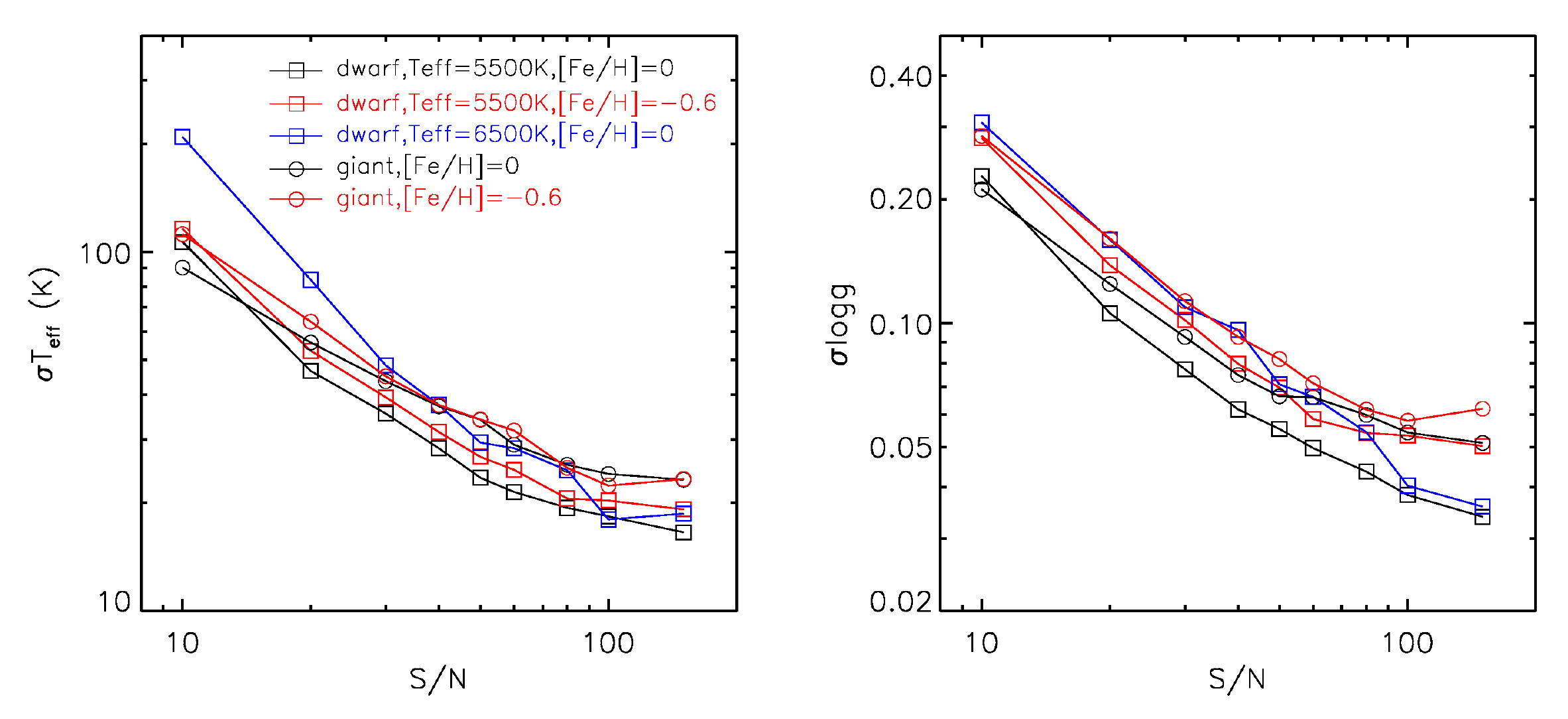}
\caption{Internal precision of $T_{\rm eff}$ and $\log g$ derived from repeat observations as a function of S/N. The internal precision is defined as the {\it rms} standard deviation of the repeat observations. Each dot is deduced from $\mathcal{O}(1000)$ stars with repeat observations. Stars are classified as giants if they have $T_{\rm eff}<5500$\,K and $\log g<4.1$. Here we only show results for the recommended $T_{\rm eff}$ and $\log g$, i.e., those derived with the LAMOST--APOGEE training set (Table\,\ref{table:table4}). Internal precision of $T_{\rm eff}$ is better than 60\,K for stars with ${\rm S/N}>20$, and the internal precision of $\log g$ is better than 0.1\,dex at ${\rm S/N}>30$.}
\label{fig:Fig9}
\end{figure*}

Fig.\,\ref{fig:Fig9} presents the internal precision of $T_{\rm eff}$ and $\log g$ deduced from repeat observations. To show the results as a function of S/N, we consider only repeat observations that have comparable S/N ($\Delta{\rm S/N}<5$). We define the internal precision to be the dispersion of differences between repeat observations divided by $\sqrt{2}$, considering that both observations could have contributed to the dispersion by a similar amount. The figure shows that the results agree with the expected $\sigma\propto(\rm S/N)^{-1}$ trend \citep[e.g.][]{Ting2017b}. At S/N$\,=\,$20, we attain a precision of $\sim$60\,K in $T_{\rm eff}$ and $\sim$0.12\,dex in $\log g$ for both giants and G-type ($T_{\rm eff}$ = 5500\,K) dwarfs at solar metallicity. The internal precision reaches $\sim$20\,K and $\sim$0.05\,dex for high S/N ($\sim$100) observations. For more metal-poor or hotter stars, the uncertainties are slightly larger. Specifically, for ${\rm [Fe/H]}=-0.6$\,dex, the uncertainties are $\sim$10\,K larger in $T_{\rm eff}$ and $\sim$0.02\,dex larger in $\log g$ compared to their solar metallicity counterparts. For dwarfs with $T_{\rm eff}=6500$\,K, the uncertainties are $\sim$40\,K larger in $T_{\rm eff}$ and $\sim$0.05\,dex larger in $\log g$ than those with $T_{\rm eff}=5500$\,K at S/N$\,=\,$20. 

\begin{figure*}
\centering
\includegraphics[width=160mm]{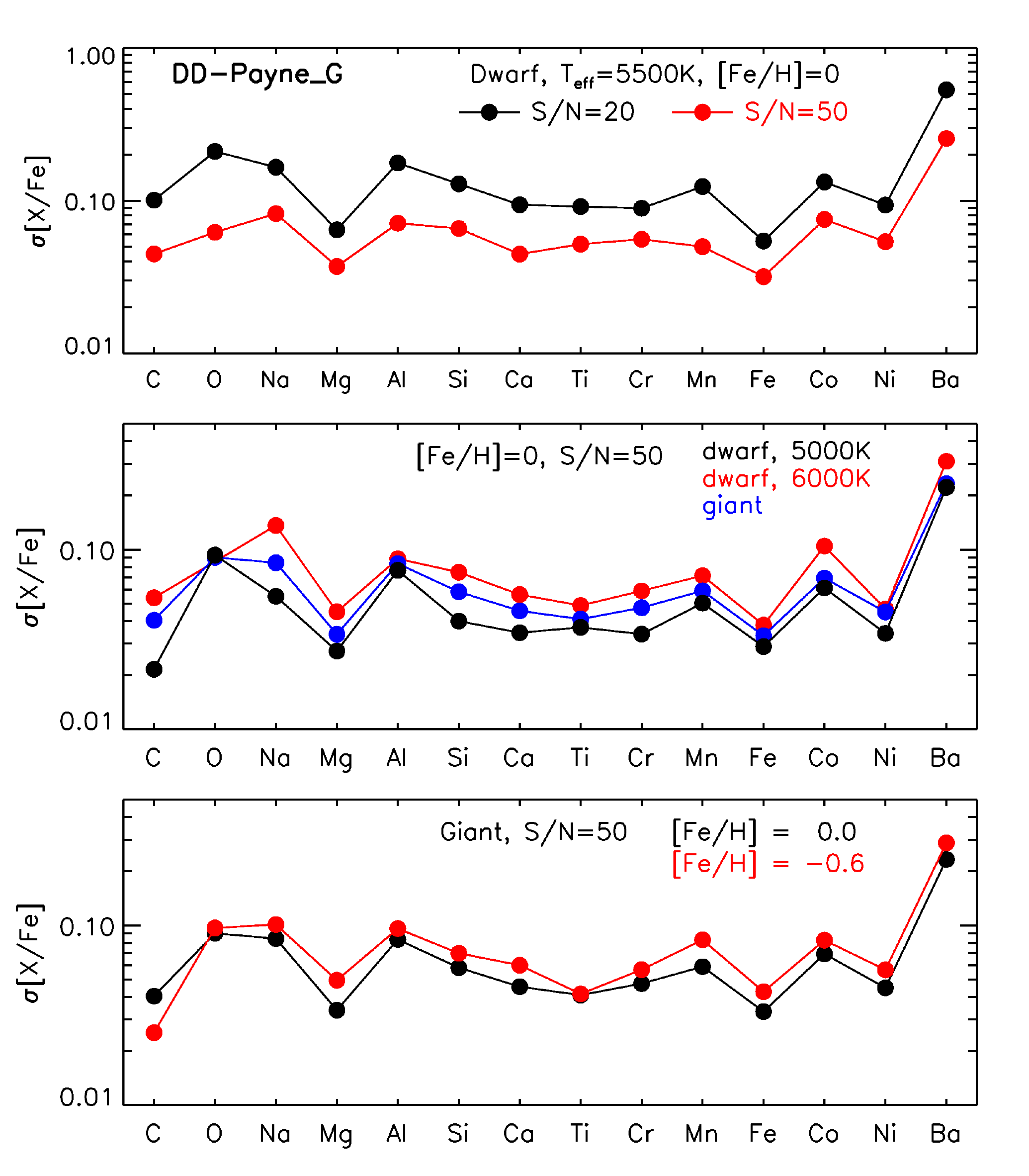}
\caption{Internal precision of elemental abundances deduced from repeat observations when the LAMOST--GALAH training set is adopted. The internal precision is defined as the {\it rms} standard deviation of the repeat observations. The {\em top panel} shows how the precision varies as a function of S/N for G-dwarfs. The {\em middle panel} shows how the precision varies as a function of $T_{\rm eff}$ for dwarfs, as well as varies between dwarfs and giants. The {\em bottom panel} shows the variation with respect to metallicity for giants. The results suggest that for stars with ${\rm S/N}>50$, the internal abundance precision for most elements, except for Ba, is $\lesssim$0.1\,dex. In particular, for C, Mg, Ca, Ti, Cr, Fe, and Ni, an internal precision of $\lesssim$0.05\,dex is achieved.}
\label{fig:Fig10}
\end{figure*}

Fig.\,\ref{fig:Fig10} shows the internal precision of abundances deduced from repeat observations, adopting the LAMOST--GALAH training set. The figure demonstrates that, at S/N$\,=\,$50, for most of the elements except for Ba, the internal abundance precision is better than 0.1\,dex for both giants and dwarf. In particular, we attain an internal precision of $\sim$0.03\,dex for [Fe/H] and [Mg/Fe], and $\sim$0.05\,dex for [X/Fe] of a few elements such as C, Ca, Ti, Cr, Mn, and Ni. The precision decreases slightly with increasing $T_{\rm eff}$ or decreasing [Fe/H], but they are generally better than 0.1\,dex over a wide range of $T_{\rm eff}$ and [Fe/H] at S/N$\,>\,$50. For [Ba/Fe], however, the internal precision is only 0.2--0.3\,dex at S/N$\,=\,$50, a consequence of the fact that there are only a few weak Ba features at the LAMOST resolution. 

\begin{figure*}
\centering
\includegraphics[width=160mm]{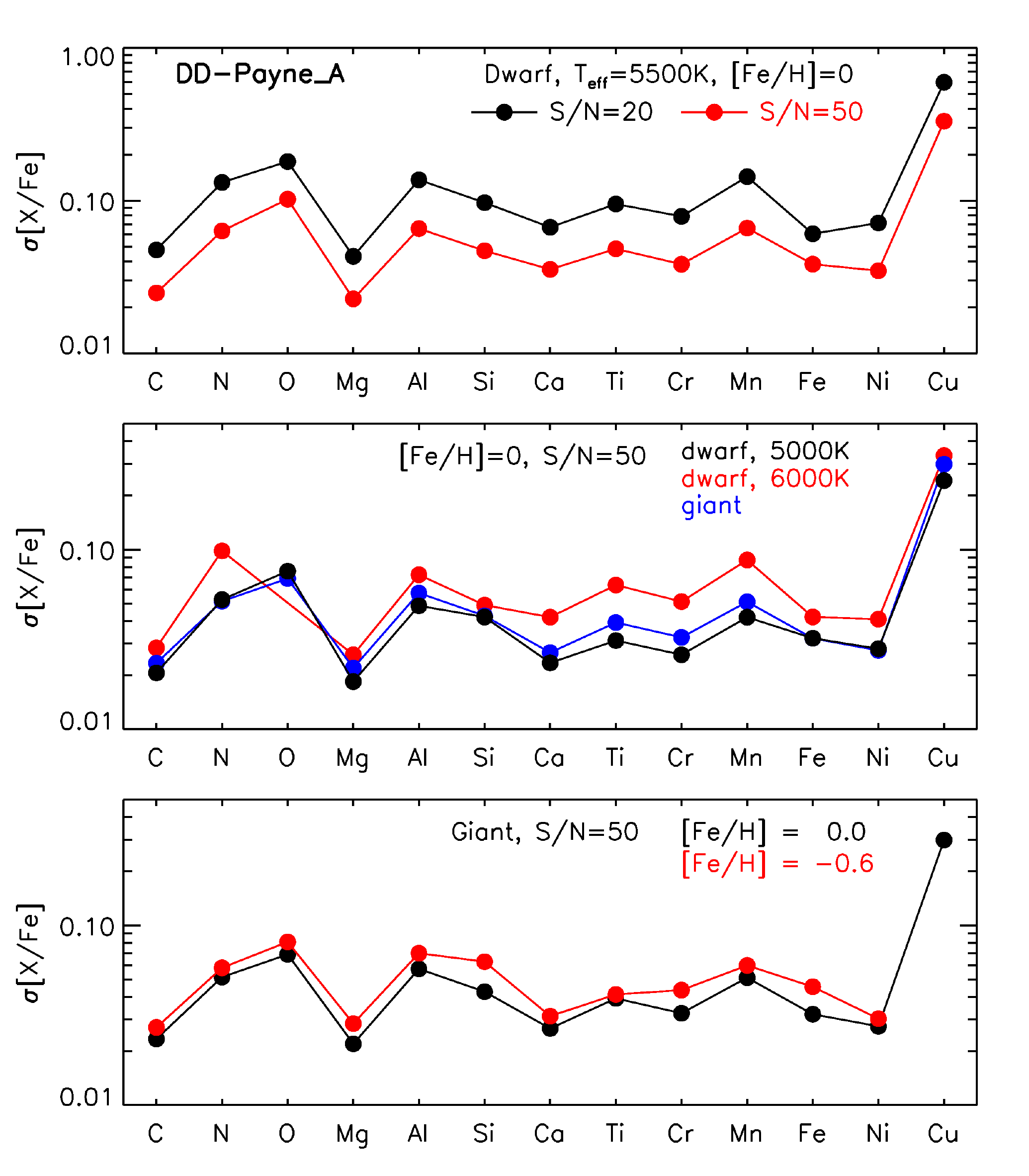}
\caption{Similar to Fig.\,\ref{fig:Fig10}, but for the internal precision of elemental abundances derived using the LAMOST--APOGEE training set. For stars with ${\rm S/N}>50$, the internal abundance precision for most elements, except for Cu, is $\lesssim$0.05\,dex.}
\label{fig:Fig11}
\end{figure*}

Similar to Fig.\,\ref{fig:Fig10}, Fig.\,\ref{fig:Fig11} shows the internal precision for the LAMOST abundances derived using the LAMOST--APOGEE training set. The internal precision is at the same level or slightly better than those derived through the LAMOST--GALAH training set as shown in Fig.\,\ref{fig:Fig10}. For most elements, the abundance precision reaches 0.05\,dex in a wide range of parameter space at S/N$\,=\,$50. Furthermore, for a few elements such as C, Mg, Ca, Ti, Cr, Fe, and Ni, we attain a precision of $\sim$0.03\,dex for both giants and dwarfs cooler than $T_{\rm eff}=5500$\,K. The Cu abundance, however, is less precise, with a typical precision of $\sim$0.3\,dex due to a lack of prominent Cu features in the LAMOST spectra. 

In summary, we attain an internal abundance precision of 0.05\,dex for most elements from the LAMOST spectra at S/N$\,>\, 50$, an encouraging result considering the LAMOST spectral resolution is only $R\,\simeq\,1800$. The precisions are approaching but still within the theoretical limits suggested by \citet{Ting2017b}. For [Cu/Fe] and [Ba/Fe], the uncertainties are relatively large, and they should be used with caution. Although not shown here, we have verified that [Ba/Fe] and [Cu/Fe] estimates -- albeit with large uncertainties -- are valuable in picking out peculiar stars with extreme abundances, e.g., Ba-rich stars.

\subsection{Assessment of systematic errors in the abundance determinations} \label{systematics}
Validations in the above sections demonstrate that the $DD$--$Payne$ has tied the LAMOST abundances to the GALAH/APOGEE scale. However, on the flip side, it also implies that the labels inherit systematic errors of GALAH and APOGEE. Systematic errors are almost unavoidable since the current abundance determinations rely on stellar atmospheric models built with imperfect astrophysical assumptions (e.g. 1D, LTE) and atomic and molecular databases. Furthermore, systematics also arise due to various spectral fitting techniques employed, for instance, the determination of continuum. It is impossible to accurately assess the systematic errors of the abundances as we are lacking of sufficient calibration stars with well-known ground-truth abundances. Nonetheless, meaningful insights could be gained by comparing abundances or their derivatives from APOGEE and GALAH as they are derived with different techniques in different wavelength windows using different stellar atmospheric models. 

\begin{figure*}
\centering
\includegraphics[width=160mm]{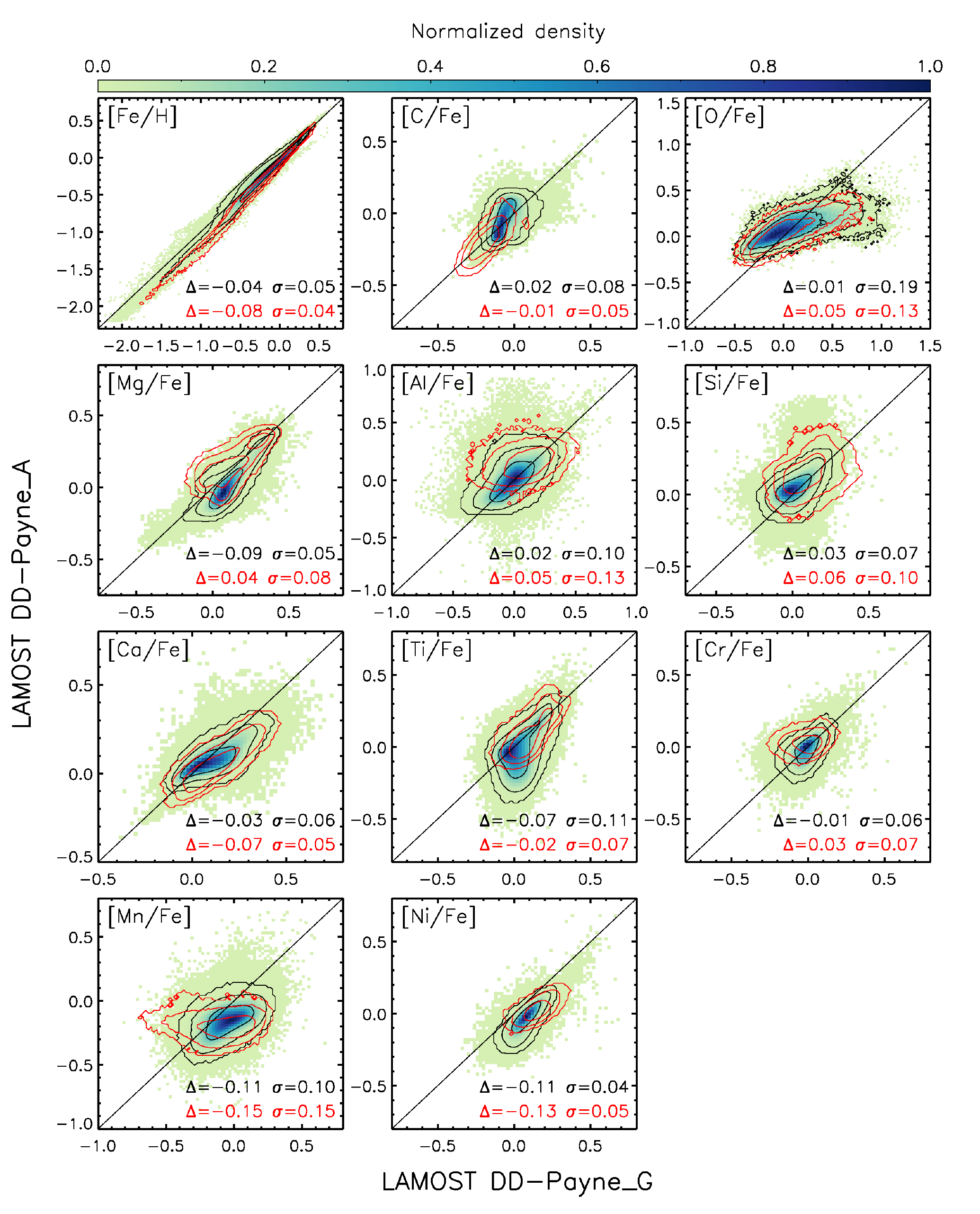}
\caption{Comparison of the $DD$--$Payne$-derived LAMOST abundances, by adopting either the LAMOST--GALAH training set (labeled as ``LAMOST DD-Payne\_G") or the LAMOST--APOGEE training set (labeled as ``LAMOST DD-Payne\_A"). The data points show all the dwarf stars, overlaid by the contours that enclose 68.3\%, 95.4\% and 99.7\% of the dwarfs (black) and the giants (red). The median difference and standard deviation between two different sets of LAMOST inferred labels in this study are marked in each panel with the same color coding as the contours. The dispersion is generally small in most cases, indicating high internal precision. However, there are significant systematic differences ($\geq$0.1\,dex) for a few elements, such as Fe, Mg, Mn, and Ni. These systematics also differ between the giants and the dwarfs. These trends are consistent with a direct comparison of GALAH DR2 and APOGEE--$Payne$ abundances (see Fig.\,\ref{fig:Fig25} in Appendix), supporting that, these systematics are caused by the discrepancy between the GALAH and APOGEE abundance scales, and our inferences inherit the systematics of the two training sets. Due to this discrepancy, for each elemental abundance in our combined and recommended catalog, we will recommend which inference to adopt based on other tests.}
\label{fig:Fig12}
\end{figure*}

Fig.\,\ref{fig:Fig12} demonstrates the comparison between LAMOST abundances derived using the LAMOST--GALAH training set and those derived using the LAMOST--APOGEE training set for common elements. We only show the results using LAMOST DR5 stars with ${\rm S/N}>50$ as they are more robustly determined. For convenience, throughout this paper, we will adopt the notation  ``[X/Fe]\_G" and ``[X/Fe]\_A" to refer to the abundances derived using the LAMOST--GALAH and LAMOST--APOGEE training sets, respectively.

The figure vividly demonstrates that systematic differences are common for the majority of elements when adopting different training data. Moreover, dwarfs and giants also show different trends. Typical values of systematic differences are comparable to the dispersion, which are $\sim$0.1\,dex. In particular, the [Fe/H] values derived from the LAMOST--APOGEE training set are systematically lower than the values from LAMOST--GALAH by 0.04\,dex for dwarfs and 0.08\,dex for giants. We also found that the [Fe/H] differences of dwarfs are strongly $T_{\rm eff}$-dependent. For relatively hot ($T_{\rm eff}>6300$\,K) stars, the APOGEE values can be higher than the GALAH stars, the opposite trend as for the cooler stars. 

As for [C/Fe], although the mean difference for the overall sample is small ($<$0.02\,dex for both dwarfs and giants), ${\rm [C/Fe]\_A }$ exhibits broader distribution than ${\rm [C/Fe]\_G}$ because of a systematic trend. Similarly, although the overall difference of ${\rm [O/Fe]}$ is small, the ${\rm [O/Fe]\_A}$ and ${\rm [O/Fe]\_G}$ also exhibit a weak systematic trend. ${\rm [Mg/Fe]\_A}$, on the other hand, is 0.09\,dex lower for dwarfs and 0.04\,dex higher for giants when compared to ${\rm [Mg/Fe]\_G}$. We also found that the APOGEE--$Payne$ [Mg/Fe] values for dwarfs show non-negligible $T_{\rm eff}$-dependence --- stars with higher $T_{\rm eff}$ have lower [Mg/Fe]. Such $T_{\rm eff}$-dependence is not prominent in the GALAH [Mg/Fe]. The origin of such a difference is unclear, but we note that obtaining reliable elemental abundances for dwarf stars from the H-band APOGEE spectra has been challenging. For APOGEE--$Payne$, the dwarf abundances are known to be less precise than that of the giants \citep{Ting2019}. 

\begin{figure*}
\centering
\includegraphics[width=160mm]{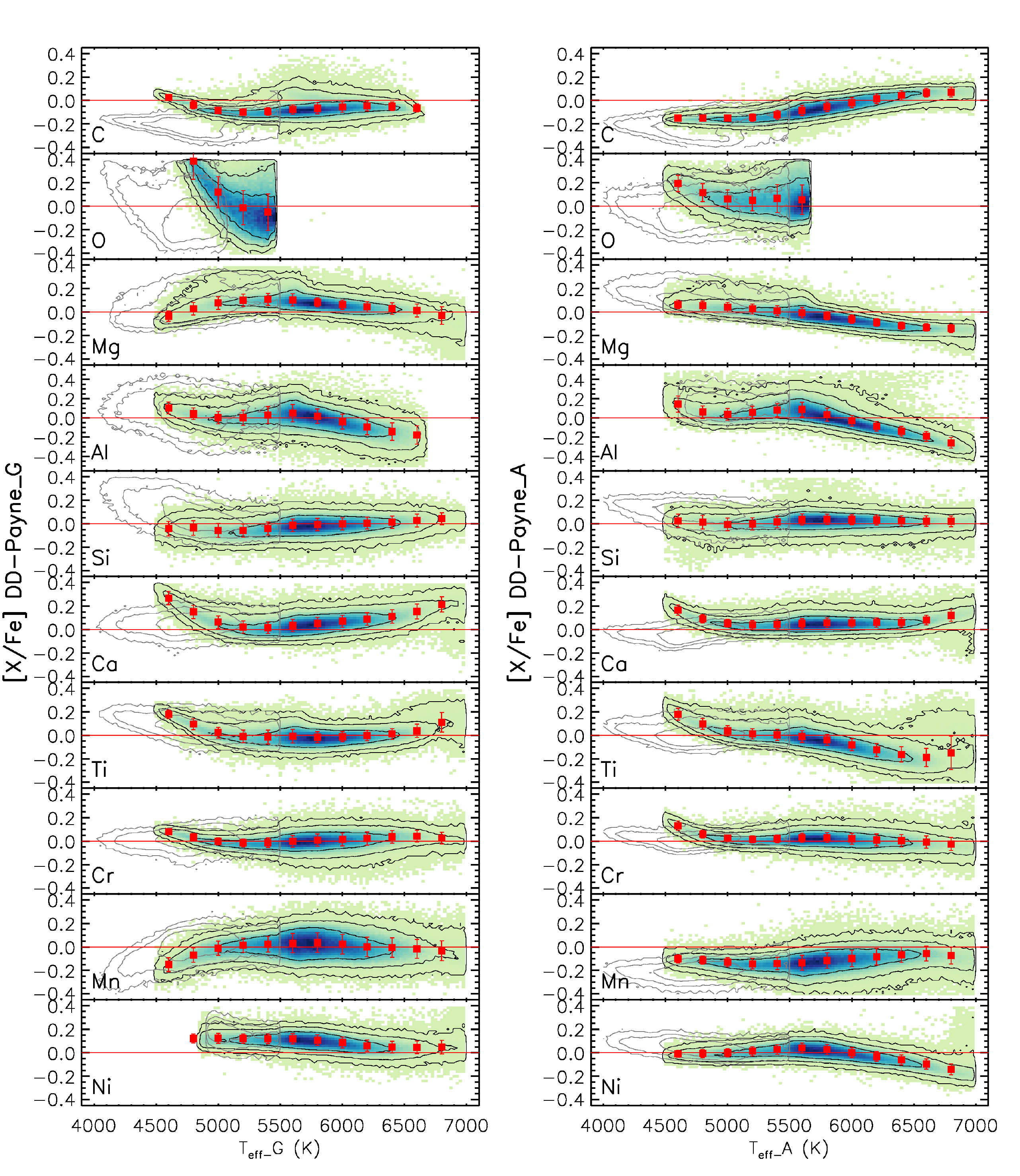}
\caption{Inferred LAMOST stellar abundances as a function of $T_{\rm eff}$ for stars with solar metallicity ($-0.2<{\rm [Fe/H]}<0.2$). The {\em left} column shows the results from the LAMOST--GALAH training set, and the {\em right} column the LAMOST--APOGEE training set. The background and contours represent the same subsets as in Fig.~\ref{fig:Fig12}.
The red symbols illustrate the median values and standard deviations at different $T_{\rm eff}$ of the dwarfs. We discard stars that are not physically determined (see Fig.~\ref{fig:Fig2}), which leads to different $T_{\rm eff}$ ranges in different panels. In most cases, there is no strong $T_{\rm eff}$-abundance trend, demonstrating that our abundance determination is robust. Nonetheless, [Mg/Fe], [Al/Fe], [Ti/Fe], and [Ni/Fe] determined with the LAMOST--APOGEE training set, as well as [Al/Fe] and [Ca/Fe] determined with LAMOST--GALAH training set, show a non-negligible trend with $T_{\rm eff}$. We verified that these systematic trends are inherited from the APOGEE--$Payne$ and GALAH DR2 values in the training set. A similar plot for the GALAH DR2 and APOGEE--$Payne$ abundances is presented in the Appendix (Fig.\,\ref{fig:Fig26}) to support this idea.}
\label{fig:Fig13}
\end{figure*}

For dwarfs, the ${\rm [Al/Fe]\_A}$ and ${\rm [Al/Fe]\_G}$ values are in good agreement, but the ${\rm [Al/Fe]\_A}$ value of giants is slightly higher ($\sim$0.05\,dex) than ${\rm [Al/Fe]\_G}$. Similar systematics are also observed for [Si/Fe] and [Cr/Fe]. The ${\rm [Ca/Fe]\_A}$ value for giants is $\sim$0.07\,dex systematically lower than ${\rm [Ca/Fe]\_G}$, while for dwarfs, the deviation arises mainly for stars with enhanced ${\rm [Ca/Fe]\_G}$, leading to $-0.03$\,dex difference for the overall dwarf sample. For ${\rm [Ti/Fe]}$, there is a significant fraction of dwarf stars whose ${\rm [Ti/Fe]\_G}$ is approximately solar value but the corresponding ${\rm [Ti/Fe]\_A}$ value is lower, leading to a $-0.07$\,dex difference on average. This bias is caused by a strong temperature trend of ${\rm [Ti/Fe]\_A}$ for stars with $T_{\rm eff}>5500$\,K (see Fig.\,\ref{fig:Fig13}). For giant stars, the ${\rm [Ti/Fe]\_A}$ value is consistent with the ${\rm [Ti/Fe]\_G}$ value. The ${\rm [Mn/Fe]\_A}$ value, on the other hand, is 0.1--0.2\,dex systematically lower than the ${\rm [Mn/Fe]\_G}$ value for both dwarfs and giants; Similarly for [Ni/Fe]. Finally, we note that these biases are not due to our spectral fitting method. A direct comparison of the APOGEE--$Payne$ and the GALAH DR2 abundances for a sample of 500 common stars between APOGEE and GALAH is presented in the Appendix (Fig.\,\ref{fig:Fig25}). It shows consistent patterns with those presented in Fig.\,\ref{fig:Fig12}, demonstrating that $DD$--$Payne$ merely inherits the systematic errors from the training sets.

To understand the systematic errors, we further examine the abundance trend as a function of $T_{\rm eff}$. The complication here, however, is that the abundance trend as a function of $T_{\rm eff}$ does not necessarily signify systematic errors of our determinations. The trend could also potentially be caused by the intrinsic variations of abundances.  For example, stars with different $T_{\rm eff}$ (hence brightness and distance) could probe different stellar populations. Furthermore, we also expect to observe a $T_{\rm eff}$ trend for dwarf stars due to stellar atomic diffusion. It is known that the measured photospheric abundances might not coincide with the initial stellar abundances at birth \citep[e.g.][]{Choi2016, Dotter2017, Deal2018}. Even for a population with a single age and the same initial abundances, stars with different masses (and hence $T_{\rm eff}$) can suffer different degrees of atomic diffusion \citep[e.g.][]{Korn2007, Lind2009, Onehag2014, Gao2018, Souto2019}. To limit the latter complication, we opt to study the $T_{\rm eff}$ trend of abundance ratios [X/Fe] rather than [X/H]. The atomic diffusion of an element X and Fe should partially cancel out, and we expect that atomic diffusion is unlikely to be the dominant source for the observed $T_{\rm eff}$-abundance trend. To minimize the impact of stellar populations, we further restrict our sample stars to a narrow [Fe/H] range of $-0.2<{\rm [Fe/H]}<0.2$\,dex. We caution however that the impact of the stellar population can still play a role as hotter stars generally have younger ages. Therefore, the $T_{\rm eff}$-abundance trend should only serve as a guide on the quality of our abundance determinations.
 
With all these caveats in mind, we examine the $T_{\rm eff}$-abundance trend in Fig.\,\ref{fig:Fig13}. We note that since we have discarded stars whose abundances are estimated via astrophysical correlations, the effective range of $T_{\rm eff}$ can vary for different elements. The figure shows that for Si and Cr, both abundances exhibit only a negligible trend in dwarf stars. However, for the other elements, a non-negligible trend with $T_{\rm eff}$ is present in either the result from the LAMOST--GALAH training set (O, Ca), the result from the LAMOST--APOGEE training set (C, Ti, Mn), or in the results from both training sets (Mg, Al, Ni). The trends are the most prominent for stars with either $T_{\rm eff}>5800$\,K or $T_{\rm eff}<5000$\,K. For Mg, Mn, and Ni, the systematic offset between abundance ratios from the two training sets are also clearly visible, as was shown in Fig.\,\ref{fig:Fig12}. We verified that these systematics are not due to the limitation of our method. In the Appendix (Fig.\,\ref{fig:Fig26}), we will present a similar examination directly comparing the GALAH DR2 and APOGEE--$Payne$ abundances using their common stars. We find similar trends as in Fig.\,\ref{fig:Fig13}, suggesting that the $T_{\rm eff}$-abundance trends, whatever their causes might be, are mostly inherited from the training sets. 

\section{The LAMOST DR5 abundance catalog} \label{lamostdr5abundancecatalog}
In light of the non-negligible systematics as described in the previous section, we provide a set of recommended labels by combing results from the LAMOST--GALAH and LAMOST--APOGEE training sets. For completeness, we publish also the individual results in separate supplement catalogs. For each star and label, we provide the measurement uncertainty by scaling the fitting uncertainty to the standard deviation as probed by the repeat observations (Section\,\ref{repeatobservation}). We further provide a few flags to guide the quality of our catalog, including (1) two flags which assess the $\chi^2$ value of the fit; (2) a flag which describes the similarity between gradient spectra from the $DD$--$Payne$ and those from the Kurucz models to evaluate if the label is derived from expected spectral features; (3) a flag which identifies if the object is in a binary/multiple stellar system. Table\,\ref{table:table3} presents a description of the columns in the catalog. The full catalog can be downloaded via \href{url}{http://dr5.lamost.org/doc/vac}. In the following, we will present some key aspects of the catalog. 
\begin{table*}
\caption{Descriptions for the LAMOST $DD$--$Payne$ abundance catalog.\tablenotemark{1}}
\label{table:table3}
\begin{tabular}{ll}
\hline
 Field &    Description  \\
\hline        
 specid &   LAMOST spectra ID in the format of ``date-planid-spid-fiberid" \\
 ra &   Right ascension from the LAMOST DR5 catalog\tablenotemark{2} (J2000; deg)\\
 dec &  Declination from the LAMOST DR5 catalog\tablenotemark{2} (J2000; deg)  \\
 snr\_u/g/r/i/z &  Spectral signal-to-noise ratio per pixel in SDSS u/g/r/i/z-band \\
 rv  & Radial velocity from LAMOST DR5 (km/s) \\
 rv\_err &  Uncertainty in radial velocity (km/s) \\
 $T_{\rm eff}$ &  Effective temperature (K)   \\
 $T_{\rm eff}$\_err  & Uncertainty in $T_{\rm eff}$ (K)\\
 $T_{\rm eff}$\_flag &  A quality flag\tablenotemark{3} for $T_{\rm eff}$ based on the examination of the $DD$--$Payne$ gradient spectra $\partial{f_\lambda}/\partial{T_{\rm eff}}$ \\
 $T_{\rm eff}$\_gradcorr  & Correlation coefficients of $\partial{f_\lambda}/\partial{T_{\rm eff}}$ between the $DD$--$Payne$ and the Kurucz models \\
$\log g$ &  Surface gravity  \\
$\log g$\_err & Uncertainty in $\log g$  \\
$\log g$\_flag & A quality flag\tablenotemark{3} for $\log g$ based on the examination of the $DD$--$Payne$ gradient spectra $\partial{f_\lambda}/\partial\log g$\\
 $\log g$\_gradcorr &  Correlation coefficients of $\partial{f_\lambda}/\partial\log g$ between the $DD$--$Payne$ and the Kurucz model \\
 $V_{\rm mic}$  & Micro-turbulent velocity (km/s)  \\
 $V_{\rm mic}$\_err &  Uncertainty in $V_{\rm mic}$ (km/s)  \\
 $V_{\rm mic}$\_flag &  A quality flag\tablenotemark{3} for $V_{\rm mic}$ based on the examination of the $DD$--$Payne$ gradient spectra $\partial{f_\lambda}/\partial{V_{\rm mic}}$\\
 $V_{\rm mic}$\_gradcorr &  Correlation coefficients of $\partial{f_\lambda}/\partial{V_{\rm mic}}$ between the $DD$--$Payne$ and the Kurucz model \\
 {\rm [Fe/H]} & Iron abundance (dex) \\
 {\rm [Fe/H]}\_err &  Uncertainty in {\rm [Fe/H]} (dex)  \\
 {\rm [Fe/H]}\_flag &  A quality flag\tablenotemark{3} for {\rm [Fe/H]} based on the examination of the $DD$--$Payne$ gradient spectra $\partial{f_\lambda}/\partial{\rm [Fe/H]}$ \\
 {\rm [Fe/H]}\_gradcorr & Correlation coefficients of $\partial{f_\lambda}/\partial{\rm [Fe/H]}$ between the $DD$--$Payne$ and the Kurucz model \\
 {\rm [$\alpha$/Fe]} & $\alpha$-element to iron abundance ratio\tablenotemark{4} \\
 {\rm [$\alpha$/Fe]\_err} & Uncertainty in {\rm [$\alpha$/Fe]} \\
 {\rm [X/Fe]} &  Element-to-iron abundance ratio (dex)  \\
 {\rm [X/Fe]}\_err & Uncertainty in {\rm [X/Fe]} (dex) \\
 {\rm [X/Fe]}\_flag & A quality flag\tablenotemark{3} for {\rm [X/Fe]} based on the examination of the $DD$--$Payne$ gradient spectra $\partial{f_\lambda}/\partial{\rm [X/Fe]}$ \\
 {\rm [X/Fe]}\_gradcorr & Correlation coefficients of gradient spectra $\partial{f_\lambda}/\partial{\rm [X/Fe]}$ between the $DD$--$Payne$ and the Kurucz model \\
 chi2 &  Reduced $\chi^2$ of the spectral fit \\
 chi2ratio & $\chi^2$ excess with respect to the typical value of stars with the same S/N, $T_{\rm eff}$, $\log g$ and [Fe/H] \\
 qflag\_chi2  &  Quality flag based on ``chi2ratio" which takes the values ``good" and ``bad" \\
 chi2\_na & Similar to ``chi2" but for the Na\,{\sc i} $\lambda$5891, 5896{\AA} lines computed using the $\lambda$5880--5910{\AA} segment \\
 chi2ratio\_na  & Similar to ``chi2ratio" but for the Na\,{\sc i} $\lambda$5891,5896{\AA} lines  \\
 qflag\_chi2na  &  Quality flag based on ``chi2ratio\_na" which takes the values ``good" and ``bad" \\
 dsnr\_parallax  & Excess in spectroscopic parallax with respect to the Gaia astrometric parallax \\
 qflag\_singlestar  & Flag to indicate if the object is a single star (``YES") or a binary/multiple system (``NO") \\
 uqflag & Flag to indicate repeat visits; uqflag = 1 means unique star, uqflag = 2, 3, ..., $n$ indicates the $n$th repeat visit \\
  &  For stars with repeat visits, the uqflag is sorted by the spectral S/N, with uqflag = 1 having the highest S/N \\
 starid  & A unique ID for each unique star based on its RA and Dec, in the format of ``Sdddmmss$\pm$ddmmss" \\ 
 subclass  & Stellar subclass from LAMOST DR5 \\
 filename & Name of the LAMOST spectral fits file \\
 \hline
\end{tabular}
\tablenotetext{1}{The full electronic catalog can be found at \href{url}{http://dr5.lamost.org/doc/vac}.}
\tablenotetext{2}{The LAMOST DR5 catalog as well as the spectra can be found at \href{url}{http://dr5.lamost.org}.}
\tablenotetext{3}{The flag takes the values 0 and 1. Flag = 0 means that the label estimates could be affected by astrophysical correlations, flag = 1 (reliable) otherwise.}
\tablenotetext{4}{The [$\alpha$/Fe] is defined as a weighted mean of 
 [Mg/Fe], [Si/Fe], [Ca/Fe] and [Ti/Fe], see Section\,\ref{labelspace}.}
\end{table*}

\subsection{Recommended Labels} \label{recommendlabel}
We provide a set of recommended labels by combing results from the two training sets. The recommendation is based mainly on the assessment of systematic errors in Section\,\ref{systematics}. Table\,\ref{table:table4} presents our recommendation for the high-resolution source of the training set for each label. In summary, for $T_{\rm eff}$, $\log g$, $V_{\rm mic}$, [Fe/H], [C/Fe], [N/Fe], [O/Fe], [Ca/Fe], [Ni/Fe], and [Cu/Fe], we adopt the results derived from the LAMOST--APOGEE training set; while for other labels, namely, [Na/Fe], [Mg/Fe], [Al/Fe], [Si/Fe], [Ti/Fe], [Cr/Fe], [Mn/Fe], [Co/Fe], and [Ba/Fe], results from the LAMOST--GALAH training set are adopted. 
\begin{table*}
\centering
\caption{Training sets for the recommended $DD$--$Payne$ stellar labels.}
\label{table:table4}
\begin{tabular}{lllllllllll}
\hline
\textbf{Label} & {$T_{\rm eff}$} & $\log g$ & {$V_{\rm mic}$} & {\rm [Fe/H]} & [C/Fe] & [N/Fe] & [O/Fe] & [Na/Fe] & [Mg/Fe] & [Al/Fe]  \\
\hline
\textbf{Source} & APOGEE & APOGEE & APOGEE & APOGEE & APOGEE & APOGEE & APOGEE & GALAH & GALAH & GALAH   \\ 
\hline
\textbf{Label} & [Si/Fe] & {\rm [Ca/Fe]} & [Ti/Fe] & [Cr/Fe] & [Mn/Fe] & [Co/Fe] & [Ni/Fe] & [Cu/Fe] & [Ba/Fe]  &  \\
\hline
\textbf{Source} & GALAH & APOGEE & GALAH & GALAH & GALAH & GALAH & APOGEE & APOGEE & GALAH &   \\
\hline
\end{tabular}
\begin{tablenotes}
\item[]{APOGEE = The $Payne$ results for the APOGEE DR14 sample \citep{Ting2019}.}  
\item[]{GALAH = The GALAH DR2 catalog \citep{Buder2018}.}
\end{tablenotes}
\end{table*}

In particular, for stellar parameters ($T_{\rm eff}$, $\log g$ and [Fe/H], we recommend results from the LAMOST--APOGEE training set as it contains more overlapping stars than the LAMOST--GALAH training set, and thus have better coverage in the stellar parameter space (Section\,\ref{lamostapogee-trainingset}). Because GALAH does not provide estimates for nitrogen, the results from the LAMOST--APOGEE training set are adopted for [C/Fe], [N/Fe] and [O/Fe]. This choice is to ensure that in our final catalog the C, N, and O are self-consistently estimated from the same training set. We recommend [Ca/Fe] from the LAMOST--APOGEE training set because the results exhibit a weaker trend with $T_{\rm eff}$ (Fig.\,\ref{fig:Fig13}). We also recommend [Ni/Fe] from the LAMOST--APOGEE training set. [Ni/Fe] from the LAMOST--GALAH training set for giants exhibits weaker correlations between the gradient spectra, and is more susceptible to astrophysical correlations. Similarly, [Cu/Fe] from the LAMOST--APOGEE training set is recommended because the values from the LAMOST--GALAH training set are more susceptible to astrophysical correlations. For other labels, the LAMOST--GALAH results are recommended. The choices are prompted either by the fact that they are only measured in the GALAH (optical) spectra or, as presented in Fig.\,\ref{fig:Fig13}, the LAMOST--GALAH results demonstrate a weaker $T_{\rm eff}$-abundance trend.

However, we emphasize that the choice does not necessarily mean that all the recommended labels  have the better internal precision or accuracy. For instance, the [Mg/Fe] and [Si/Fe] values from the LAMOST--APOGEE training set in fact have better internal precision (Figs.\,\ref{fig:Fig6} and \ref{fig:Fig7}). Also, as discussed, the strong $T_{\rm eff}$--[Mg/Fe] trend in results from the LAMOST--APOGEE training set might be physical due to stellar population effects. Moreover, a combination of the two training sets could introduce some level of inconsistency. For example, the abundances of the $\alpha$-elements, Mg, Si, Ca and Ti no longer come from a unified analysis, so there could be mismatch in abundance scales among them. To sum up, while we provide a recommended catalog for convenience, we urge users to make their own choices according to their scientific interests. To facilitate this option, we therefore also provide two individual catalogs that contain labels derived from the two training sets alongside with the recommended combined catalog.

\subsection{Stellar distribution in label space} \label{labelspace}
The recommended catalog contains label estimates for 5,979,381 unique stars from 8,162,566 spectra. About three quarters of the sample stars have ${\rm S/N_{\rm pix}}>30$ in at least one of the SDSS $g$, $r$, and $i$ bands. For 95 percent of the sample stars, the $DD$--$Payne$ has carried out good spectral fitting, i.e., qflag\_chi2 = ``good". Among them, 4,259,093 stars have physically sensible abundance estimates for at least 10 elements.   

\begin{figure*}
\centering
\includegraphics[width=180mm]{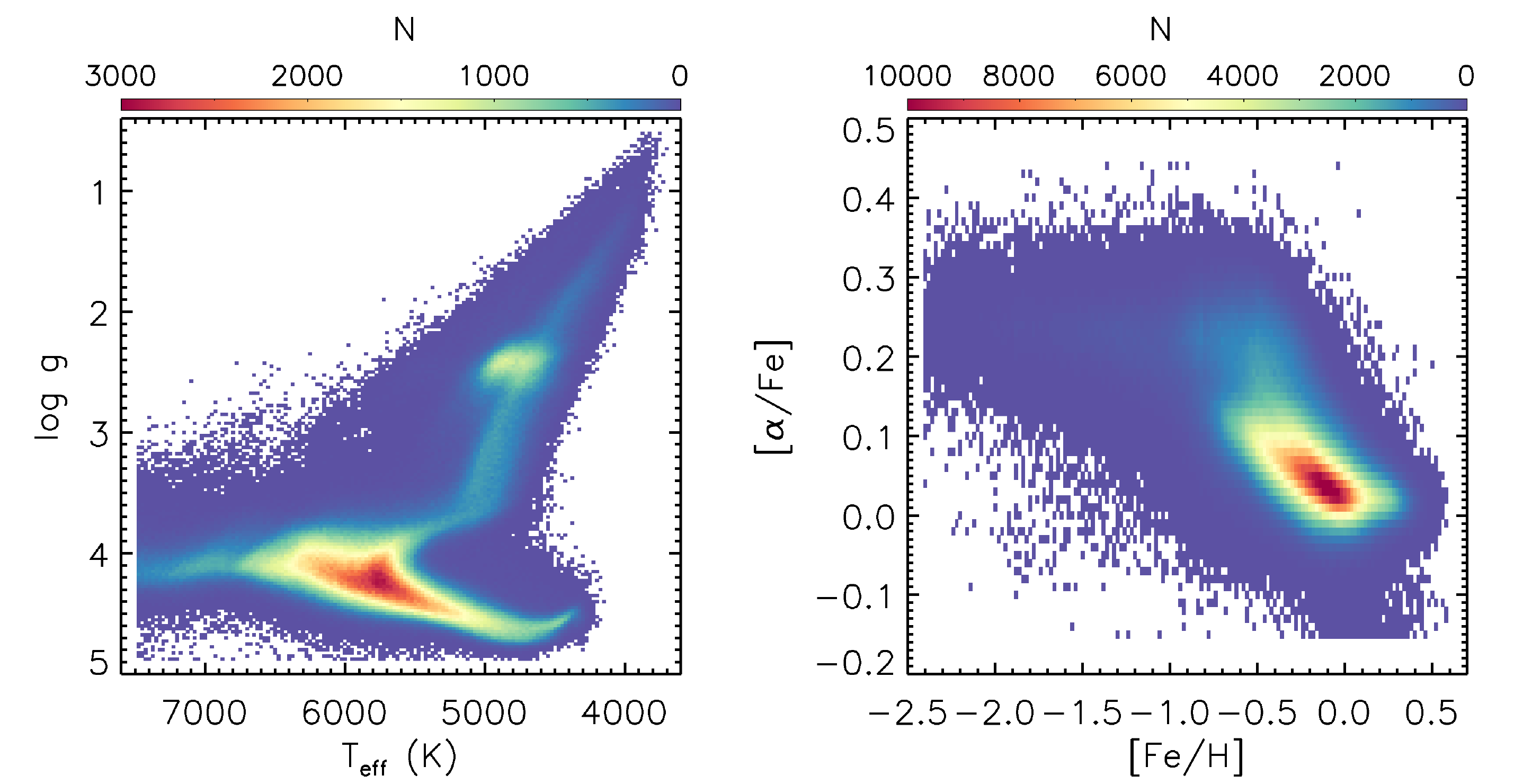}
\caption{The $T_{\rm eff}$--$\log g$ and [Fe/H]--[$\alpha$/Fe] diagrams for a sample of 2,932,585 LAMOST DR5 stars that have good S/N ($>$30) and decent $DD$--$Payne$ fits (qflag\_chi2 = ``good"). Both figures are color-coded by the stellar number density. We define [$\alpha$/Fe] to be the average of [Mg/Fe], [Si/Fe], [Ca/Fe] and [Ti/Fe] weighted by their inverse variance.  
On the {\em left}, while dwarf stars dominate the sample in the $T_{\rm eff}$--$\log g$ diagram, the red giant branch, and red clump, are also clearly visible, demonstrating the excellent quality of the $DD$--$Payne$ determined stellar parameters. On the {\em right}, the LAMOST sample occupies mostly the low-$\alpha$ sequence in the  [Fe/H]--[$\alpha$/Fe] diagram because, as already shown in the {\em left} panel, a large fraction of the LAMOST sample are consist of nearby dwarf stars in the thin disk. Nonetheless, the $DD$-$Payne$ values also recover the well-known high-$\alpha$ sequence, supporting the idea that the $\alpha$-abundances are determined through {\it ab initio} spectral features, and are not merely recovered through astrophysical correlations (e.g., through the correlation with [Fe/H]). Although the sample stars are cut off at ${\rm [Fe/H]}=-2.5$\,dex for illustration, the [Fe/H] values in the catalog may reach $-4.0$\,dex at the metal-poor end.}
\label{fig:Fig14}
\end{figure*}

Fig.\,\ref{fig:Fig14} shows the distribution of stellar number density in the $T_{\rm eff}$--$\log g$ (Kiel) diagram and in the [Fe/H]--[$\alpha$/Fe] diagram. For the former, the left panel demonstrates that our derived LAMOST stellar parameters are precise --- the main sequence, red giant branch, as well as the red clump are clearly distinguishable in this Kiel diagram.

\begin{figure*}
\centering
\includegraphics[width=180mm]{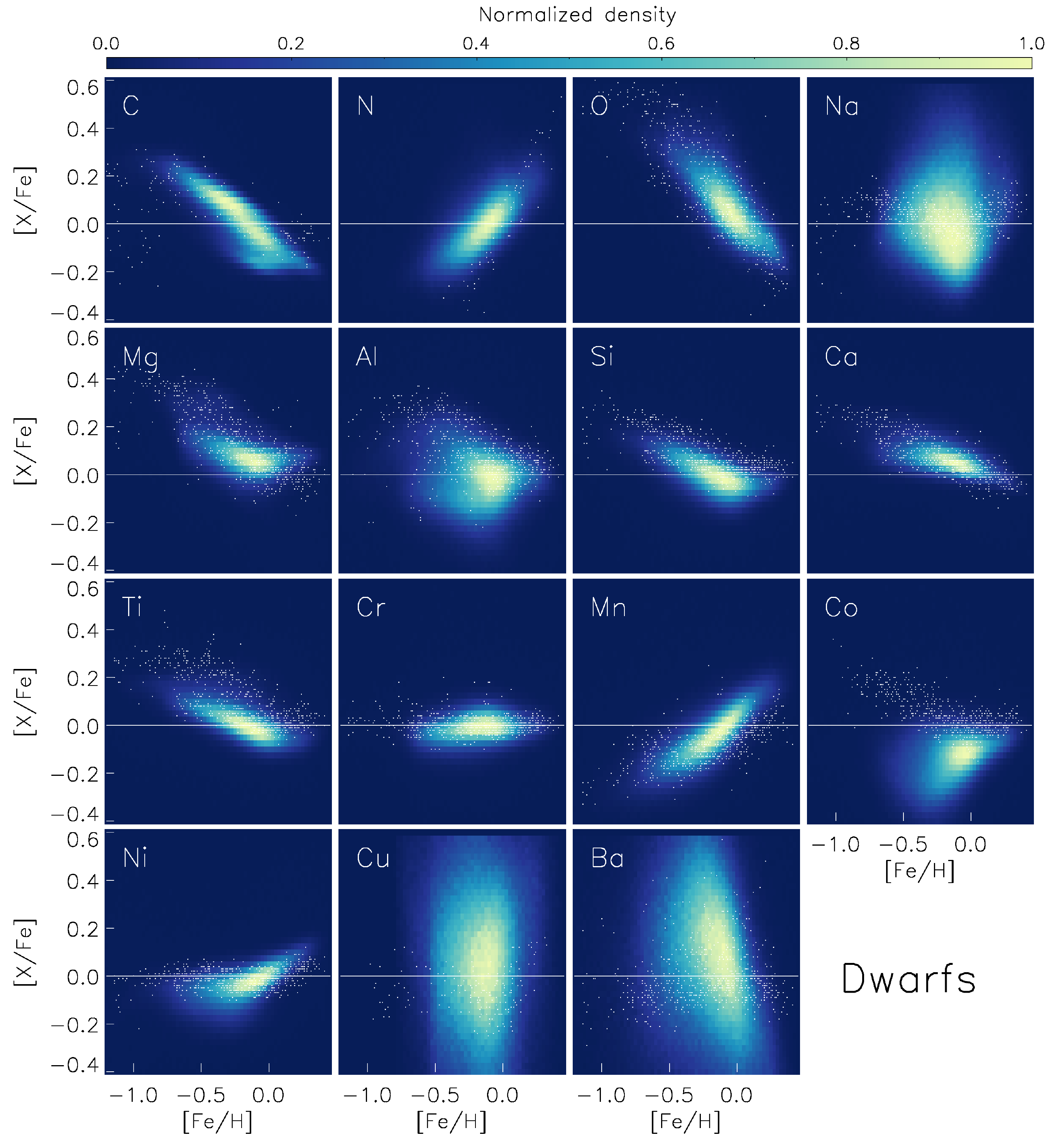}
\caption{Elemental abundance distributions of dwarf stars in the [X/Fe]--[Fe/H] plane. All subplots are color-coded with the stellar number density. Only the recommended set of abundances are shown. Since stars with [X/Fe]\_flag = 0 are discarded, different elements can have slightly different [Fe/H] range. The white dots are literature results from high-resolution spectroscopy, namely \citet{Mishenina2011} for Cu, \citet{Bensby2014} for O, Na, Mg, Al, Si, Ca, Ti, Cr, Ni, and Ba, \citet{Nissen2014} for C and O, \citet{Battistini2015} for Mn and Co, \citet{Suarez-Andres2016} for N, and \citet{Zhao2016} for C and Cu. Note that, depending on the sampling strategy, some literature values probed different Galactic populations, and the comparison is not one-to-one. Nonetheless, for most elements, the overall $DD$--$Payne$ abundance trend is consistent with the high-resolution results, except for Co --- the $DD$--$Payne$ results show an opposite trend, likely a consequence of the lack of good Co abundances for our training stars (see Section\,\ref{lamostgalah-trainingset}). For Cu and Ba, the abundance uncertainties from LAMOST are significantly larger than those from the high-resolution results due to the lack of strong spectral features at the LAMOST resolution.}
\label{fig:Fig15}
\end{figure*}

On top of that, the right panel of Fig.\,\ref{fig:Fig14} demonstrates that our abundances of $\alpha$-elements are also reliably determined. The [$\alpha$/Fe] is defined as the weighted mean of [Mg/Fe], [Si/Fe], [Ca/Fe], and [Ti/Fe], specifically, 
\begin{equation}
{\rm [\alpha/Fe]} = \frac{\sum_{\rm X}\,{\omega_{\rm X}\cdot{\rm [X/Fe]}}}{\sum_{\rm X}\,{\omega_{\rm X}}}, 
\end{equation}
where X=Mg, Si, Ca, Ti, and $\omega_{\rm X} = 1/\sigma_{\rm [X/Fe]}^2$.
To ensure that the [$\alpha$/Fe] is uniformly defined for all stars, we opt not to apply any cut on the correlation coefficients of gradient spectra between the $DD$--$Payne$ and the Kurucz model for abundances of the individual $\alpha$-elements. Defining [$\alpha$/Fe] uniformly is important because the abundance of individual $\alpha$-elements (Mg, Si, Ca, Ti) might not be in the same scale. For example, the values of [Mg/Fe] and [Ca/Fe] for dwarfs are systematically higher than [Si/Fe] and [Ti/Fe] (Fig.\,\ref{fig:Fig15}). We caution that for stars of which the abundances of individual $\alpha$-elements are not physically estimated, for instance, the metal-poor stars with ${\rm [Fe/H]}<-1.0$\,dex (Fig.\,\ref{fig:Fig2}), the [$\alpha$/Fe] may be susceptible to astrophysical correlations. The [Fe/H]--[$\alpha$/Fe] diagram shows that, as expected, our sample is dominated by stars in the thin disk sequence. But importantly, the thick disk sequence at $-1<{\rm [Fe/H]}<-0.5$\,dex is also clearly visible, demonstrating that [$\alpha$/Fe] of stars in this [Fe/H] range are not merely inferred from the astrophysical correlations (e.g., the [Fe/H]--[$\alpha$/Fe] correlation). 

Although the sample is dominated by disk stars with ${\rm [Fe/H]}>-1.0$\,dex, there is also a considerably large number of metal-poor stars. For example, there are 15,773 very metal-poor stars with ${\rm [Fe/H]}<-2.0$\,dex that have qflag\_chi2 = ``good" and ${\rm S/N_g}>20$. At the metal-poor end, the stars may have an [Fe/H] as low as $-4.0$\,dex. As discussed in Section\,\ref{lamostapogee-trainingset}, due to the limitation of the training set, labels for stars with ${\rm [Fe/H]}<-1.5$\,dex are derived by extrapolation, and thus should be used with cautious. Nonetheless, we declare that, with appropriate quality cut, the [Fe/H] estimates at the metal-poor end are reliable, at least for selecting metal-poor star candidates.
Note however that, the boundary for [X/Fe] values derived from the LAMOST--APOGEE training set may show a discontinuity at ${\rm [Fe/H]}\sim-1$\,dex. This is caused by the fact that, in order to achieve an extrapolation to ${\rm [Fe/H]}<-4.0$\,dex, we have allowed the $DD$--$Payne$ to extrapolate in a larger volume when running the LAMOST--APOGEE metal-poor training set. In principle, this defect can be fixed by re-running the overall LAMOST--APOGEE training set to enlarge the volume for extrapolation. However, considering that this is only a marginal effect, we leave it unchanged in the catalog.

\begin{figure*}
\centering
\includegraphics[width=180mm]{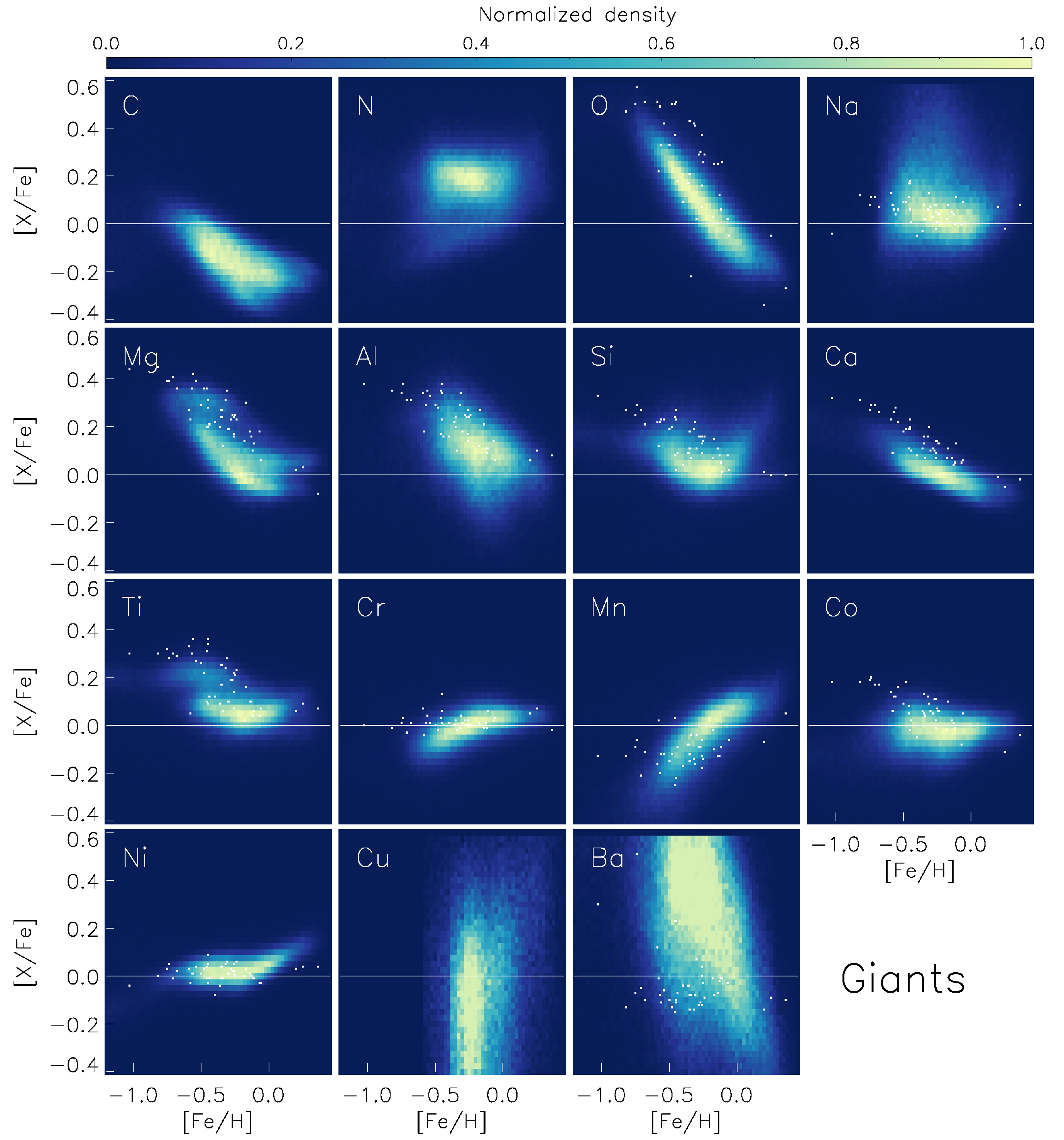}
\caption{Same as Fig.\,15, but for the giant stars ($T_{\rm eff}<5500$\,K and $\log g<4.1$). The white dots are high-resolution results for giants from the same literatures as in Fig.\,15.}
\label{fig:Fig16}
\end{figure*}

Figs.\,\ref{fig:Fig15} and \ref{fig:Fig16} show the stellar density distributions in abundance space for dwarfs and giants, respectively. The symbols overplotted show the literature values. As a whole, the abundance trends are encouraging and are consistent with literature results from high-resolution studies. We do not expect a perfect match, considering that the LAMOST sample has different selection function from the literature samples. Nevertheless, our [Co/Fe] for dwarfs shows an opposite trend with the literature results, likely a consequence of the lack of good Co abundance for our training sets (see Section\,\ref{lamostgalah-trainingset}). Finally, our [Cu/Fe] and [Ba/Fe] show large spreads. As discussed, Cu and Ba have only a few weak spectral features in the LAMOST spectra. Therefore, their derived abundances have large uncertainties and should be used with caution.

\subsection{Estimating Uncertainties} \label{uncertainty}
Our catalog provides uncertainty estimates for individual labels of each star. The uncertainties are estimated by scaling the formal fitting uncertainties from the $\chi^2$ fits to match the standard deviations of repeat observations. The scaling is necessary because quoting only the formal statistical fitting uncertainties will almost certainly underestimate the uncertainties. The formal uncertainties reflect only the reported flux uncertainties but do not account for other sources of error, such as the fiber-to-fiber variation of the line spread function and its time variation \citep{Xiang2015b}. Furthermore, scattered light, poor fiber-flat fielding, and background subtraction could all further introduce additional systematic errors that are not accounted for. 

Due to the systematics mentioned above, label uncertainties estimated using repeat observations, are usually larger than the formal fitting uncertainties. The difference is especially prominent at high S/N because the systematic errors in the spectra dominate over the photon noise. We found that uncertainties from repeat observations are $\sim$1.2 times larger than the fitting uncertainties at S/N$\,=\,$20, and $\sim$4 times larger at S/N$\,=\,$200. Moreover, the difference also depends on $T_{\rm eff}$, and [Fe/H]. To account for that, we adopt a 3-order polynomial to model the scaling factor as a function of S/N, $T_{\rm eff}$ and [Fe/H], and we build models for dwarfs and giants separately. The scale factor is defined to be the ratio of the uncertainties derived from repeat observations and the median value of the fitting uncertainties at a given S/N, $T_{\rm eff}$, and [Fe/H] bin. For each star, we then read off from the scaling model and scale the uncertainties accordingly.

\subsection{Quality flags} \label{flags}
Several flags are provided in the catalog to assess the quality of the label determinations and to identify erroneous estimates. We urge users to take these flags into account when adopting stellar labels in this study.

\subsubsection{$\chi^2$ anomalies} \label{chi2anomaly}
The first flag, labeled as ``qflag\_chi2", describes the global quality of the spectral fits. There are several reasons why a spectrum could have a bad fit (a large $\chi^2$). For example, the line spread function of a star could deviate significantly from the average. 
Furthermore, fitting stars whose parameters differ significantly from the covered range of the training sample might also show large $\chi^2$ value. However, the absolute value of $\chi^2$ is not a direct indicator of the quality of the fits. For example, we found that stars with higher S/N usually have larger $\chi^2$ value, a consequence of the fact that, as has been discussed above, the reported flux uncertainties are inadequate to account for the true uncertainties of the spectra. To account for that, we construct a 3-order polynomial model, evaluating the median and standard deviation of $\chi^2$ as a function of S/N, $T_{\rm eff}$, and [Fe/H]. We construct different models for dwarfs and giants separately. For dwarfs, only stars with $4500$ K$\,<T_{\rm eff}<7500$\,K are used to constrain the polynomial model because this is approximately the parameter range covered by the training stars. The ``chi2ratio" column in the catalog shows the value of ($\chi^2-\chi^2_{\rm median}$)/$\sigma_{\chi^2}$, where $\chi^2_{\rm median}$ and $\sigma_{\chi^2}$ are evaluated from the polynomial models. It describes the $\chi^2$ excess when compared to the typical $\chi^2$ value of stars with similar S/N, $T_{\rm eff}$, and [Fe/H].

We visually inspect the fits and declare the fits for stars with 
\begin{itemize}
    \item ${\rm S/N}<200$ and ${\rm chi2ratio}>5$
    \item $200<{\rm S/N}<300$ and ${\rm chi2ratio}>7.5$
    \item $300<{\rm S/N}<400$ and ${\rm chi2ratio}>10$
    \item $400<{\rm S/N}<500$ and ${\rm chi2ratio}>12.5$
    \item ${\rm S/N}>500$ and ${\rm chi2ratio}>15$
\end{itemize} 
to be of poor quality. We assign the ``qflag\_chi2" as ``bad" for these stars which constitute about $\sim$6 percent of all LAMOST DR5 stars. We note that this criterion is, unfortunately, somewhat arbitrary. While we find that these cuts effectively pick out stars with poor fits, the criterion could also exclude a considerable fraction of stars with reasonable estimates. 

Finally, noting that the LAMOST spectra may have low quality in the dichroic region, which could impact the determination of sodium abundance, we further introduce the flag ``qflag\_chi2na". The flag is defined in a similar way to ``qflag\_chi2" to pick out stars with bad fits for the $\lambda =\,$5880--5910\,{\AA} segment.

\subsubsection{Consistency of gradient spectra with Kurucz models} \label{consistenctofgradient}
To assess if the labels of a star are physically determined or are inferred through astrophysical correlations, we provide a specific flag, $X$\_flag, for each label $X$. The flag showcases the similarity of the $DD$--$Payne$ gradient spectrum to that of the Kurucz model. We expect that if a label is physically determined from {\it ab initio} spectral features, the $DD$--$Payne$ should predict a gradient spectrum that is similar to that of the Kurucz model (see Section\,\ref{method}). It is straightforward to evaluate the data-driven gradient spectra from the $DD$--$Payne$ neural network models as it simply requires evaluating the neural network. On the other hand, it is expensive to compute the corresponding {\em ab initio} gradient spectra from the Kurucz spectral model for all LAMOST stars evaluated at their labels. The latter requires solving for the stellar atmospheric models as well as radiative transfer. Due to this limitation, for each star, we compare the $DD$--$Payne$ gradient spectra with the Kurucz gradient spectra of a reference star in Table\,\ref{table:table1} that has the closest distance to the target star. Here we adopt the distance metric $D:=\sqrt{(\Delta T_{\rm eff}/100{\rm K})^2+(\Delta\log g/0.2)^2+(\Delta{\rm [Fe/H]}/0.1)^2}$.

The correlation coefficient describes the similarities of the two gradient spectra, and is provided in the catalog as ``$X$\_gradcorr". As introduced in Section\,\ref{method}, we derive the median value of correlation coefficients (separately for dwarfs and giants) in different $T_{\rm eff}$ and [Fe/H] bins. If the median correlation coefficient is larger than 0.5, we assign an $X$\_flag of 1 for all stars in that particular $T_{\rm eff}$ and [Fe/H] bin, otherwise $X$\_flag$\,=\,$0. Only stars with a spectral ${\rm S/N}>30$ are adopted to calculate the median correlation coefficients as they are more reliable. We note that the flags are assigned to all the stars in the $T_{\rm eff}$ and [Fe/H] bins regardless of their S/N. As a result, for a small fraction of stars with low S/N, they may have an $X$\_flag of 1 (passed) even though the correlation coefficient is smaller than 0.5. Finally, we also assign an $X$\_flag of 0 (failed) to stars with $T_{\rm eff}>7500$\,K or stars with $T_{\rm eff}<4500$\,K and $\log g>3.8$, as these stellar parameters are beyond the covered range of the training set, and hence are less reliable.

In the recommended catalog, all elemental abundances (except [Fe/H]) are assigned $-999.0$ if they failed in the examination of the correlation coefficient ($X$\_flag = 0). However, all the basic stellar parameters ($T_{\rm eff}$, $\log g$, [Fe/H]) are reported regardless of their $X$\_flag. For each star, we also provide an [$\alpha$/Fe] estimate according to the definition in Section\,\ref{recommendlabel}. In the supplementary catalogs, where the original results of the $DD$--$Payne$ labels derived from both the LAMOST--APOGEE and LAMOST--GALAH training sets are presented, we provide the labels for all stars regardless of their $X$\_flag for completeness. But users should apply the flags in individual catalogs with care when necessary.

\subsubsection{Binary and multiple star systems} \label{binary}
About half of the stars are expected to be in binary or multiple systems \citep[see e.g.,][]{Duchene2013, Gao2014, Yuan2015c}. For binary and multiple stars, abundance determination could be biased if they are treated like single stars. Although dedicated work has been carried out to investigate the impact of binary on stellar parameters $T_{\rm eff}$, $\log g$, [Fe/H] \citep[e.g.,][]{El-Badry2018}, a comprehensive analysis of the impact on detailed elemental abundances is still largely absent. We will show in this section that our elemental abundances estimates are mostly robust even if they might be in binary or multiple systems. To study that, we first need to figure out the subset of the LAMOST sample that might be in binary or multiple systems. In a separate effort, we have developed a method to effectively eliminate many main-sequence binary/multiple stars from the LAMOST database (Xiang et al. in prep). Here we will briefly summarize our method.

The method is based on the comparison of $Gaia$ astrometric parallax with luminosity parallax. The latter is inferred from the distance modulus using absolute magnitudes deduced from the LAMOST spectra with the $DD$--$Payne$. Stellar absolute magnitudes (luminosity) can be accurately and precisely derived from the spectra as the latter are characterized by stellar atmospheric parameters, e.g., $T_{\rm eff}$, $\log g$ and [Fe/H], which are tightly related with luminosity \citep[e.g.,][]{Xiang2017a}. For binary/multiple systems, we expect their luminosity parallax to be underestimated because the apparent magnitudes from the photometry are over-luminous, whereas the absolute magnitudes from the LAMOST spectra are found to be nearly identical or slightly fainter with respect to that of the primary component \citep[for example, see][]{El-Badry2018}. More quantitatively, for equal mass binaries, the luminosity parallax will be underestimated by 35\%. Such a difference is easily distinguishable because the $Gaia$ parallax has a precision better than 5\% for most stars in our sample. We found that our method is efficient in picking out binary/multiple stars with mass ratios larger than 0.6.

We provide a tag ``snr\_dparallax", defined as ($\omega_{\rm sp} - \omega_{\rm Gaia}$)/$\sqrt{\sigma^2_{\omega_{\rm sp}} +\sigma^2_{\omega_{\rm Gaia}}}$, to describe the likelihood of a star to be in a binary or multiple system. Here $\omega_{\rm sp}$ and $\omega_{\rm Gaia}$ are the spectroscopic parallax and Gaia DR2 parallax, respectively, $\sigma^2_{\omega_{\rm sp}}$ and $\sigma^2_{\omega_{\rm Gaia}}$ are their error estimates. The ``snr\_dparallax" indicates the deviation significance of the spectroscopic parallax from the $Gaia$ astrometric parallax. If the ``snr\_dparallax" of an object is larger than 3$\sigma$, we classify it to be a binary or multiple system and assign ``flag\_singlestar" as ``NO," otherwise ``flag\_singlestar" $=$ ``YES." For subgiant and giant stars with ($T_{\rm eff}<5500$\,K and $\log g<3.6$) or stars with $T_{\rm eff}>7000$\,K, we do not provide any binary tag. The subgiant or giant star is likely to outshine any companion that is not of the equal mass. It is therefore hard to make any conclusion if it is in a binary or multiple system with single epoch spectra. While for stars with $T_{\rm eff}>7000$\,K, our current results have large uncertainties due to the limitation of training set. 

\begin{figure}
\centering
\includegraphics[width=85mm]{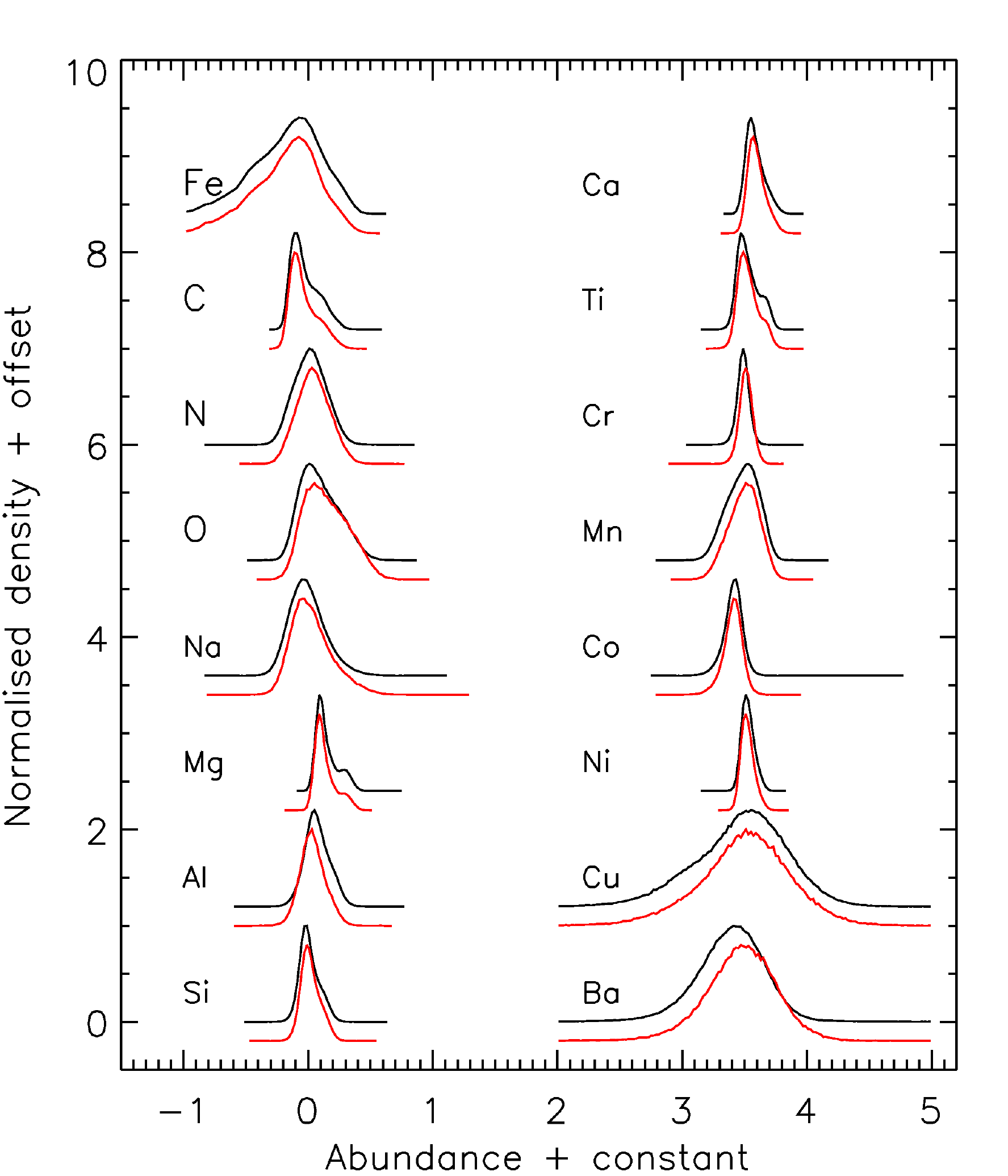}
\caption{Comparison of the [Fe/H] and [X/Fe] distribution between presumed single stars (black) and binary/multiple stars (red). Shown are stars with $5200<T_{\rm eff}<5800$\,K and ${\rm S/N}>50$. Since both subsets should trace the same Galactic chemical evolution, we expect the two distributions to be similar. As shown, the two distributions agree well with each other, suggesting that the abundance determination is also likely to be robust for stars in binary/multiple systems. }
\label{fig:Fig17}
\end{figure}

Since we always treat stars as single stars in our fits in this study, it is important to make sure that our fits are not significantly biased for stars in multiple systems. A complete analysis of the impact requires that we fit a mixture of models to individual stars, as was done with $The$ $Payne$ in \citet{El-Badry2018}, but this is clearly beyond the scope of this paper. Nonetheless, in Fig.\,\ref{fig:Fig17} we compare the abundance distributions of single stars and binary/multiple stars with $5200<T_{\rm eff}<5800$\,K in our sample, as the abundance determination and binary identification are most robust in this temperature range. The figure shows that the abundance distributions for all elements of the single stars and of the binary/multiple stars only show marginal differences. Therefore, we conclude that, statistically speaking, the current abundance determinations for binary stars are not severely biased. We defer a complete analysis to future studies.

\section{Summary} \label{summary}
In this study, we presented stellar parameters ($T_{\rm eff}$, $\log g$, $V_{\rm mic}$) and abundances of 16 elements (C, N, O, Na, Mg, Al, Si, Ca, Ti, Cr, Mn, Fe, Co, Ni, Cu, and Ba) from 8 million low-resolution ($R\simeq1800$) spectra of LAMOST DR5. While other preamble studies have shown that it is plausible, this is the first attempt to derive multiple elemental abundances at such low resolution ($R\simeq1800$) for the full LAMOST catalog, a sample size that is an order of magnitude larger than other high-resolution survey counterparts. The catalog is made fully available electronically.

To maximally extract effective information from the LAMOST low-resolution spectra, we adopt a technique which we dubbed the $DD$--$Payne$ (the data-driven $Payne$). The $DD$--$Payne$ combines the key ideas from two approaches proposed recently. At its core, it follows {\it The Payne} which allows for a fast and accurate interpolation, as well as a data-driven approach similar to $The$ $Cannon$. Such a hybrid approach, as proposed in \citet{Ting2017a}, mitigates some fundamental limitations faced when attempting to extract precise stellar labels from low-resolution spectra: On one hand, {\em ab initio} spectral models as was done in {\it The Payne} might not fully resemble the observed spectra, especially at low resolution, due to model systematics; on the other hand, adopting a data-driven model might risk inferring stellar labels from astrophysical correlations.

To mitigate these limitations, we adopt the common stars between LAMOST and GALAH/APOGEE as the training set to build a data-driven model. On top of that, the {\em ab initio} gradient spectra, i.e., the response of spectral flux to the change of stellar labels, calculated from the Kurucz models are used as the physical constraints to regularize the training process. We further compare the gradient spectra predicted by the $DD$--$Payne$ for all LAMOST DR5 stars with the Kurucz gradient spectra. We ensure that, for each star, the labels are measured through physical spectral features rather than merely through astrophysical correlations of the stellar labels.

We verified our results through cross-validation of the LAMOST--GALAH and LAMOST--APOGEE common stars as well as repeat observations of the same stars in LAMOST. We demonstrate that, even at low resolution, we can attain precise stellar parameters at low S/N. In particular, we estimated that at ${\rm S/N}_{\rm pix} = 20$, the typical internal uncertainties of the $DD$--$Payne$ estimates are about 60\,K in $T_{\rm eff}$ and 0.1\,dex in $\log g$. Our study also shows that, while obtaining reliable elemental abundances remains challenging at low S/N for low-resolution spectra, precise abundances are possible at S/N$\,\simeq\,$50. For ${\rm S/N}_{\rm pix}>50$, we estimated that the typical internal uncertainties of the $DD$--$Payne$ abundances are about 0.05\,dex for [Fe/H], [Mg/Fe], [Ca/Fe], [Ti/Fe], [Cr/Fe], and [Ni/Fe],  0.1\,dex for [C/Fe], [N/Fe], [O/Fe], [Na/Fe], [Al/Fe], [Si/Fe], [Mn/Fe], and [Co/Fe], and 0.2--0.3\,dex for [Cu/Fe] and [Ba/Fe]. Alongside the measurements, we also provide the uncertainty estimates for individual stellar labels of all stars in the catalog.

Due to the data-driven nature of the approach, we demonstrate that our sample inherits the systematic errors of the training set. We examine the potential systematic errors by comparing the $DD$--$Payne$ results derived from the two independent training sets (GALAH and APOGEE). 
In light of this limitation, we recommend a set of stellar labels that have smaller systematic errors and combine the results from the LAMOST--GALAH and the LAMOST--APOGEE training sets. Several flags and quantities are also provided to indicate the quality of the label determinations, which includes the quality of the spectral fits and the consistency of gradient spectra between the $DD$--$Payne$ and the Kurucz models. We also provide a flag to identify  binary/multiple systems from this huge LAMOST stellar sample.

Finally, our method is general and can be applied to other surveys. We demonstrated that it is possible to obtain precise elemental abundances from low-resolution spectra. Obtaining high-resolution spectra is often prohibitively expensive for the Galactic halo as well as dwarf galaxies around the  Milky Way. With LSST on the horizon, fully characterizing the halo and dwarf galaxies will only become more important in the near future. Our method paves the way of doing so by maximally extracting information from low-resolution spectra, like those will soon be collected by DESI.

\vspace{7mm} \noindent {\bf Acknowledgments}{
This work is based on data acquired through the Guoshoujing Telescope. Guoshoujing Telescope (the Large Sky Area Multi-Object Fiber Spectroscopic Telescope; LAMOST) is a National Major Scientific Project built by the Chinese Academy of Sciences. Funding for the project has been provided by the National Development and Reform Commission. LAMOST is operated and managed by the National Astronomical Observatories, Chinese Academy of Sciences.

This work has made use of the data from the GALAH and APOGEE (SDSS-IV) surveys. The GALAH data are acquired through the Australian Astronomical Observatory,
under programmes: A/2013B/13 (The GALAH pilot survey); A/2014A/25, A/2015A/19, and A2017A/18 (The GALAH survey). Funding for the Sloan Digital Sky Survey IV has been provided by the Alfred P. Sloan Foundation, the U.S. Department of Energy, Office of Science, and the participating institutions. 

This work has also made use of data from the European Space Agency (ESA) mission Gaia, processed by the Gaia Data Processing and Analysis Consortium (DPAC). Funding for the DPAC has been provided by national institutions, in particular the institutions participating in the Gaia Multilateral Agreement.

The authors thank A. M. Amarsi for useful discussions. We also thank Kah Fee Ng for his careful reading of the manuscript. YST is supported by the NASA Hubble Fellowship grant HST-HF2-51425.001.}

\bibliographystyle{mn2e}
\bibliography{reference.bib}

\appendix

\section{Gaia$+$Isochrone Calibration of the Training Labels $T_{\rm eff}$ and $\log g$} \label{appendixA}
Information beyond the normalized spectra, such as Gaia parallaxes and multi-band photometry can improve the accuracy and precision of basic stellar parameters. However, Gaia parallax information was not incorporated in GALAH DR2. Therefore, we decided to recalibrate $T_{\rm eff}$ and $\log g$ values of our training stars by making full use of this extra information with a Bayesian approach.  Note that we only recalibrate the $T_{\rm eff}$ and $\log g$ of the training stars, leaving the training abundances unchanged from GALAH DR2 and APOGEE--$Payne$. As such, the abundances still inherit any systematic patterns of abundances in GALAH DR2 and APOGEE--$Payne$. The full results for the whole LAMOST $\times$ Gaia catalog will be presented in future studies, but in the following we will briefly describe the method.

We adopt the spectroscopic $T_{\rm eff}$ and $\log g$, the multi-band photometry in $V$, $g$, $r$, $i$, $G$, $BP$, $RP$, $J$, $H$, $K_{\rm s}$, $W1$ and $W2$ as well as the Gaia DR2 parallax as observables when generating a likelihood function. Here the $g$, $r$ and $i$ photometries are a combination of the Xuyi Schmidt Telescope Photometric Survey of the Galactic Anticentre \citep[XSTPS-GAC;][]{Liu2014, Zhang2014, Yuan2015}, which is the LAMOST input catalog \citep{Liu2014, Yuan2015}, the Sloan Digital Sky Survey \citep[SDSS;][]{York2000, Abazajian2009, Alam2015} for high Galactic latitudes, and the AAVSO Photometric All-Sky Survey \citep[APASS;][]{Munari2014} for bright stars. The $V$-band photometry is also from the APASS survey. The $G$, $BP$ and $RP$ photometries are from Gaia DR2 \citep{Brown2018, Evans2018}. The $J$, $H$ and $K_{\rm s}$ photometries are from the Two Micron All Sky Survey \citep[2MASS;][]{Skrutskie2006}, and the $W1$ and $W2$ photometries are from the Wide-field Infrared Survey Explorer \citep[WISE;][]{Wright2010}. We adopt the spectroscopic [Fe/H] and [$\alpha$/Fe] as a prior. 

We sample the stellar age, initial mass, [Fe/H] and [$\alpha$/Fe] (which can then be converted into a corrected $T_{\rm eff}$ and $\log g$), and use stellar isochrones to convert them into stellar observables. More precisely, given the observation $\mathbf{x}$, the posterior probability of $\mathbf{\theta}$ can be sampled via
\begin{equation}
 P(\theta | x) \propto P(x | \theta) \times P(\theta),
\end{equation}
where $\mathbf{\theta}$ represents the fundamental stellar parameters age, initial mass, [Fe/H] and [$\alpha$/Fe].
$P(x | \theta)$ is the likelihood function of the observations which are assumed to have Gaussian uncertainties. We have
\begin{equation}
 P(x | \theta)  \propto \Pi_{i=1}^N {\rm exp}\left(-\frac{(\mathbf{x} - \mathbf{s})^2}{2\sigma_{\mathbf{x}}^2}\right),
\end{equation}
where $N$ is the total number of observables, $\mathbf{s}$ is the observable prediction from stellar isochrones given by $\mathbf{\theta}$, $\sigma_\mathbf{x}$ is the measurement uncertainties of the observables. 

For uncertainties of spectroscopic stellar parameters, we adopt the reporeted values in GALAH DR2 for the LAMOST--GALAH training set. For APOGEE--$Payne$, we assume an uncertainty of 90\,K in $T_{\rm eff}$, 0.1\,dex in $\log g$ and 0.1\,dex in [Fe/H], which is based on their external validation \citep{Ting2019}. Considering that the synthetic magnitudes in the isochrones could be systematically deviated from the photometric magnitudes of the surveys due to uncertainties of the inferred passbands \citep[e.g.,][]{Evans2018}, the parameter estimation may suffer from systematic errors that cannot be well accounted for by using only the reported uncertainty of the photometric magnitudes. We therefore assign an uncertainty of 0.02\,mag to all the synthetic magnitudes of the isochrones to reduce possible systematic errors of the parameter estimation. This is especially important for the Gaia photometric bands because even a small mismatch between the synthetic and photometric magnitudes may cause large systematics in the posterior distribution owing to the very small errors in the Gaia photometry. We have adopted a global zero-point correction of 29\,$\mu$as to the Gaia parallax \citep{Lindegren2018}. There are evidence showing that the zero point is a function of magnitude, color, and sky position \citep{Lindegren2018, Luri2018, Leung2019b}. Considering that our training stars are very bright ($G<14$\,mag), a global zero-point correction seems to be plausible.  

For the prior, $P(\theta)$, we adopt a Kroupa IMF mass prior, a flat metallicity prior, and an age prior assuming the star formation history derived in \citet{Xiang2018}. As for stellar isochrones, we adopt the Dartmouth Stellar evolution database \citep[DESP;][]{Dotter2008} for dwarfs ($\log g>3.6$) as it covers a wide range of stellar parameter space. In particular, it provides a model grid with different [$\alpha$/Fe] from $-0.2$ to 0.8\,dex. We found the $\alpha$-enhancement could have an impact of $\sim$100\,K on $T_{\rm eff}$ estimation for main-sequence turnoff stars, and thus should not be ignored. For giant stars ($\log g\leq3.6$ and $T_{\rm eff}<5600$\,K), we adopt the PARSEC isochrones \citep{Bressan2012} because they include also $H_{\rm e}$-burning sequences and later evolutionary stages. Note that for stars with solar [$\alpha$/Fe], we found the DESP and PARSEC isochrones give consistent $T_{\rm eff}$  for dwarf stars, with difference smaller than 20\,K. A mixture of $T_{\rm eff}$ from the two sets of isochrones does not cause significant discontinuity in the $T_{\rm eff}$--$\log g$ diagram. 

The multi-band photometric data are dereddened using $E_{B-V}$ estimated through the spectroscopy-based star pair method \citep{Yuan2013, Yuan2015}. Stars with the same stellar parameters should have the same colors \citep{Yuan2015b}. On this basis, we select a control sample constitutes of high latitude stars that have well-known $E_{B-V}$. This sample allows us to build a non-parametric model between stellar parameters and multi-band (dereddened) colors \citep{Yuan2013, Yuan2015}, which can then be used to estimate the extinction for other stars that are in more extincted region.
The $E_{B-V}$ are then converted to extinction for different photometric bands using a $T_{\rm eff}$, $\log g$, and [Fe/H]-dependent extinction coefficients that are based on the Fitzpatrick extinction curve \citep{Fitzpatrick1999} and the Kurucz model spectra \citep{Castelli2003}. 

\begin{figure*}
\centering
\includegraphics[width=180mm]{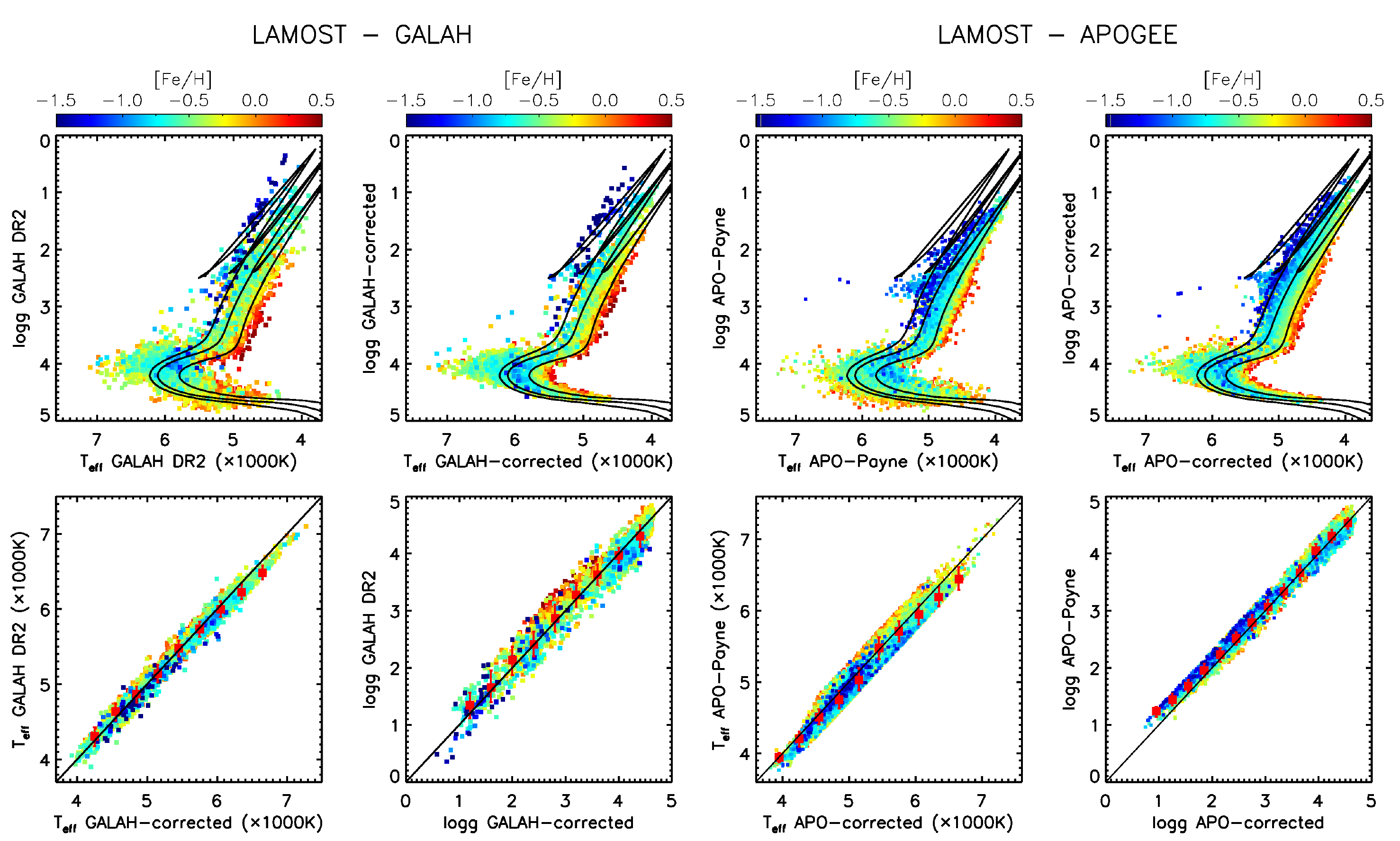}
\caption{Comparison of $T_{\rm eff}$ and $\log g$ for the GALAH DR2 ($left$ panels) and APOGEE--$Payne$ ($right$ panels) catalog values with the corrected values used in this work. The correction adopts a Bayesian framework with $Gaia$ parallax and multi-band photometry as extra constraints. In the top panels, PARSEC isochrones of ([Fe/H] = 0, $\tau$ = 10\,Gyr), ([Fe/H] = $-0.5$, $\tau$ = 10\,Gyr), and ([Fe/H] = $-1.0$, $\tau$ = 12\,Gyr) are shown. In the bottom panels, the red symbols indicate the median and standard deviations in each $T_{\rm eff}$/$\log g$ bin. The corrected values generally fit better with the isochrones, especially at the metal-poor end.}
\label{fig:Fig18}
\end{figure*}

Fig.\,\ref{fig:Fig18} shows the comparisons of $T_{\rm eff}$ and $\log g$ between the ones derived with our method and that of GALAH DR2 and APOGEE--$Payne$. Overall, the figure demonstrates that our results agree better with stellar isochrones. The GALAH DR2 $T_{\rm eff}$ values are consistent with ours at the cool end ($\lesssim$6000\,K) with a dispersion of 60\,K. However, for stars with $T_{\rm eff}$ higher than 6000\,K, there is a non-negligible systematic trend --- the GALAH DR2 values are lower than ours, and the difference reaches 100--200\,K at $T_{\rm eff}\sim7000$\,K. At the same time, the GALAH DR2 $\log g$ values are lower than our estimates, indicating that the systematics are at least partially caused by the $T_{\rm eff}$--$\log g$ degeneracy. Similarly, for giant stars with super-solar metallicity, $T_{\rm eff}$ from GALAH DR2 is $\sim$100\,K higher than ours, and $\log g$ is 0.3--0.5\,dex higher as well. All these biases suggest that corrections, as presented in this section, are necessary to bring the training set stellar parameters to the same scale as the stellar isochrones. 

The figure shows that the APOGEE--$Payne$ stellar parameters for both dwarfs and giants are consistent with stellar isochrones. This is encouraging considering that the APOGEE--$Payne$ values are directly from the infrared APOGEE spectra without any external calibration \citep{Ting2019}. Nevertheless, there are still several noticeable systematic differences between the APOGEE--$Payne$ results and ours. The most prominent difference shows up for the metal-poor ${\rm [Fe/H]}<-1.0$ stars. For metal-poor giant stars, our results show lower values by up to 0.5\,dex for $\log g$. Nonetheless, the two $T_{\rm eff}$ are consistent with each other. For metal-poor dwarf stars, our $T_{\rm eff}$ is higher than the APOGEE--$Payne$ results by $\sim$300\,K, and our $\log g$ is slightly higher by $\sim$0.1\,dex. Another noticeable difference is the $T_{\rm eff}$ for relatively hot stars ($6000<T_{\rm eff}<6700$\,K). The APOGEE--$Payne$ $T_{\rm eff}$ is systematically lower than ours. At $T_{\rm eff}\sim6500$\,K, the median difference reaches $\sim$200\,K, with a significant ($\sim200$\,K) dispersion. At $T_{\rm eff}\sim7000$\,K, the temperature difference subsides. For metal-rich giant stars, we observe a $\sim$100\,K difference between our $T_{\rm eff}$ estimates and the APOGEE--$Payne$ results, There is also a moderate difference in $\log g$. We emphasize that although we expect our $\log g$ to be more accurate because the APOGEE--$Payne$ values, in some cases, deviate quite a bit from the isochrones, the same cannot be said for $T_{\rm eff}$. Our $\log g$ is more consistent with the stellar isochrones because we explicitly tied our stellar parameters to the stellar isochrone scale. In terms of $T_{\rm eff}$, there are still mismatches at the level of $\sim$100\,K between different temperature scales, such as the widely used IRFM method \citep{Casagrande2010} and the interferometry-based method \citep{Huang2015}. It has also been suggested that the mismatch is also metallicity dependent \citep[e.g.][]{Xiang2017}.

\section{$DD$--$Payne$ gradient spectra} \label{appendixB}
In this section, we will show some examples of gradient spectra predicted by the $DD$--$Payne$ of our LAMOST sample to supplement Fig.\,\ref{fig:Fig1}. Figs.\,\ref{fig:Fig19}--\ref{fig:Fig21} show examples in which the $DD$--$Payne$ gradient spectra are consistent with the Kurucz gradient spectra, with a correlation coefficient of $\sim$0.9. Figs.\,\ref{fig:Fig22}--\ref{fig:Fig24} illustrate cases where the correlation coefficient is weaker, $\sim$0.6. We note that for individual stars, the correlation could be different for individual elements (see Fig.\,\ref{fig:Fig2} in the main text). For example, in Figs.\,\ref{fig:Fig19}--\ref{fig:Fig21}, the $DD$--$Payne$ reproduce the Kurucz gradient spectra well for C, N, O, Na, Mg, Al, Si, Ca, Ti, Cr, Mn, Co, Ni, Cu, and Ba. The agreement for Sc, V, Zn, Y, and Eu is weaker, but there is still a decent agreement. Figs.\,\ref{fig:Fig22}--\ref{fig:Fig24} further suggest that even with a correlation coefficient of $\sim$0.6, most of the features in the Kurucz gradient spectra are properly reproduced by the $DD$--$Payne$. This has prompted our choice of making the correlation cut at 0.5, above which we deem the elemental abundances to be physically determined. 
\begin{figure*}
\centering
\includegraphics[width=180mm]{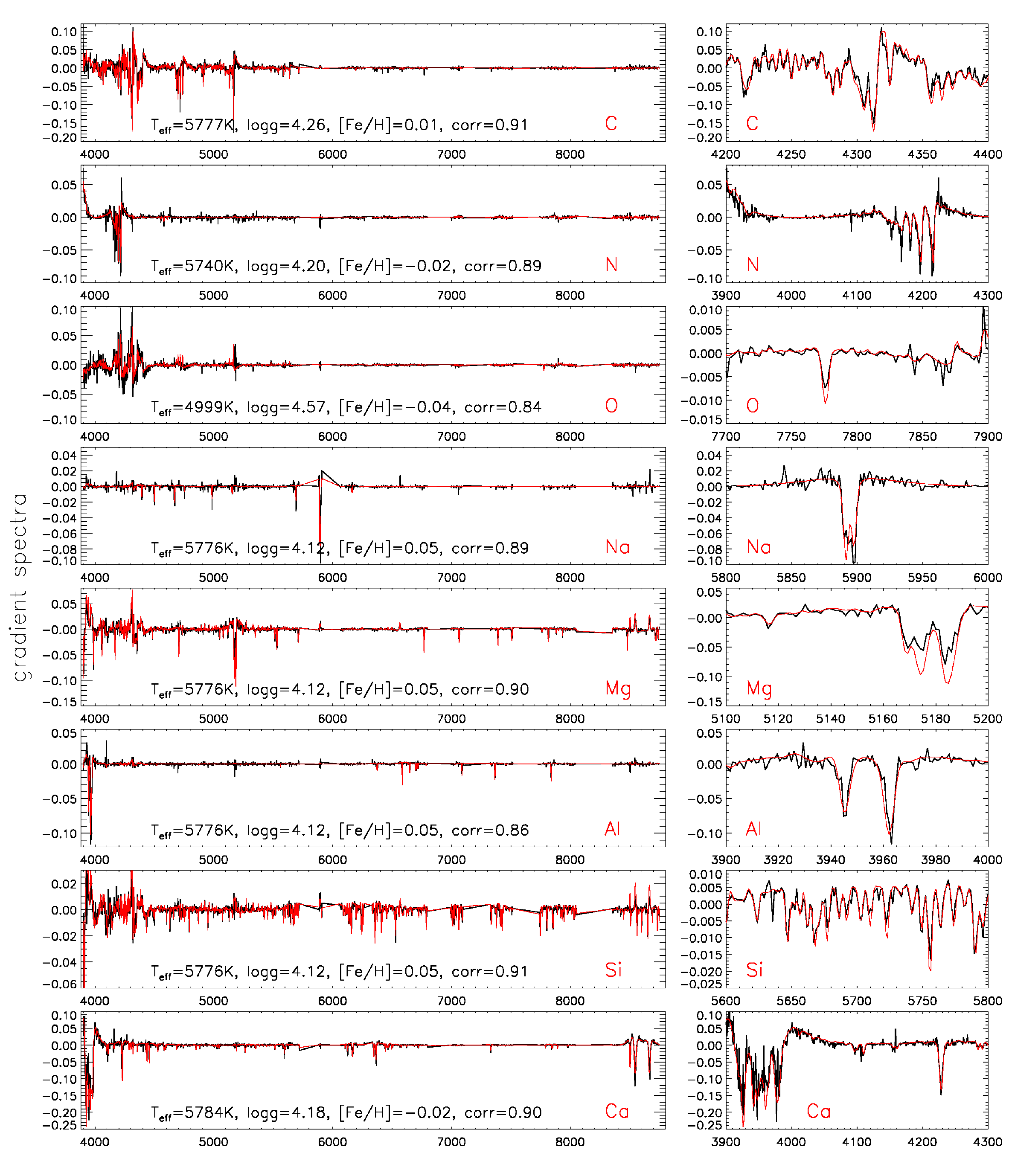}
\caption{Comparison of the $DD$--$Payne$ empirical gradient spectra (black) to the {\it ab initio} calculations from the Kurucz models (red). The {\em left} panels show the overview of the full optical range, and the {\em right} panels zoom in on some of the most prominent features for each element. We present the result for a star of which the two gradient spectra have correlation coefficients of $\sim$0.9 for all elements. For the Kurucz models, we adopt a reference star in Table\,\ref{table:table1} that has the closest distance in the stellar parameter space
to the observed star. In each panel, the stellar parameters of the star, as well as the correlation coefficients between the two gradient spectra, are marked.}
\label{fig:Fig19}
\end{figure*}

\begin{figure*}
\centering
\includegraphics[width=180mm]{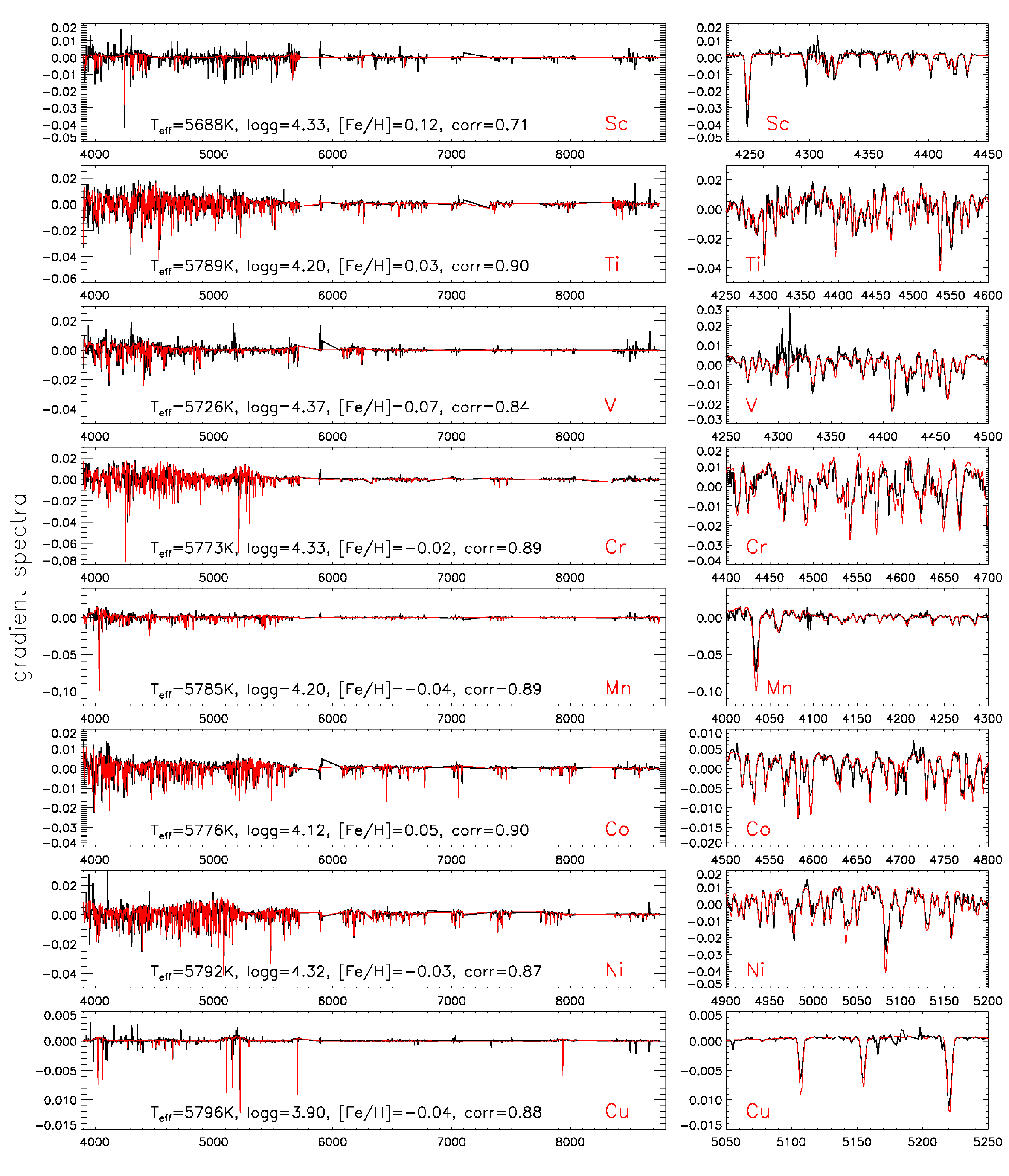}
\caption{Continuation for Fig.\,19.}
\label{fig:Fig20}
\end{figure*}

\begin{figure*}
\centering
\includegraphics[width=180mm]{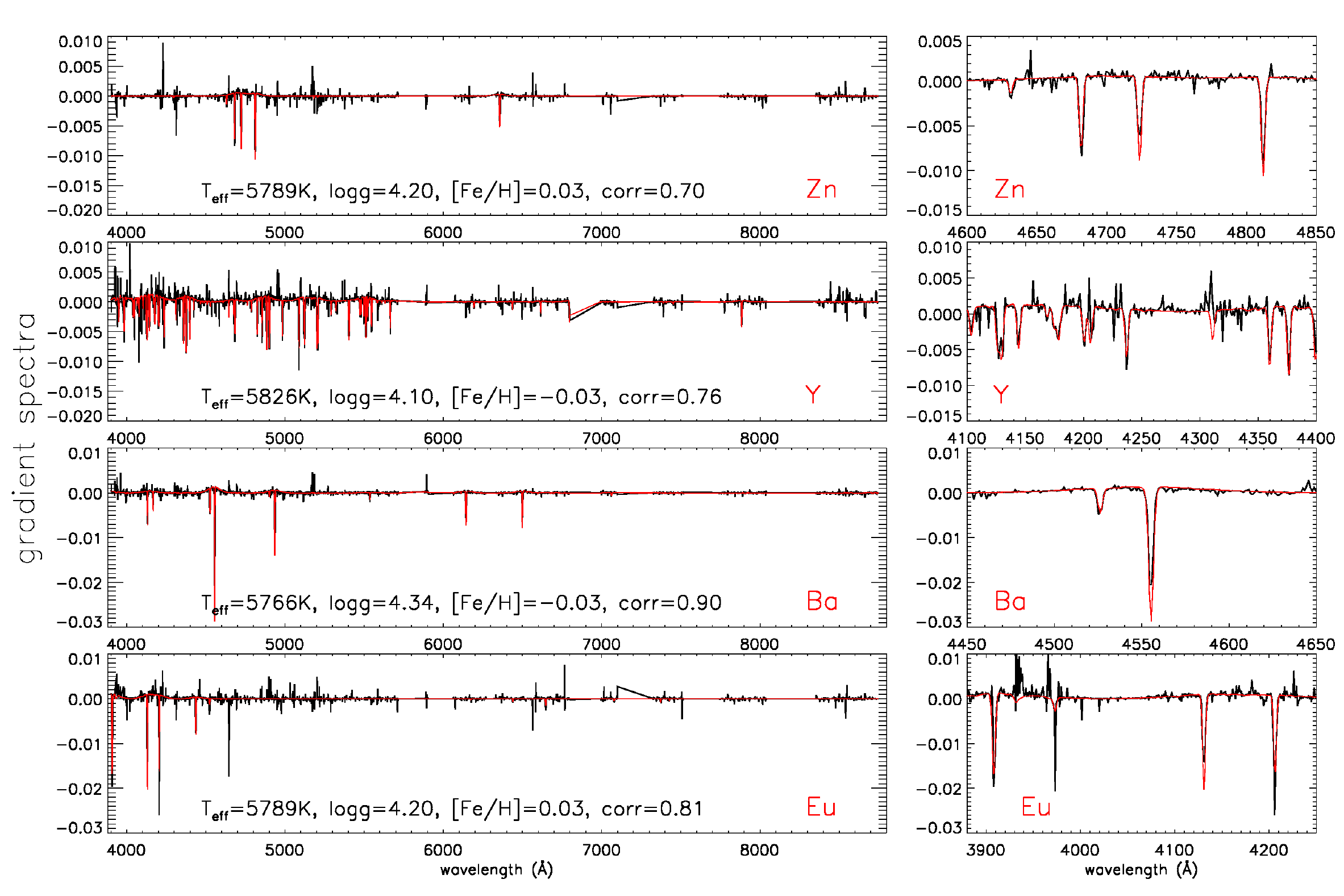}
\caption{Continuation for Fig.\,19.}
\label{fig:Fig21}
\end{figure*}

%%\appendix
\begin{figure*}
\centering
\includegraphics[width=180mm]{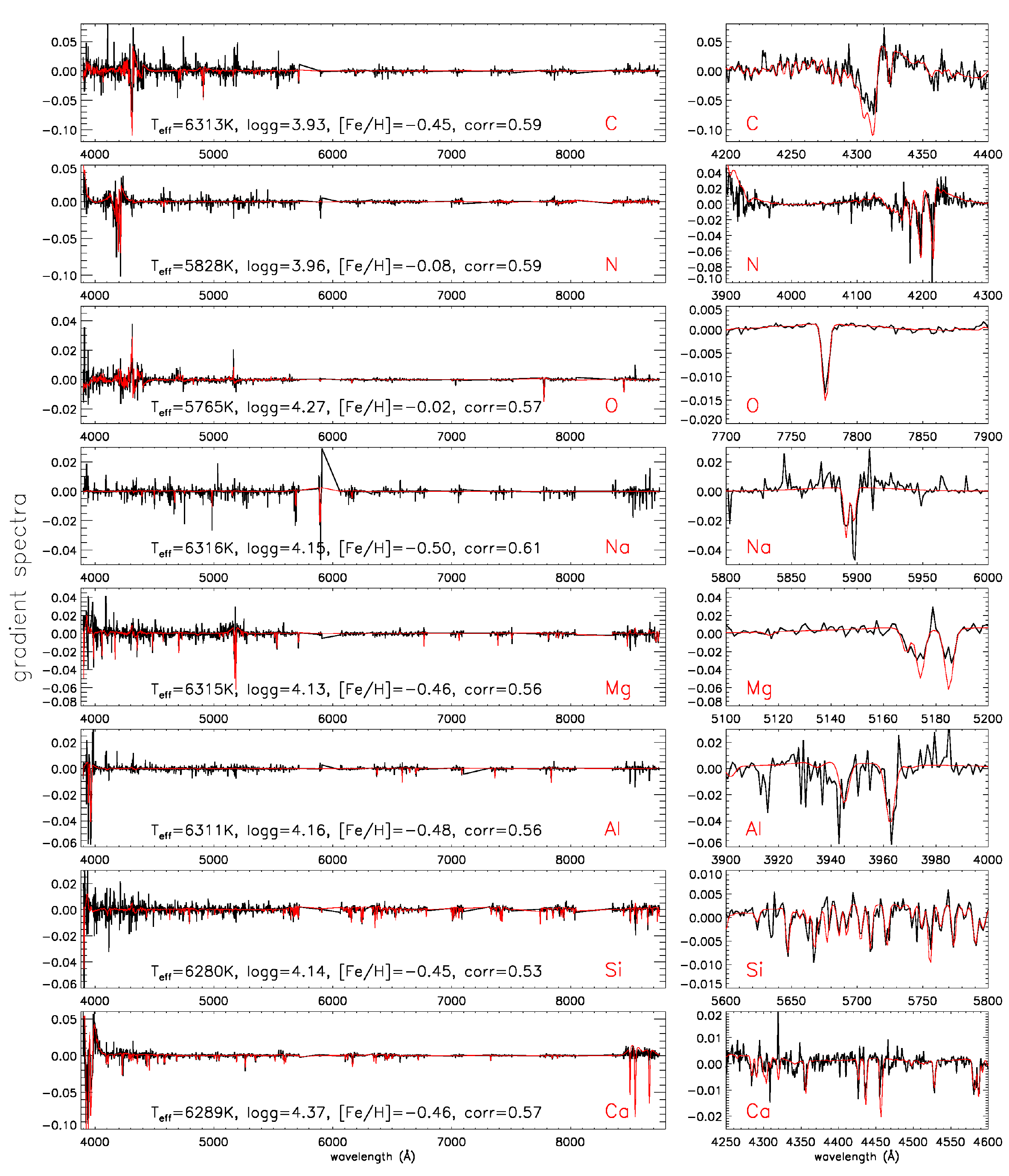}
\caption{Similar to Fig.\,19, but here we show the result for a star that has weaker correlation coefficients ($\sim$0.6). This is a borderline case in our flagging procedure -- above this correlation, we deem the determination of the elemental abundances is reliable, in the sense that $DD$-$Payne$ draws information from physical spectral features instead of astrophysical correlations.}
\label{fig:Fig22}
\end{figure*}

\begin{figure*}
\centering
\includegraphics[width=180mm]{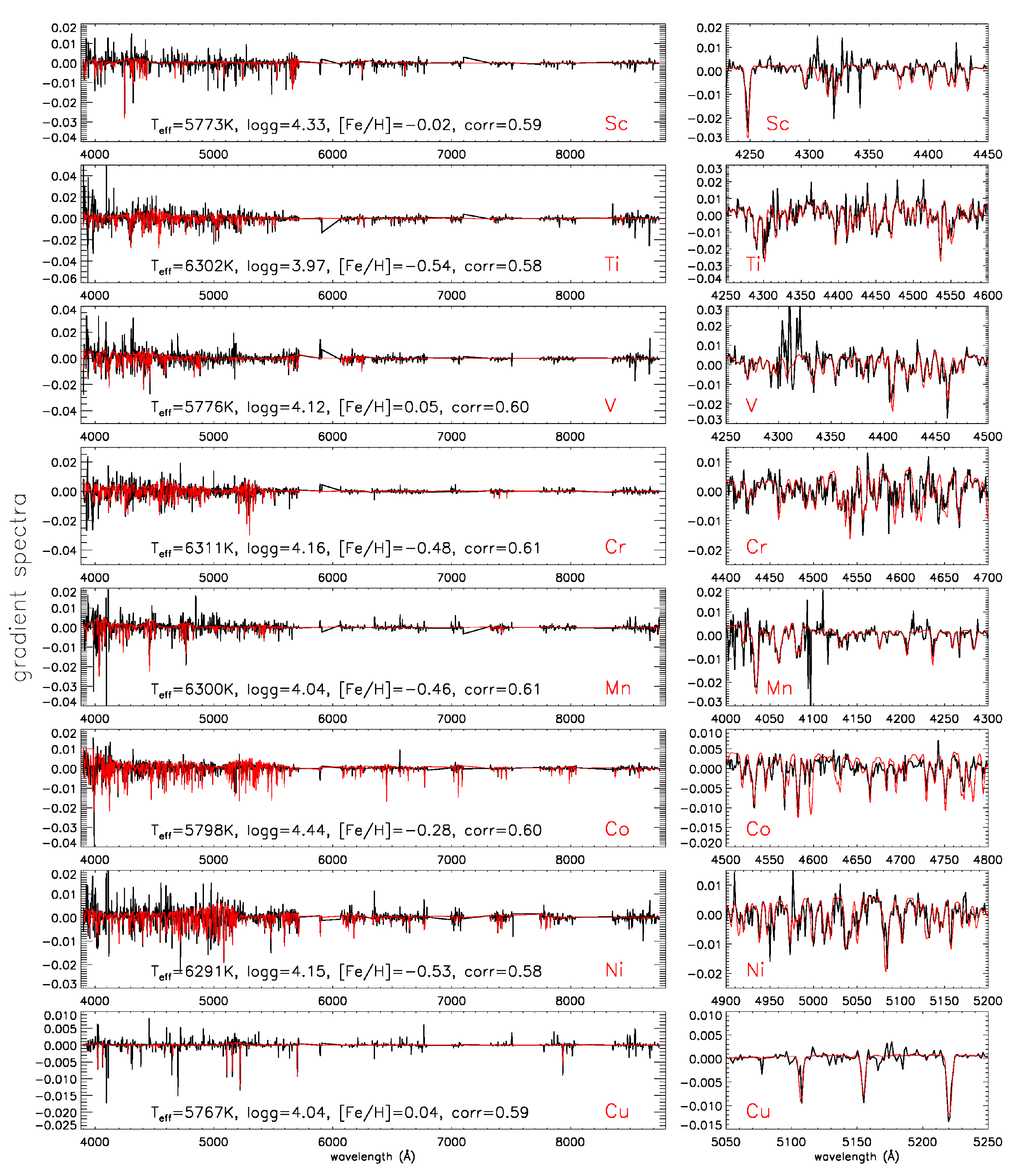}
\caption{Continuation for Fig.\,22.}
\label{fig:Fig23}
\end{figure*}

\begin{figure*}
\centering
\includegraphics[width=180mm]{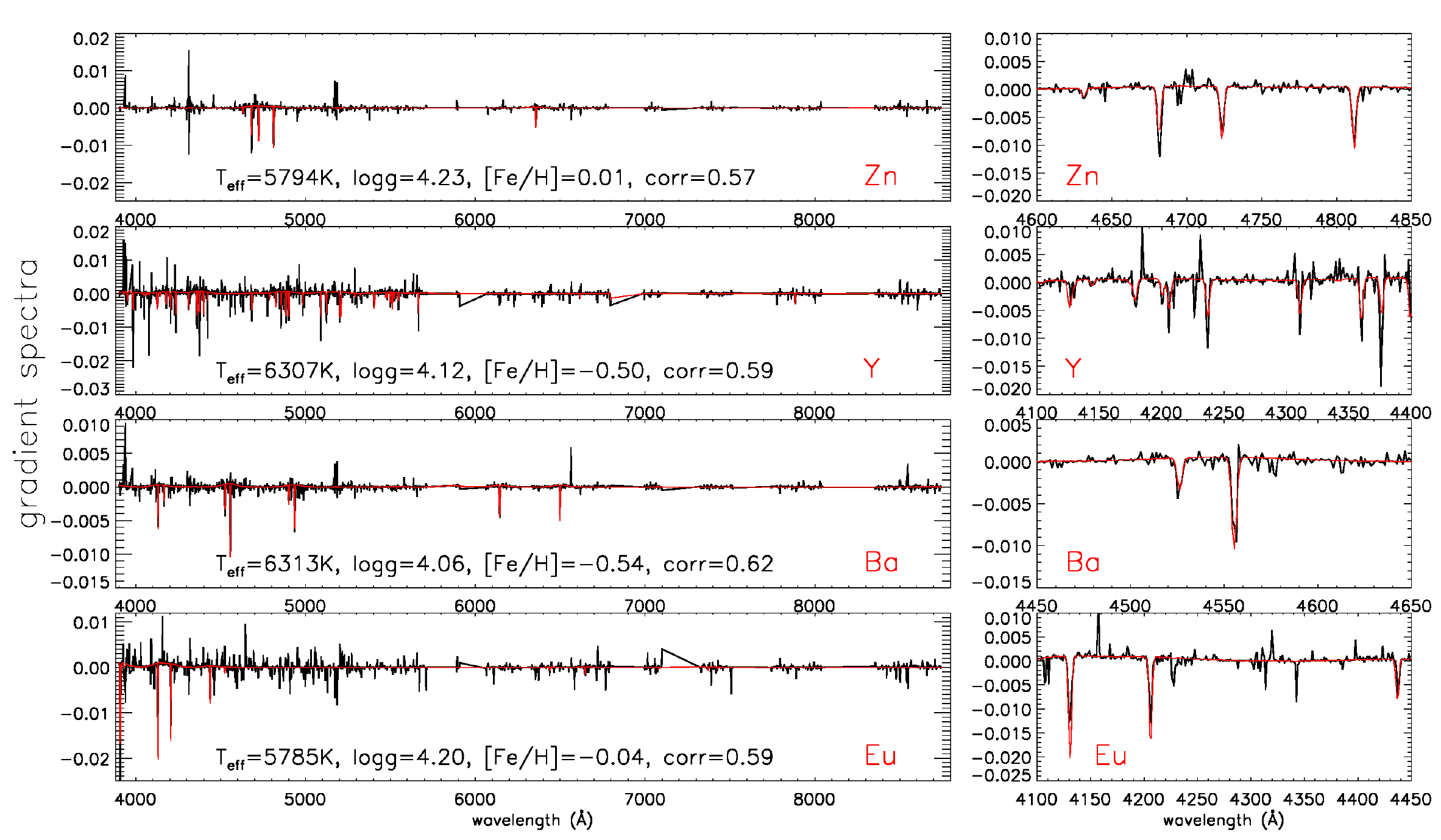}
\caption{Continuation for Fig.\,22.}
\label{fig:Fig24}
\end{figure*}

\section{Comparison of GALAH DR2 and APOGEE--$Payne$ abundances} \label{appendixC}
\begin{figure}
\centering
\includegraphics[width=160mm]{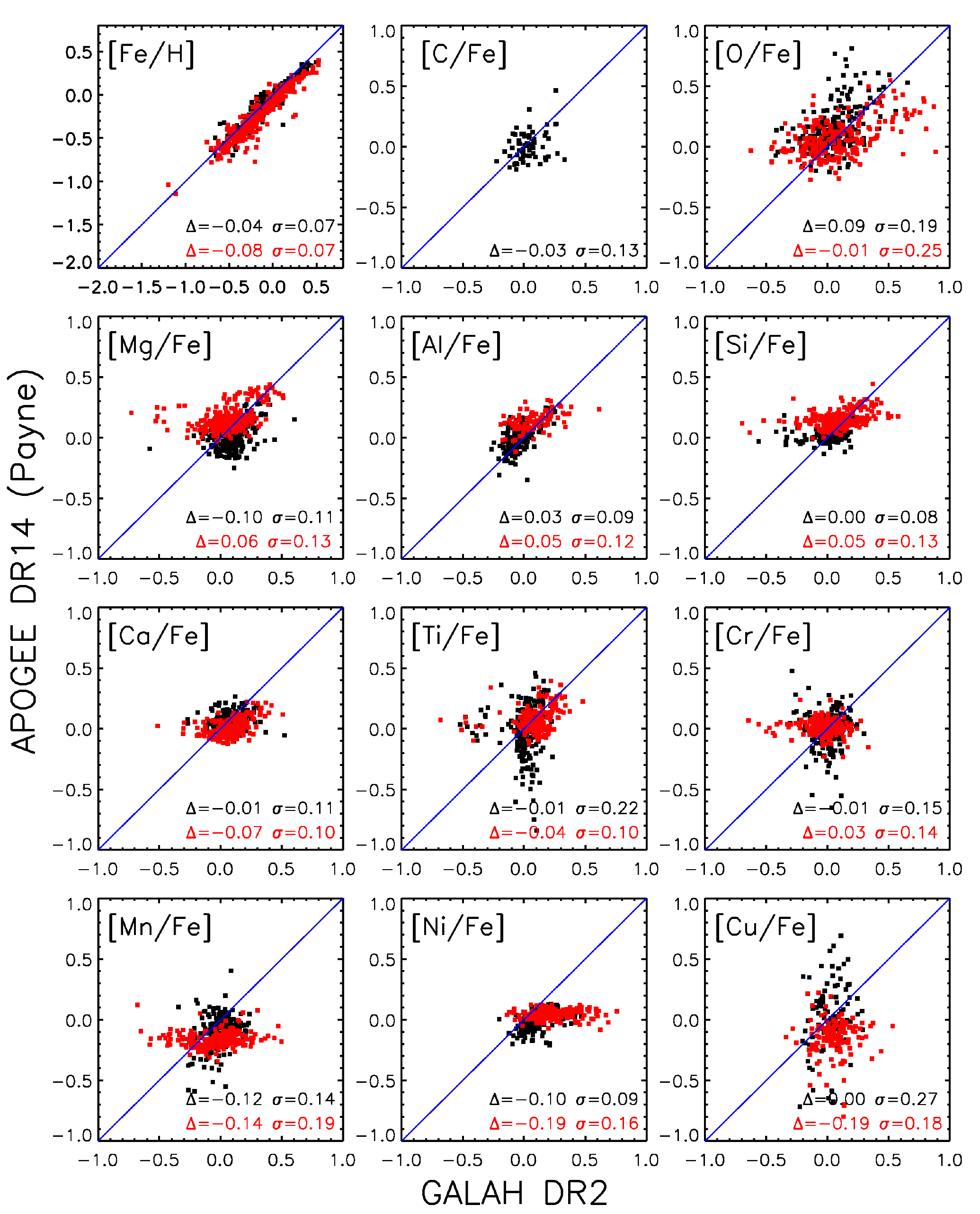}
\caption{Comparison of the GALAH DR2 and APOGEE-$Payne$ abundances for the common stars between GALAH and APOGEE. Dwarfs (black) and giants (red) are shown in different colors. The numbers in each panel mark the median and standard deviation. We select only stars with reliable GALAH (flag $=$ 0) and APOGEE--$Payne$ (quality flag $=$ ``good'') values. The figure demonstrates that, for a few elements, GALAH and APOGEE-Payne can have non-negligible systematic differences. Such systematic patterns are the root of the deviations as seen in Fig.\,\ref{fig:Fig9} -- when adopting different training sets, the $DD$--$Payne$ can yield different results. Thus careful choice has to be made for each element.} 
\label{fig:Fig25}
\end{figure}

\begin{figure}
\centering
\includegraphics[width=160mm]{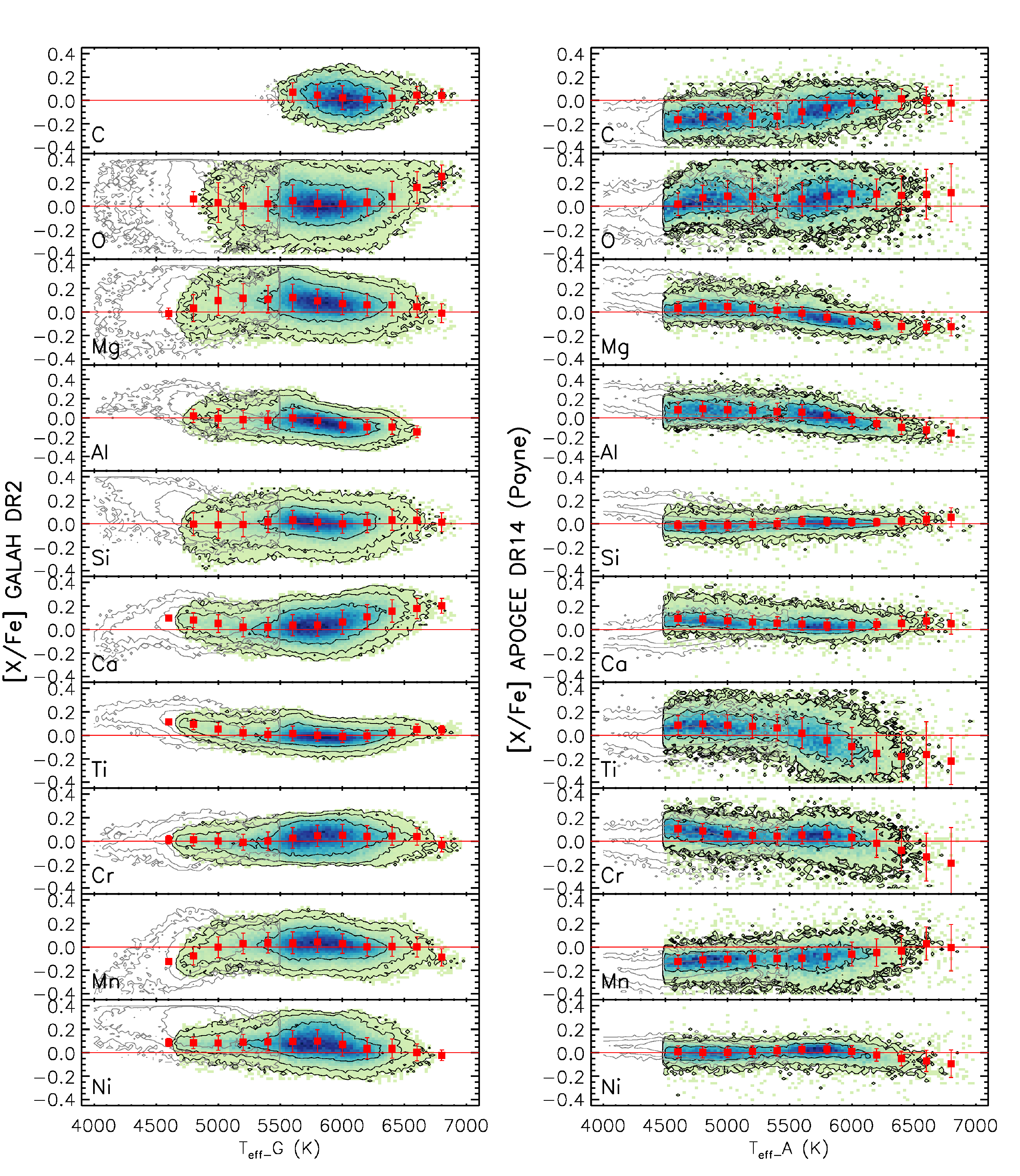}
\caption{GALAH DR2 and APOGEE--$Payne$ abundances as a function of $T_{\rm eff}$ for stars with solar metallicity ($-0.2<{\rm [Fe/H]}<0.2$). We select only stars with reliable GALAH (flag == 0) and APOGEE--$Payne$ (quality flag == ``good'') determination. The figure demonstrates that, for some elements, the GALAH and APOGEE--$Payne$ can have non-negligible residual $T_{\rm eff}$-abundance trend, which subsequently causes our $DD$--$Payne$ LAMOST values to inherit a similar $T_{\rm eff}$-abundance trend (see Fig.\,10 in the main text).} 
\label{fig:Fig26}
\end{figure}

We have shown in this study that, depending on the choice of the training set, the derived LAMOST elemental abundances could be different. This leaves the question if the $DD$--$Payne$ is introducing unexpected systematics. Fig.\,\ref{fig:Fig25} shows a comparison of GALAH DR2 and APOGEE--$Payne$ abundances. It demonstrates that there is noticeable systematic offsets and trends for several elements between these two surveys, such as Fe, Mg, Mn, and Ni. More importantly, these patterns are consistent with the systematics shown between the LAMOST abundances derived from the LAMOST--GALAH and LAMOST--APOGEE training sets (Fig.\,\ref{fig:Fig12}). It demonstrates that the LAMOST $DD$--$Payne$ abundances simply inherit systematic errors of the training sets, and the $DD$--$Payne$ method itself does not introduce the systematics shown in the main text. Similarly, in Fig.\,\ref{fig:Fig26}, we show the GALAH DR2 and APOGEE--$Payne$ abundances as a function of $T_{\rm eff}$. The trends are also consistent with those of the LAMOST $DD$--$Payne$ results, demonstrating that the $T_{\rm eff}$-abundance systematics as shown in the main text are also inherited from the training sets. They are not due to the limitation of our method. 

\end{document}